\documentclass[twocolumn]{aastex62}
\usepackage{hyperref}

\usepackage{graphicx}
\usepackage{color}
\usepackage{float}
\usepackage{ulem}
\usepackage[euler]{textgreek}
\usepackage{amsmath}
\usepackage{ifxetex,ifluatex}
\usepackage{microtype}
\usepackage[english]{babel}
\usepackage{cleveref}
\usepackage{natbib}
\usepackage{amsmath}
\usepackage{lineno}

\usepackage{ulem} 
\DeclareUnicodeCharacter{2212}{-}

\received{2021 April 6}
\revised{2021 December 1}
\accepted{2021 December 2}

\shorttitle{NANOGrav Polarimetry}
\shortauthors{Wahl et al.}

\graphicspath{{./}{figures/}}

\begin{document}

\title{The NANOGrav 12.5-Year Data Set: Polarimetry and Faraday Rotation Measures from Observations of Millisecond Pulsars with the Green Bank Telescope}

\correspondingauthor{H.M. Wahl}
\email{hmw0023@mix.wvu.edu}

\author[0000-0001-9678-0299]{H. M. Wahl}
\affiliation{Department of Physics and Astronomy, West Virginia University, P.O. Box 6315, Morgantown, WV 26506, USA}
\affiliation{Center for Gravitational Waves and Cosmology, West Virginia University, Chestnut Ridge Research Building, Morgantown, WV 26505, USA}

\author[0000-0001-7697-7422]{M.A. McLaughlin}
\affiliation{Department of Physics and Astronomy, West Virginia University, P.O. Box 6315, Morgantown, WV 26506, USA}
\affiliation{Center for Gravitational Waves and Cosmology, West Virginia University, Chestnut Ridge Research Building, Morgantown, WV 26505, USA}

\author{P. A. Gentile}
\affiliation{Department of Physics and Astronomy, West Virginia University, P.O. Box 6315, Morgantown, WV 26506, USA}
\affiliation{Center for Gravitational Waves and Cosmology, West Virginia University, Chestnut Ridge Research Building, Morgantown, WV 26505, USA}

\author[0000-0001-6607-3710]{M.\,L.\,Jones}
\affiliation{Center for Gravitation, Cosmology and Astrophysics, Department of Physics, University of Wisconsin-Milwaukee, P.O. Box 413, Milwaukee, WI 53201, USA}

\author[0000-0002-6730-3298]{R.\,Spiewak}
\affiliation{Jodrell Bank Centre for Astrophysics, Department of Physics and Astronomy, University of Manchester, Manchester M13 9PL, UK}
\affiliation{Centre for Astrophysics and Supercomputing, Swinburne University of Technology, P.O. Box 218, Hawthorn, Victoria 3122, Australia}

\author{Z.\,Arzoumanian}
\affiliation{X-Ray Astrophysics Laboratory, NASA Goddard Space Flight Center, Code 662, Greenbelt, MD 20771, USA}

\author{K.\,Crowter}
\affiliation{Department of Physics and Astronomy, University of British Columbia, 6224 Agricultural Road, Vancouver, BC V6T 1Z1, Canada}

\author{P.\,B.\,Demorest}
\affiliation{National Radio Astronomy Observatory, 1003 Lopezville Rd., Socorro, NM 87801, USA}

\author{M.\,E.\,DeCesar}
\affiliation{Department of Physics, Lafayette College, Easton, PA 18042, USA}
\affiliation{AAAS, STPF, ORISE Fellow hosted by the U.S. Department of Energy}

\author[0000-0001-8885-6388]{T.\,Dolch}
\affiliation{Department of Physics, Hillsdale College, 33 E. College Street, Hillsdale, MI 49242, USA}
\affiliation{Eureka Scientific, Inc.  2452 Delmer Street, Suite 100, Oakland, CA 94602-3017}

\author{J.\,A.\,Ellis}
\affiliation{Infinia ML, 202 Rigsbee Avenue, Durham NC, 27701}

\author{R.\,D.\,Ferdman}
\affiliation{School of Chemistry, University of East Anglia, Norwich, NR4 7TJ, United Kingdom}

\author{E.\,C.\,Ferrara}
\affiliation{NASA Goddard Space Flight Center, Greenbelt, MD 20771, USA}

\author[0000-0001-8384-5049]{E.\,Fonseca}
\affiliation{Department of Physics, McGill University, 3600 University St., Montreal, QC H3A 2T8, Canada}

\author{N.\ Garver-Daniels}
\affiliation{Department of Physics and Astronomy, West Virginia University, P.O. Box 6315, Morgantown, WV 26506, USA}
\affiliation{Center for Gravitational Waves and Cosmology, West Virginia University, Chestnut Ridge Research Building, Morgantown, WV 26505, USA}

\author{G.\, Jones}
\affiliation{Department of Astronomy, Columbia University, 550 W. 120th St. New York, NY 10027, USA}

\author[0000-0003-0721-651X]{M.\,T.\,Lam}
\affiliation{School of Physics and Astronomy, Rochester Institute of Technology, Rochester, NY 14623, USA}
\affiliation{Laboratory for Multiwavelength Astrophysics, Rochester Institute of Technology, Rochester, NY 14623, USA}

\author{L.\,Levin}
\affiliation{Jodrell Bank Centre for Astrophysics, School of Physics and Astronomy, The University of Manchester, Manchester M13 9PL, UK}

\author[0000-0003-0771-6581]{N.\,Lewandowska}
\affiliation{Department of Physics and Astronomy, West Virginia University, P.O. Box 6315, Morgantown, WV 26506, USA}
\affiliation{Center for Gravitational Waves and Cosmology, West Virginia University, Chestnut Ridge Research Building, Morgantown, WV 26505, USA}

\author{D.\,R.\,Lorimer}
\affiliation{Department of Physics and Astronomy, West Virginia University, P.O. Box 6315, Morgantown, WV 26506, USA}
\affiliation{Center for Gravitational Waves and Cosmology, West Virginia University, Chestnut Ridge Research Building, Morgantown, WV 26505, USA}

\author{R.\,S.\,Lynch}
\affiliation{Green Bank Observatory, P.O. Box 2, Green Bank, WV 24944, USA}

\author{D.\,R.\,Madison}
\affiliation{Department of Physics and Astronomy, West Virginia University, P.O. Box 6315, Morgantown, WV 26506, USA}
\affiliation{Center for Gravitational Waves and Cosmology, West Virginia University, Chestnut Ridge Research Building, Morgantown, WV 26505, USA}
\affiliation{Department of Physics, University of the Pacific, 3601 Pacific Avenue, Stockton, CA 95211, USA}

\author[0000-0002-3616-5160]{C.\,Ng}
\affiliation{Dunlap Institute for Astronomy and Astrophysics, University of Toronto, 50 St. George St., Toronto, ON M5S 3H4, Canada}

\author[0000-0002-6709-2566]{D.\,J.\,Nice}
\affiliation{Department of Physics, Lafayette College, Easton, PA 18042, USA}

\author[0000-0001-5465-2889]{T.\,T.\,Pennucci}
\altaffiliation{NANOGrav Physics Frontiers Center Postdoctoral Fellow}
\affiliation{National Radio Astronomy Observatory, 520 Edgemont Road, Charlottesville, VA 22903, USA}
\affiliation{Institute of Physics, Eötvös Loránd University, Pázmány P. s. 1/A, 1117 Budapest, Hungary }

\author[0000-0001-5799-9714]{S.\,M.\,Ransom}
\affiliation{National Radio Astronomy Observatory, 520 Edgemont Road, Charlottesville, VA 22903, USA}

\author{P.\ Ray}
\affiliation{Space Science Division, Naval Research Laboratory, Washington, DC 20375-5352, USA}

\author[0000-0001-9784-8670]{I.\,H.\,Stairs}
\affiliation{Department of Physics and Astronomy, University of British Columbia, 6224 Agricultural Road, Vancouver, BC V6T 1Z1, Canada}

\author{K.\,Stovall}
\affiliation{National Radio Astronomy Observatory, 1003 Lopezville Rd., Socorro, NM 87801, USA}

\author{J.\,K.\,Swiggum}
\altaffiliation{NANOGrav Physics Frontiers Center Postdoctoral Fellow}
\affiliation{Department of Physics, Lafayette College, Easton, PA 18042, USA}

\author{W.\,W.\,Zhu}
\affiliation{National Astronomical Observatories, Chinese Academy of Science, 20A Datun Road, Chaoyang District, Beijing 100012, China}

\begin{abstract}
In this work, we present polarization profiles for 23 millisecond pulsars observed at 820\,MHz and 1500\,MHz with the Green Bank Telescope as part of the NANOGrav pulsar timing array. We calibrate the data using Mueller matrix solutions calculated from observations of PSRs B1929+10 and J1022+1001. We discuss the polarization profiles, which can be used to constrain pulsar emission geometry, and present both the first published radio polarization profiles for nine pulsars and the discovery of very low intensity average profile components (``microcomponents") in four pulsars.  Using the Faraday rotation measures, we measure for each pulsar and use it to calculate the Galactic magnetic field parallel to the line of sight for different lines of sight through the interstellar medium. We fit for linear and sinusoidal trends in time in the dispersion measure and Galactic magnetic field and detect magnetic field variations with a period of one year in some pulsars, but overall find that the variations in these parameters are more consistent with a stochastic origin.

\end{abstract}


\section{Introduction} \label{sec:intro}

Pulsars are highly-magnetized, rapidly-rotating neutron stars that emit electromagnetic radiation that sweeps across our line of sight as they rotate. In addition to being laboratories for study themselves, pulsars are useful in probing the properties of the interstellar medium (ISM). As the radio waves from a pulsar traverse the Galaxy, they experience Faraday rotation, which is a frequency-dependent rotation of the polarization position angle by the Galactic magnetic field. Faraday rotation changes the angle of linear polarization by an angle

\begin{equation}\label{1}
\beta=\frac{e^{3} \lambda^{2}}{2 \pi m_{e}^{2} c^{4}} \int_{0}^{d} n_{e}(l) B_{ \|}(l)  d l~, 
\end{equation}

\noindent where \textit{e} is the charge of the electron,  \textit{\textlambda} is the wavelength of the radio waves, \textit{m$_{e}$} is the mass of the electron, \textit{c} is the speed of light, \textit{$n_{e}$} is the free electron density along a line of sight \textit{l}, \textit{d} is the pulsar distance, and \textit{B$_{ \|}$} is the estimate of the electron-density-weighted average (Galactic) magnetic field (in cgs units). The degree to which the pulsar's radio waves are rotated is called the rotation measure (RM), where

\begin{equation}
\textrm{RM} = \frac{\beta}{\lambda^{2}}~.
\label{eq:2}
\end{equation} 

\noindent We can also measure directly from radio observations the dispersion measure (DM), which is the integrated free electron density along the line of sight 

\begin{equation}\label{2}
\text{\rm DM}=\int_{0}^{d} n_{e}(l) dl~
 \end{equation}

\noindent and varies with the observational frequency as $\nu^{-2}$.
We can then calculate the parallel component of the magnetic field along the line of sight using both the DM and RM as

\begin{equation}
\left\langle B_{ \|}\right\rangle=1.23 \frac{\mathrm{RM}}{\mathrm{DM}} \mu \mathrm{G}~,
\label{eq:bparallel}
\end{equation}

\noindent where RM is in rad m$^{-2}$ and DM is in pc cm$^{-3}$. 

When the radio waves reach the receiver on the telescope, the telescope's response alters the components of the waves. These components can be described by the Stokes vector

\begin{equation}
S=\left[\begin{array}{l}{I} \\ {Q} \\ {U} \\ {V}\end{array}\right] ~,
\end{equation}

\noindent where Stokes \textit{I} is the total intensity, Stokes \textit{Q} and Stokes \textit{U} form the linear polarization $L=\sqrt{Q^{2}+U^{2}}$, and Stokes \textit{V} is the circular polarization intensity. Using the International Astronomical Union (IAU)'s circular polarization sign convention, right-handed circular polarization is positive (corresponding to a clockwise rotation of the position angle) and left-handed circular polarization is negative (corresponding to a counterclockwise rotation of the position angle) \citep{Stokes}. The total amount of polarized emission can be described by the latter three Stokes parameters, $P=\sqrt{L^{2}+V^{2}}$.

The orientation of the linearly polarized radio waves emanating from the pulsar can be described by the position angle (PA) of the linearly polarized emission:

$$
\Psi=0.5 \tan ^{-1} \frac{U}{Q}.
$$

The polarization angle is quoted using the IAU convention with the polarization angle increasing in the counterclockwise direction. We can solve for the telescope's response to the incoming radio waves
\begin{equation}
S_{\rm m e a s}=M S_{\rm s r c}\,,
\end{equation}

\noindent where \textit{S$_{\rm m e a s}$} is the Stokes vector measured at the telescope, \textit{S$_{\rm s r c}$} is the Stokes vector of the incoming radio waves, and \textit{M} is the Mueller matrix, which depends on the ellipticity of the receiver arms, non-orthoganality of the receivers, the differential gain, and the differential phase of the receiver (see \citealt{mueller} for more details).

By observing a strongly-polarized source through a series of telescope azimuth angles, Mueller matrix elements for a given telescope and observing system can be determined. The Mueller matrix can then be used to correct other observational data and recover the intrinsic Stokes parameters of the source under observation. We can determine the Mueller matrix for a certain receiver by taking a long observation of one pulsar and tracking it across the sky. By doing this for multiple epochs, we can judge the stability of the receiver by observing how the solutions change over a long period of time.

Pulsars are highly-polarized sources, and the position angle can vary across the pulse phase. For many pulsars, this follows an S-shaped curve, interpreted through the rotating vector model \citep[RVM;][]{RVM} as the observer's line of sight traversing a conal emission beam, with radio emission originating from the open magnetic field lines.
The position angle is measured with respect to the magnetic axis such that it will rotate through the pulse by at most 180$^{\circ}$. 

Numerous polarization studies on millisecond pulsars (MSPs) \citep{msp1, msp2, Stairs1999} have demonstrated that most MSPs have  more complex position angle curves which 
are notoriously difficult to fit to this model \citep{craig14} \citep{Stairs1999}, and \citep{ord_msp}. This is due to the recycled nature of MSPs, creating a complicated field configuration and a reduction in the magnetic field strength, resulting in much smaller period derivatives than canonical pulsars.

Millisecond pulsars also feature emission over a large portion of the profile, with more complex profiles and less profile evolution with frequency than canonical pulsars \citep{msp1}. Geometric arguments imply that pulse widths should vary as the inverse square root of the period \citep{paper6}.
In addition, the beams of millisecond pulsars are wider than those of canonical pulsars due to emission that is produced farther out in the magnetosphere.
This is supported by recent studies with the NICER telescope, which show that MSPs radio profiles could originate in the outer edge of the beam instead of from the core of the emission beam \citep{nicer}.

\citet{Gentile2018} published fully-calibrated polarization profiles at 430 MHz, 1400 MHz, and 2300 MHz for 29 MSPs based on the NANOGrav 11-year data set \citep{11yr} using the Arecibo Telescope. As expected, analysis of these profiles showed position angles that are generally inconsistent with the RVM. They also found microcomponents, which they defined as pulse components with peak intensities much lower than the total pulse peak intensity, in three pulsars.

In this paper, we present polarization profiles\footnote{These data are available to be downloaded from \href{data.nanograv.org/polarization}{data.nanograv.org/polarization}.} of 23 MSPs observed with the Green Bank Telescope (GBT) at both 820\,MHz and 1400 MHz as part of the NANOGrav 12.5-year data set \citep{12halfyr}. 
We measure how the rotation and dispersion measures, and hence  $\langle B_{\parallel}\rangle$ (i.e. from Equation Eqn.~\ref{eq:bparallel}), vary over the course of the data set. 

In \S\ref{sec:obs}, we detail the observations. In \S\ref{sec:analysis} we discuss the polarimetric calibration, Faraday rotation fits, ionospheric corrections, and magnetic field calculations. In \S\ref{sec:results_discussion}, we detail the results of the calibration and discuss the pulse profiles (comparison to published profiles, microcomponents, and frequency evolution/emission geometry), variations in  dispersion measure and magnetic field, correlations with spindown parameters, polarization fractions, and implications for timing. In \S\ref{sec:conclusions}, we conclude the work.

\section{Observations} \label{sec:obs}

\begin{deluxetable*}{cccc|cccc|cccc|ccccc}
\tablecaption{Number and timespan of observations for each pulsar.\label{tbl-1} }
\tablehead{
 &\multicolumn{3}{c}{820 MHz} &  \multicolumn{3}{c}{1500 MHz}   \\
 \colhead{Pulsar}  & \colhead{Start} & \colhead{End} & \colhead{\# of Obs} & \colhead{Start} &\colhead{End} & \colhead{\# of Obs}
 }
\startdata
J0340+4130	&	55972	&	56726	&	29	&	55972	&	56728	&	24	\\
J0613$-$0200 	&	55278	&	56727	&	46	&	55275	&	56733	&	47	\\
J0636+5128 	&	56677	&	56727	&	3	&	56640	&	56729	&	3	\\
J0645+5158 	&	55704	&	56706	&	28	&	55892	&	56736	&	25	\\
J0740+6620  & --- & --- & --- & 56640 & 56736 & 4 \\
J0931$-$1902 & 56387 & 56727 & 8 & 56351 & 56703 & 12 \\
J1012+5307 	& 55278	& 56706	&	50	&	55275	&	56431	&	39	\\
J1024$-$0719 	&	55278	&	56727	&	47	&	55275	&	56703	&	52	\\
J1125+7819 	&	56675	&	56735	&	3	&	56640	&	56736	&	5	\\
J1455$-$3330 	&	55278	&	56709	&	37	&	55773	&	56706	&	27	\\
J1600$-$3053 	&	55641	&	56709	&	35	&	55639	&	56733	&	43	\\
J1614$-$2230 	&	55307	&	56709	&	51	&	55265	&	56733	&	63	\\
J1643$-$1224 &55278	&56709&	50	&	55275	&	56733	&	54	\\
J1713+0747 	&	55278	&	56709	&	55	&	55275	&	56733	&	82	\\
J1744$-$1134 	&	55278	&	56735	&	30	&	55275	&	56736	&	51	\\
J1747$-$4036	&	56270	&	56703	&	15	&	56034	&	56733	&	22	\\
J1832$-$0836 	&	56407	&	56675	&	8	&	56367	&	56736	&	14	\\
J1909$-$3744 	&	55278	&	56725	&	55	&	55275	&	56733	&	76	\\
J1918$-$0642 	&	55307	&	56735	&	48	&	55429	&	56733	&	51	\\
B1937+21 	&	55278	&	56709	&	48	&	55305	&	56676	&	41	\\
J2010$-$1323 	&	55278	&	56709	&	50	&	55275	&	56733	&	52	\\
J2145$-$0750 	&	55278	&	56709	&	46	&	55275	&	56736	&	47	\\
J2302+4442	&	56003	&	56726	&	32	&	55972	&	56728	&	29	\\
\enddata
\tablecomments{These numbers reflect only the data used in the analysis; the outliers have been removed.}
\end{deluxetable*}

We present a subset of the NANOGrav 12.5-year data set taken between MJDs 55265 and 56739 (2010 March 10 and 2014 March 23) at 820\,MHz and 1500\,MHz with the GUPPI instrument \citep{guppi}. We analyze observations of 23 pulsars, two of which overlap with \citet{Gentile2018} (PSRs J1713+0747 and B1937+21). Most pulsars were observed on a monthly cadence, with the exception of PSRs J1713+0747 and J1909$-$3744, which were observed weekly starting in 2013. The data were taken with the GUPPI backend at 820 MHz and 1.4 GHz with bandwidths of 200 MHz and 800 MHz, respectively. The data were coherently dedispersed, with  frequency resolution of 1.56 MHz, and on average each observation lasted around 25 minutes. Table \ref{tbl-1} shows the data timespan and number of observations for each pulsar at each frequency.

The data were run through the standard NANOGrav radio frequency interference (RFI) excision pipeline; for each frequency channel, the minimum and maximum values in the off-pulse region were found and any channels for which this value was an outlier relative to the surrounding channels were zapped (see \citealt{12halfyr} for more details). 

NANOGrav measures a DM at nearly every observation epoch. The value is then recorded as a DMX parameter in TEMPO, where DMX is the difference between the accepted DM and the measured DM at each epoch \citep{Jones2017}.

NANOGrav timing observations with the GUPPI data acquisition instrument began in 2010 March and continued into 2020. However, a technical problem arose in 2014 March, making all data collected after this date unsuitable for polarimetric work.  The problem was instability in the time alignment of the digitizers for the X- and Y-polarizations of the telescope signal.  This corrupted the polarization cross-products and made it impossible to recover full Stokes parameters from these data.  The power in the two individual polarizations was uncorrupted, and well-calibrated total intensity measurements could still be derived, allowing for the use of these data in timing even without full Stokes parameter information. This instability only affects the polarization of the observations, should not affect the total intensity and therefore the timing after 2014.

\begin{figure*}[ht] 
\centering
\includegraphics[height=200mm,trim=3cm 0cm 4cm 0 cm,angle=270,  clip=true]{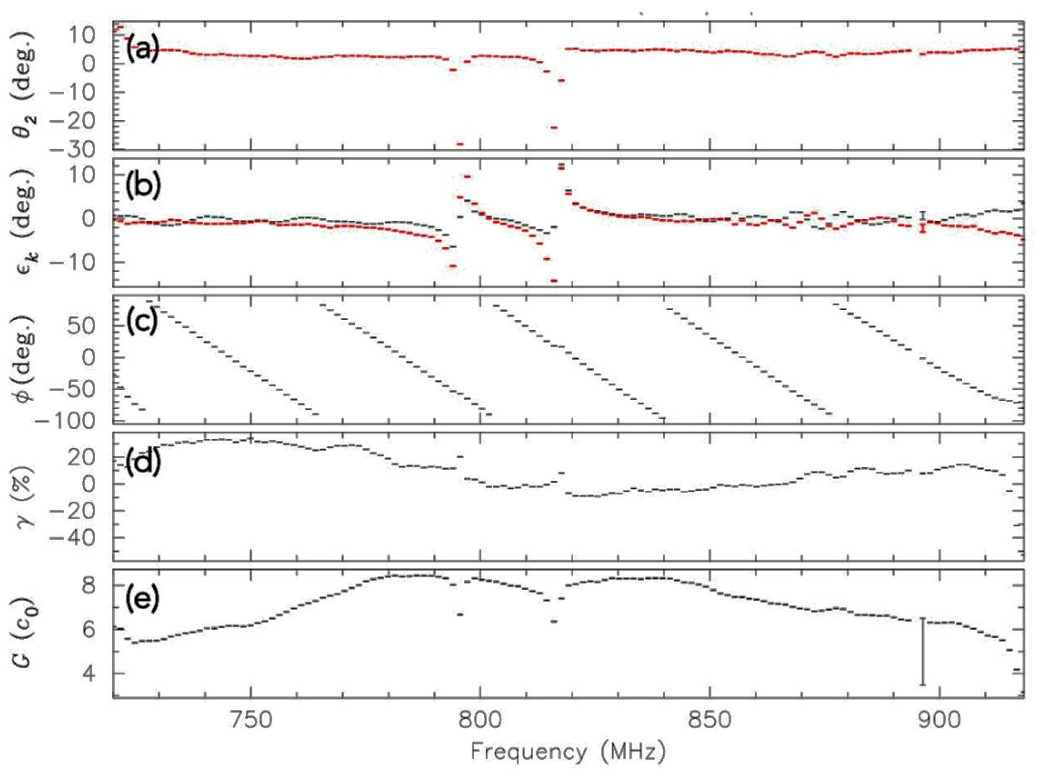}
\caption{An example solution (MJD 56244) used to calibrate the data. Panel (a) shows the degree of cross-coupling between receivers, (b) shows the ellipticity of the receivers (the two colors show the two polarizations; $\theta_{1}$ is assumed to be zero so black is not shown in the above panel.), (c) shows the differential phase, (d) shows the differential gain, and (e) shows the absolute gain of the receiver (specified in units of the square root of the reference flux density). (See  \citealt{solution} for more details of this procedure).}  
\label{fig-1}
\end{figure*}

\section{Data Analysis}\label{sec:analysis}

\subsection{Calibration Method}
All NANOGrav observations go through a basic polarization calibration procedure. At the telescope, a 25-Hz broadband signal is generated at a noise diode and injected into the receiver. At the beginning of each observation, this artificial noise signal is split into two polarization signal paths and measured with the pulsar backend.

A calibration scan is taken for every NANOGrav observation. The noise signals themselves, and also the power in both X- and Y-polarizations, are calibrated by observations on and off a bright, unpolarized source (for the GBT, this is quasar B1442+101; \citealt{9yr}). 

A set of four scans: pulsar, noise diode (which is the off-quasar scan), pulsar and noise diode, and quasar and noise diode, are used to obtain flux and polarization calibration solutions. A noise diode is observed with  every pulsar scan but  B1442+101 is observed once per each multi-day observing session at each frequency. This constitutes the standard calibration scheme, which is applied to all NANOGrav observations. While likely sufficient for timing purposes, in order to study the polarization in detail, more rigorous and precise polarization calibration is needed.

In this analysis, we used long-track observations of two pulsars to calculate  Mueller matrix solutions. For our 820\,MHz data, we used observations of PSR B1929+10 acquired by Kramer et al. (in prep.) for the double pulsar, which were shared with NANOGrav. This pulsar is known for being very bright and has well-known polarization characteristics. We solved for the Mueller matrix at  six epochs  (MJDs 56244, 56419, 56608, 56793, 56984, and 57890) and used the solution closest to the epoch of each pulsar observation to calibrate the 820\,MHz data. The solutions produced calibrated profiles for PSR~B1929+10 that matched those in the literature \citep[e.g.,][]{Stairs1999, Dai2015, Gentile2018} for every epoch, which suggests that our solutions accurately calibrated the data. See  \citet{solution} for full details of our calibration procedure.

At 1500\,MHz, we used a single long-track observation of PSR J1022+1001 taken on MJD 55670 (2011 April 19)  to calculate a Mueller matrix solution. Note that while PSR J1022+1001 has been found to show  pulse profile variations by at most a few percent over the course of a year \citep{J1022-1}, we do not expect this to affect our observations, as the solution was derived from an observation of this pulsar on a single day. After calibrating all of the data with the single solution, we found that the profiles were similar to both those in the literature \citep[e.g.,][]{Stairs1999, Dai2015} and to each other, suggesting that using a single solution for multiple epochs produces accurately-calibrated profiles.

Figure \ref{fig-1} shows an example of an 820\,MHz solution used to calibrate our data. Panel (a) shows $\theta$, the degree of cross-coupling between the receivers. Panel (b) shows $\epsilon$, which indicates how much Stokes \textit{Q} has leaked into Stokes \textit{V}. The slight leakage of one Stokes parameter into another is caused by a small amount of non-orthogonality in the receivers. Panel (c) shows $\phi$, the differential phase, which quantifies the mixing of the Stokes \textit{U} and \textit{V} parameters. Panel (d) shows $\gamma$, the differential gain. Ideally, $\gamma = 0$; in our data set, this parameter is consistent with zero for nearly all the epochs, with only slight offsets. Finally, Panel (e) shows \textit{G}, the absolute gain for the receiver. As described earlier, we measured  six independent realizations of the Mueller matrix as a function of frequency at 820\,MHz at six different epochs.  These realizations were generally consistent with each other.

\subsection{Fitting for Faraday Rotation}

\begin{figure*}[ht] 
\centering
\includegraphics[height=150mm,angle=90,clip=true]{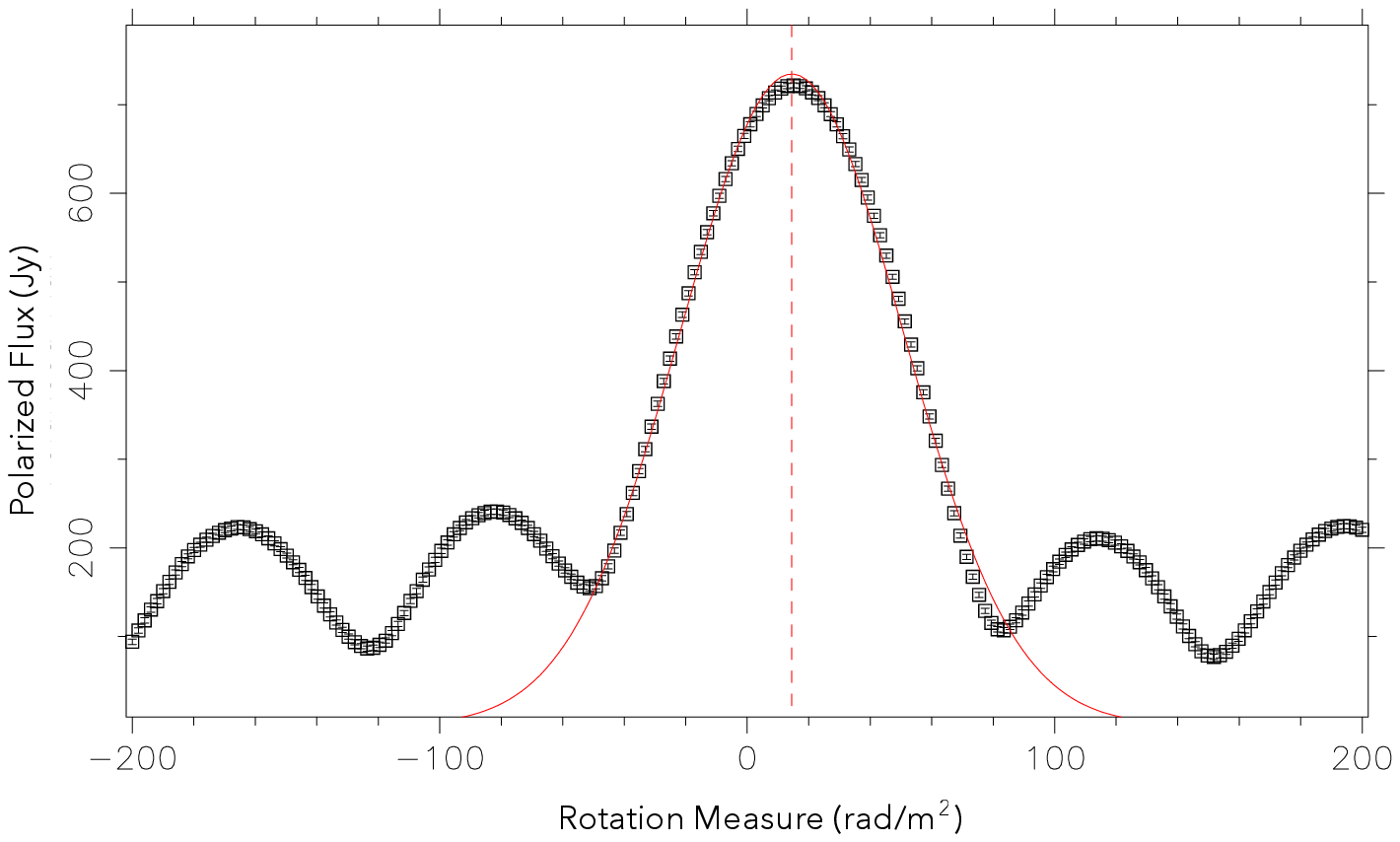}
\vspace{-1 cm}
\caption{An example of an RM fitting An example of an \textit{rmfit} output. The black boxes indicate the linearly polarized flux as a function of rotation measure for PSR~J1713+0747 at 820~MHz. The red curve illustrates the best-fit Gaussian, and the vertical red line denotes the brute force method RM estimate.}  
\label{fig-2}
\end{figure*}

To fit for Faraday rotation and calculate RMs, we used the \textit{rmfit} feature of PSRCHIVE \citep{PSRCHIVE}, specifically the brute force method followed by the iterative position angle refinement technique. The iterative position angle refinement begins by using the brute force method to find the RM at which the linear polarization is maximized by fitting a Gaussian to the linear polarization vs. RM curve and using the centroid of the function as the best RM. We first re-binned the profiles to four frequency channels and 512 pulse phase bins at 820 MHz and 16 frequency channels and 512 pulse phase bins at 1500 MHz. We then searched in a range of --200 to 200 rad/m$^2$ with 200 steps for the majority of pulsars. An example of a fit is shown in Figure \ref{fig-2}. Because of its location behind an HII region \citep{stellahii}, PSR J1643$-$1224 has a large RM $\sim$ $-$308 rad/m$^2$ \citep{Yan2011RM}, so it requires finer frequency resolution to track the shift of PA with frequency. We therefore did not bin- or frequency-scrunch (which left us with the full 2048 bins and 512 channels) and searched from --550 to --150 rad/m$^2$ with 200 steps for the RM.  

Once we had calculated an initial RM from the brute force method, we applied position angle refinement, which compares the position angles measured from the integrated profiles in the two halves of the band, and the weighted differential polarization angle ($\Delta$PA) is  computed between the two halves of the band, using only the pulse phase bins in which the linear polarization is more than 3$\sigma$ above the off-pulse noise. If $\Delta$PA is larger than its uncertainty, the data are corrected for Faraday rotation using that RM and $\Delta$PA between the two bands is estimated again. This process is repeated until $\Delta$PA is smaller than its uncertainty, at which point the final RM is reported. This produces a more accurate RM estimate and uncertainty than the brute force method alone.

If an RM was not able to be fit with these parameters, we removed the profile from further analysis. For the most part, the number of observations taken out for this reason was relatively small (\textless 15\% of the total number of observations for each pulsar) but for PSRs J0645+5158, J0740+6620, J1455$-$3330, J1747$-$4036, and J1832$-$0836, the percentage removed was 25\%, 43\%, 32\%, 31\%, and 22\%, respectively.

To ensure there were no outliers in RM values due to instrumental effects or miscalibration, we calculated the mean and RMS variations of the RMs for each pulsar and then removed data with  RMs that were more than three standard deviations away from the mean from further analysis. After the first cut, a new mean was calculated and anything more than 3$\sigma$ away from that value was cut. This  process was repeated three times. Epochs with outlier RMs are  not present in the combined (composite) profiles (Figures \ref{fig-5}--\ref{fig-16}) and were not used in the variability analyses. Most outliers showed up on specific days at both frequencies.

In addition to the method described above, we inspected the profiles by eye and eliminated any that looked noticeably different from the others. Criteria for this removal include incorrect handedness of the polarization, unusual variations in the profile baseline, and severe deviation from the composite profile on one epoch, all artifacts of a technical/instrumental problem with the observation.

Nearly all of the data sets that required outlier removal had $\textless$17\% of  observations removed. The  exceptions were PSRs J0740+6620 (for which we excised 33\% of the 1500\,MHz data), J0931$-$1092 (which has 33\% removed at 820 MHz), and J1832$-$0836 (which has 25\% removed at 820 MHz). These high percentage are due to the small number of total observations relative to the number of excised observations.

Though most outliers point to instrumental effects, high RMs that occur when a pulsar's line of sight passes close to the Sun may be due to a contribution from the solar magnetic field. We compared the epochs of the outliers we identified with those at which the relevant pulsar has the smallest  elongation (the angle between the Sun and the pulsar). We also searched for outlier RMs at epochs at which DM peaks were detected. We find two such points for one pulsar, PSR J1614$-$2230, that are close to minimum elongation, when our line of sight to the pulsar passes closest to the Sun. See Section \ref{5.1.3} for an in-depth discussion of these points.

\subsection{Ionospheric Corrections}

As the radio waves from the pulsar travel along our line of sight, they pass through the magnetic field of the Earth's ionosphere, which contributes a non-negligible amount to the measured RM. Therefore it must be subtracted in order to study the Galactic magnetic field. We used the ionFR \citep{ionFRpaper} code, which uses publicly available GPS-derived total electron content CODE maps with a 2-hour time resolution for each day of observations, along with the eleventh release of the International Geomagnetic Reference Field, which covers the period when our data were taken. The code calculates the contribution of the ionosphere to the RM along the line of sight and takes into account the time of day of the observation, telescope location, and sky coordinates of the pulsar to get an accurate measurement for each hour of the day. We 
subtracted the ionospheric correction for the closest hour to the mid-point of each observation and were left with the RM due to the magnetic field of the ISM. 

Systematic uncertainties have been associated with this method, including a daily and yearly time dependence, with corrected RMs found to be accurate to 0.06--0.07 rad/m$^2$\citep{ion_greater_yearly} . Therefore, we do not expect systematic uncertainties to be important for our RM measurements, given that the RM errors we derive are higher than this level (see Table \ref{tbl-2}).

\begin{deluxetable*}{cccccccccc}
\tabletypesize{\small}
\tablecaption{Properties of each pulsar and derived quantities. The uncorrected RM at each frequency is the average of all of the measurements at each frequency, while the error is the standard deviation of all of the measurements divided by the square root of the number of measurements. The corrected RM at each frequency is the average of all of the measurements after the ionospheric correction is subtracted from each day, and the error is the standard deviation of all of those corrected measurements divided by the square root of the number of measurements. The magnetic field errors are a combination of the \textit{rmfit}, ionFR, and DMX errors.  \label{tbl-2}}
\tablewidth{0pt}
\tablehead{
& & & &\multicolumn{2}{c}{820 MHz} &  \multicolumn{2}{c}{1500 MHz} & & \\
Pulsar &&Distance & DM  & RM   & Corrected RM  & RM&   Corrected RM & B-field\\
&&(kpc)&(pc cm$^{-3}$)&(rad m$^{-2}$) & (rad m$^{-2}$)&(rad m$^{-2}$)&(rad m$^{-2}$)&($\mu$G)
}
\startdata
J0340+4130&&1.4&49.6&56.8(4)&55.0(4)&54(1)&52(1)&1.33(2)\\
J0613$−$0200&&1.1&38.8&22.2(3)&20.2(3)&19.1(5)&16.9(5)&0.59(1)\\
J0636+5128&&1.1&11.1&1(4)&--1(3)&--5(2)&--7(1)&--0.5(2)\\
J0645+5158&&1.2&18.2&--1.1(7)&--2.9(6)&2(1)&0(1)&--0.08(5)\\
J0740+6620&&0.44&15.0&---&---&--39(1)&--41(1)&--3.3(1)\\
J0931$−$1902&&0.80&41.5&--97(2)&--100(2)&--95.5(9)&--98.5(9)&--2.94(3)\\
J1012+5307&&0.76&9.0&4.0(2)&2.4(2)&4.2(2)&2.6(2)&0.34(2)\\
J1024$−$0719&&1.3&6.5&--1.1(3)&--3.5(2)&--1.6(2)&--3.9(1)&--0.70(3)\\
J1125+7819&&0.052&11.2&--28(2)&--29(1)&--26.7(6)&--28.4(7)&--3.16(8)\\
J1455$−$3330&&13.0&13.6&15.5(7)&12.0(6)&17(1)&13(1)&1.12(6)\\
J1600$−$3053&&2.0&52.3&--7.2(6)&--11.1(4)&--5.4(6)&--9.6(3)&--0.243(6)\\
J1614$−$2230&&0.67&34.5&--27.4(3)&--30.5(2)&--26.1(2)&--29.1(1)&--1.062(5)\\
J1643$−$1224&&1.4&62.4&--303.0(2)&--305.7(2)&--300.3(2)&--303.1(2)&--6.000(3)\\
J1713+0747&&1.2&16.0&10.9(3)&8.9(3)&13.2(3)&10.9(3)&0.76(2)\\
J1744$−$1134&&0.44&3.1&4.9(3)&2.3(3)&3.6(2)&0.7(1)&0.58(5)\\
J1747$−$4036&&2.8&152.7&--38(1)&--43(1)&--45(1)&--51(1)&--0.376(6)\\
J1832$−$0836&&2.9&28.2&44(2)&41(1)&41.9(8)&38.9(7)&1.74(3)\\
J1909$−$3744&&1.1&10.4&4.1(3)&--0.1(3)&2.7(3)&--1.9(2)&--0.12(2)\\
J1918$−$0642&&1.1&26.6&--59.9(9)&--62.5(8)&--54.8(4)&--57.7(3)&--2.78(2)\\
B1937+21&&6.6&71.0&9.7(2)&7.8(3)&9.3(2)&7.3(1)&0.130(2)\\
J2010$−$1323&&2.9&22.2&--2.2(5)&--4.9(4)&--5.8(4)&--8.0(3)&--0.38(2)\\
J2145$−$0750&&0.63&9.0&--0.6(4)&--3.1(4)&--1.6(3)&--4.4(3)&--0.51(3)\\
J2302+4442&&1.8&13.7&19.4(4)&17.4(3)&21.2(3)&19.3(2)&1.64(2)\\  
\enddata
\tablecomments{The uncorrected RM at each frequency is the average of all of the measurements at each frequency, while the error is the standard deviation of all of the measurements divided by the square root of the number of measurements. The corrected RM at each frequency is the average of all of the measurements after the ionospheric correction is subtracted from each day, and the error is the standard deviation of all of those corrected measurements divided by the square root of the number of measurements. The magnetic field errors are a combination of the \textit{rmfit}, ionFR, and DMX errors. The quoted errors are the uncertainties on the last digit.}
\end{deluxetable*}

\subsection{Magnetic Field Calculations}

To accurately constrain magnetic fields along the line of sight to the each pulsar (see Equation~\ref{eq:bparallel}), we need to take into account variations in DM. The NANOGrav data set includes DMX parameters, which measure how much the DM of an observation varies from some fiducial or reference DM \citep[see][]{Jones2017}. 

Table \ref{tbl-2} shows the distance to each pulsar (calculated from parallax measurements from \citealt{12halfyr}), the reference DM (obtained from the par file for each pulsar), the average RM at each frequency (both corrected and uncorrected for the ionosphere), and the average magnetic field derived from the ionosphere-corrected RMs using Eqn.~\ref{eq:bparallel}. 

The uncertainties on the RMs show that the values are broadly consistent between the two frequencies, though some are discrepant at the 1 to 2-sigma level, suggesting that the error bars on the measurements are underestimated.

The error on the magnetic field at each epoch is the error from the RM and DM added in quadrature. The  magnetic field value listed is the average over all epochs and both frequencies for each pulsar. 

\begin{figure*}
\vspace{-4.0 cm}
\begin{center}
\includegraphics[height=180mm,angle=0]{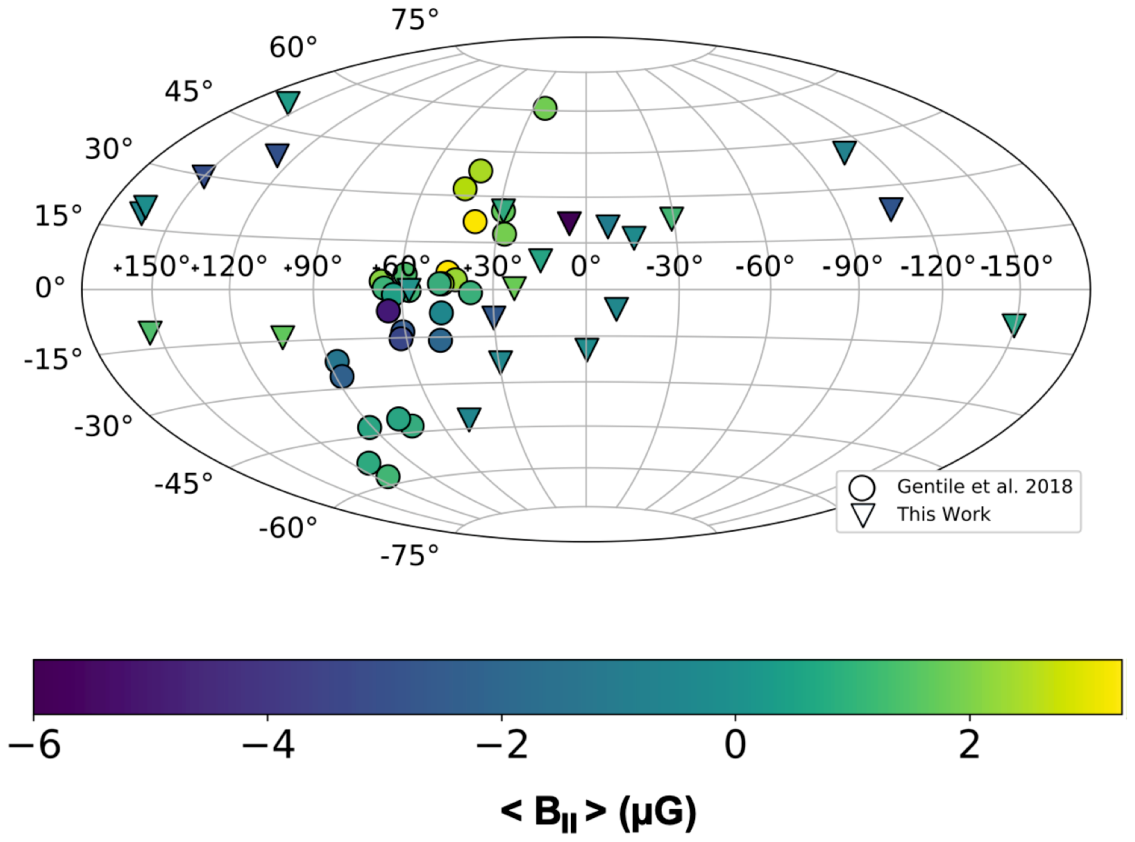}

\end{center}
\vspace{-4.0 cm}
\caption{Magnetic field values derived from pulsar Faraday rotation measures superposed on an Aitoff plot of the Galaxy. The color bar at the bottom shows the value of $\langle B_{\parallel} \rangle$ in unit of $\mu$G. Results from this work are combined with those of \citet{Gentile2018} to get a complete picture of the values around the sky. Note: the plot of \citet{Gentile2018} is incorrect in terms of the sign of the Galactic longitude of the pulsars (which is corrected here).}
\label{fig-3}
\end{figure*}

Figure \ref{fig-3} shows the value of the magnetic field of pulsars around the sky using the values from this work combined with those of \citet{Gentile2018}. The results are consistent with those of \citet{sobeyRM}, which uses pulsars and extragalactic sources in the northern sky to map the Faraday rotation measures, and hence the magnetic field of the Galaxy. For the most part, our results also match those of \citet{Gentile2018} as well as the values of \citet{dike} which uses the Long Wavelength Array to analyze polarization of pulsars below 100 MHz.

\begin{deluxetable*}{cccc|cccc|cccc|ccccc}
\tabletypesize{\small}
\tablecaption{Polarized intensity parameters. $\langle$P$\rangle$ is the phase-averaged power, $\langle$L$\rangle$ is the phase-averaged linear polarization, $\langle$V$\rangle$ is the phase-averaged circular polarization, $\langle$V$\rangle$ is the phase-averaged absolute value of the circular polarization, and $I$ is the  total intensity. The polarization fractions reported are those of the composite profiles. \label{tbl-3}}

\tablewidth{0pt}
\tablehead{
Pulsar & \multicolumn{3}{c|}{$\langle$P$\rangle$/I} & & \multicolumn{3}{c|}{$\langle$L$\rangle$/I} & & \multicolumn{3}{c|}{$\langle$V$\rangle$/I} & & \multicolumn{3}{c}{$\langle$$\mid$V$\mid$$\rangle$/I} \\
&820 MHz & &1500 MHz& & 820 MHz& & 1500 MHz&  & 820 MHz & & 1500 MHz&  & 820 MHz& & 1500 MHz
}
\startdata
J0340+4130	&	0.55	&&	0.14	&&	0.54	&&	0.12	&&	0.00	&&	0.05	&&	0.03	&&	0.07	\\
J0613$-$0200 	&	0.21	&&	0.20	&&	0.19	&&	0.19	&&	0.04	&&	--0.03	&&	0.07	&&	0.04	\\
J0636+5128 	&	0.30	&&	0.38	&&	0.28	&&	0.36	&&	0.01	&&	0.03	&&	0.05	&&	0.07	\\
J0645+5158 	&	0.22	&&	0.19	&&	0.19	&&	0.17	&&	--0.01	&&	--0.02	&&	0.07	&&	0.07	\\
J0740+6620 	&	---	&&	0.27	&&	---	&&	0.21	&&	---	&&	--0.05 &&	---	&&	0.11	\\
J0931$-$1902	&	0.29	&&	0.35	&&	0.28	&&	0.33	&&	--0.02	&&	--0.02	&&	0.06	&&	0.08	\\
J1012+5307 	&	0.66	&&	0.55	&&	0.66	&&	0.54	&&	0.00	&&	0.01	&&	0.05	&&	0.07	\\
J1024$-$0719 	&	0.57	&&	0.60	&&	0.56	&&	0.59	&&	0.00	&&	--0.05	&&	0.03	&&	0.06	\\
J1125+7819 	&	0.38	&&	0.43	&&	0.32	&&	0.38	&&	--0.04	&&	--0.02	&&	0.13	&&	0.13	\\
J1455$-$3330 	&	0.20	&&	0.19	&&	0.19	&&	0.17	&&	0.02	&&	0.04	&&	0.05	&&	0.06	\\
J1600$-$3053 	&	0.25	&&	0.32	&&	0.25	&&	0.31	&&	0.01	&&	--0.02	&&	0.02	&&	0.04	\\
J1614$-$2230 	&	0.69	&&	0.64	&&	0.69	&&	0.62	&&	--0.02	&&	--0.01	&&	0.05	&&	0.05	\\
J1643$-$1224 	&	0.21	&&	0.21	&&	0.17	&&	0.15	&&	--0.04	&&	0.03	&&	0.12	&&	0.12	\\
J1713+0747 	&	0.33	&&	0.31	&&	0.32	&&	0.30	&&	0.02	&&	0.02	&&	0.04	&&	0.03	\\
J1744$-$1134 	&	0.78	&&	0.88	&&	0.78	&&	0.87	&&	0.02	&&	--0.02	&&	0.03	&&	0.03	\\
J1747$-$4036	&	0.14	&&	0.15	&&	0.14	&&	0.13	&&	0.00	&&	0.00	&&	0.04	&&	0.06	\\
J1832$-$0836 	&	0.22	&&	0.30	&&	0.18	&&	0.27	&&	0.07	&&	0.01	&&	0.13	&&	0.10	\\
J1909$-$3744 	&	0.51	&&	0.45	&&	0.48	&&	0.43	&&	--0.14	&&	--0.13	&&	0.15	&&	0.15	\\
J1918$-$0642 	&	0.23	&&	0.21	&&	0.22	&&	0.19	&&	0.04	&&	0.05	&&	0.05	&&	0.06	\\
B1937+21 	&	0.37	&&	0.30	&&	0.36	&&	0.30	&&	--0.02	&&	0.01	&&	0.02	&&	0.02	\\
J2010$-$1323 	&	0.20	&&	0.21	&&	0.17	&&	0.19	&&	--0.05	&&	--0.01	&&	0.10	&&	0.07	\\
J2145$-$0750 	&	0.21	&&	0.19	&&	0.17	&&	0.16	&&	--0.06	&&	--0.06	&&	0.10	&&	0.07	\\
J2302+4442	&	0.56	&&	0.54	&&	0.55	&&	0.53	&&	0.02	&&	0.00	&&	0.04	&&	0.06	\\
\enddata
\end{deluxetable*}

\section{Results and Discussion}\label{sec:results_discussion}

\subsection{Pulse Profiles}

Here we present the polarization-calibrated average profiles from the method described in the previous section. Figures~\ref{fig-5}--\ref{fig-16} show the composite profiles, which were made by summing the profiles from individual epochs. The position angle, which is shown in the top panel of each composite profile, is plotted when the linear polarization is $>$3$\sigma$ above the off-pulse noise. Table \ref{tbl-3} shows the fractions of total power of the emission, average fractional linear polarization, average fractional circular polarization, and average fractional absolute circular polarization of all pulsars in the data set at 820\,MHz and 1500\,MHz, all fractions are calculated with respect to the total power. 

\begin{deluxetable*}{l|c}
\tabletypesize{\small}

\tabletypesize{\footnotesize}
\tablecaption{Previously published polarization profiles. \label{tbl-4}}
\tablehead{
\colhead{Pulsar} & \colhead{Published Polarization Profiles}
 }
\startdata
J0340+4130&820\,MHz$^{(1)}$\\\hline
J0613$-$0200&410\,MHz$^{(2)}$, 610\,MHz$^{(2)}$, 728\,MHz$^{(3)}$, 1335\,MHz$^{(4)}$, 1369\,MHz$^{(3)}$, 1369\,MHz$^{(5)}$,  1405\,MHz$^{(6)}$, 3100\,MHz$^{(3)}$\\\hline
J0636+5128&---\\ \hline
J0645+5158&---\\\hline
J0740+6620&---\\\hline
J0931$-$1902&---$^\dagger$\\\hline
J1012+5307&149\,MHz$^{(7)}$, 610\,MHz$^{(2)}$\\\hline
J1024$-$0719&728\,MHz$^{(3)}$, 1369\,MHz$^{(3)}$, 1369\,MHz$^{(5)}$, 1373\,MHz$^{(6)}$, 3100\,MHz$^{(3)}$\\\hline
J1125+7819&---\\\hline
J1455$-$3330&1300\,MHz$^{(6)}$\\\hline
J1600$-$3053&728\,MHz$^{(3)}$, 1369\,MHz$^{(3)}$, 1369\,MHz$^{(5)}$, 1373\,MHz$^{(6)}$, 3100\,MHz$^{(3)}$\\ \hline
J1614$-$2230&---$^\dagger$\\\hline
J1643$-$1224&610\,MHz$^{(2)}$, 728\,MHz$^{(3)}$, 1331\,MHz$^{(4)}$, 1369\,MHz$^{(3)}$, 1369\,MHz$^{(5)}$, 3100\,MHz$^{(3)}$\\\hline
J1713+0747 & 410\,MHz, 610\,MHz$^{(2)}$, 728\,MHz$^{(3)}$, 1369\,MHz$^{(3)}$, 1369\,MHz$^{(5)}$, 1400\,MHz$^{(8)}$, 1405\,MHz$^{(6)}$, 1414\,MHz$^{(2)}$, \\& 2100\,MHz$^{(8)}$, 3100\,MHz$^{(3)}$\\\hline
J1744$-$1134 &610\,MHz$^{(2)}$, 728\,MHz$^{(3)}$, 1341\,MHz$^{(6)}$, 1369\,MHz$^{(3)}$, 1369\,MHz$^{(5)}$, 3100\,MHz$^{(3)}$\\\hline
J1747$-$4036&---$^\dagger$\\\hline
J1832$-$0836&728\,MHz$^{(3)}$, 1369\,MHz$^{(3)}$, 1369\,MHz$^{(9)}$, 3100\,MHz$^{(3)}$\\\hline
J1909$-$3744&728\,MHz$^{(5)}$, 1369\,MHz$^{(3)}$, 1369\,MHz$^{(5)}$, 1373\,MHz$^{(6)}$, 3100\,MHz$^{(3)}$\\\hline
J1918$-$0642& ---$^\dagger$ \\\hline
B1937+21& 610\,MHz$^{(2)}$, 728\,MHz$^{(3)}$, 1369\,MHz$^{(3)}$, 1369\,MHz$^{(5)}$, 1373\,MHz$^{(6)}$, 1400\,MHz$^{(8)}$, 1414\,MHz$^{(2)}$, 2100\,MHz$^{(8)}$, 3100\,MHz$^{(3)}$\\\hline
J2010$-$1323&1373\,MHz$^{(6)}$\\\hline
J2145$-$0750&410\,MHz$^{(2)}$, 610\,MHz$^{(2)}$, 728\,MHz$^{(3)}$, 1335\,MHz$^{(4)}$, 1369\,MHz$^{(3)}$, 1369\,MHz$^{(5)}$, 1373\,MHz$^{(6)}$, 1414\,MHz$^{(2)}$, 3100\,MHz$^{(3)}$\\\hline
J2302+4442&---\\\hline
\enddata
\tablecomments{References: $^{(1)}$\citet{bangale2011}, $^{(2)}$\citet{Stairs1999}, $^{(3)}$\citet{Dai2015}, $^{(4)}$\citet{ManchesterandHan2004}, $^{(5)}$\citet{Yan2011pol}, $^{(6)}$\citet{ord_msp}, $^{(7)}$\citet{nsk}, $^{(8)}$\citet{Gentile2018}, $^{(9)}$\citet{burgay}. Note: several pulsars with no previously published polarization profiles are included in a study of MeerKAT data (Spiewak et al.~submitted); these are denoted with a dagger ($^\dagger$).}
\end{deluxetable*}
\subsubsection{Comparison to Published Polarization Profiles}
Table \ref{tbl-4} shows all previously published profiles for these pulsars. We present the first published polarization profiles at any frequency for PSRs  J0636+5128, J0645+5158, J0740+6620, J0931$-$1902, J1125+7819, J1614$-$2230, J1747$-$4036, J1918$-$0642, and J2302+4442. 
We find no major discrepancies between our profiles and those previously published. The only exception is the sign of the circular polarization; \citet{Dai2015} uses the Institute of Electrical and Electronics Engineers definition of circular polarization whereas we use the IAU convention. This results in a sign change in the circular polarization (in the IEEE convention, left-hand circular polarization is positive and right-hand circular polarization is negative, whereas the IAU convention is the opposite).

Another exception is PSR B1937+21. At 820\,MHz, the degree of linear polarization for B1937+21 shows epoch-to-epoch variability of up to $\sim$18\%in the second main structure (the interpulse) . Because the RMs matched published values, we chose to carry out the analysis with them; the average profile is also similar to the literature, so we chose to present it. We will explore the reason for this variability in future work.

Overall, our RMs also agree with those previously published. There are several ways to measure RMs from pulsar profiles, and these methods have different systematic uncertainties. Most studies, such as this work and \citet{Yan2011pol}, use the \textit{rmfit} method to calculate RMs and uncertainties, but other methods exist. For example,  \citet{sobey2019} calculate Faraday RMs through Faraday spectra, or Faraday dispersion functions, with uncertainties calculated vis the method described in \citet{rmsynthesis2}. Our RM values are consistent within a few sigma of both of those results for the pulsars analyzed by both methods. The RM errors derived through these different methods are also consistent. This is reassuring, especially as \citet{sobey2019} calculate the RMs in a different way and using different bandwidths, center frequencies, and another telescope.

\subsubsection{Microcomponents}

We detect microcomponents in the pulse profiles of seven pulsars in this work. Microcomponents were discussed in \citet{Gentile2018} and here we define them as components that are $<$3\% of the intensity of the highest peak on the average profile. Out of these seven, four pulsars have microcomponents that are detected for the first time. The microcomponents have varying degrees of polarization; for example, the microcomponents of PSR J2145$-$0750 are almost fully-polarized, whereas those of PSR J1909$-$3744 exhibit very little polarization. There is no apparent correlation between the amount of polarization in the microcomponents and that in the main pulse (i.e. the microcomponent of J2145$-$0740 is almost fully-polarized whereas the profile shows little).

Microcomponents that have been previously detected in other works have a flux density above 1.6 mJy, and all of the new ones have a flux density of less than 1.5 mJy. Because of our long data sets, which produce a very high S/N composite profiles, we are able to detect these very faint microcomponents. To ensure that the microcomponents were not an instrumental effect, we split each frequency band in half to see if the microcomponent was detected in each half. This was generally the case at both 820 MHz and 1500 MHz; the exception is J1713+0747, which exhibits a microcomponent only at 1500 MHz. This can be explained by the pulsar's very flat spectrum \citep{Dai2015}, resulting in lower S/N at lower frequencies. The tests show that microcomponents are not an anomalous instrumental artifact but are of astrophysical origin. The detection of microcomponents demonstrates that MSPs emit over a wide phase range due to their larger opening angles and emission produced further out in the magnetosphere \citep{msp2}.
 
These microcomponents make it difficult to define the duty cycle of millisecond pulsars. As noted in \citet{Gentile2018}, they may cause an overestimation of the radiometer noise in the off-pulse region which could affect flux calibration (although NANOGrav does not rely on the radiometer noise for flux calibration). If these microcomponents are present in other pulsars, they would be revealed by longer data sets (and therefore higher S/N profiles). Microcomponents are generally most prevalent in our highest S/N pulsars; higher gain telescopes like the MeerKAT telescope in South Africa would improve that S/N, allowing us to probe weaker pulsars for these microcomponents (e.g., Spiewak et al.~submitted). If not accounted for in template profiles, these mircocomponents could lead to higher uncertainties in TOA calculation. To make template profiles for TOAs, NANOGrav aligns and averages the reduced data profiles, and applies wavelet smoothing to the average profile \citep{12halfyr}. This wavelet smoothing preserves the microcomponents, and therefore they are taken into account when calculating TOAs.

\subsubsection{Frequency Evolution/Emission Geometry}
The profiles for the majority of canonical pulsars are thought to evolve in frequency according to the core double cone model of \citet{coredoublecone}. This model makes specific predictions about how the number of components in a pulsar's average profile will vary with frequency. For example, a conal single pulsar will have two components at low frequencies ($\sim$100 MHz) that will merge into one at higher frequencies ($\sim$1 GHz). \citet{msp2} show that MSPs show three types of evolution: they can evolve minimally, evolve as predicted, or evolve contrary to any prediction. In their survey, 12 pulsars evolved minimally, five as predicted, and eight against predictions (e.g., with more components at higher frequency). This suggests that the emission of MSPs does not behave like the emission of canonical pulsars. 

Frequency evolution is difficult to track in our pulsars,  as many have more than five components and multiple structures in their profile (e.g. PSR J0931$-$1902). Out of the 22 MSPs for which we have accumulated profiles at  both 820 MHz and 1500 MHz, 14 show the same number of components at both frequencies (i.e., develop minimally) and eight seem to develop  more components at higher frequencies, seemingly in contrast to the predictions of \citet{coredoublecone} and in line with the \citet{msp2} results. While some of this evolution in MSPs could be due to decreased scatter broadening (causing separate components to appear as one at low frequencies), it supports the suggestion that MSPs do not evolve like canonical pulsars. While profiles evolve with frequency for all of the MSPs studied, there is no consistent trend and the frequency evolution is less dramatic than seen for non-recycled pulsars.

\citet{johnston2008_slowpulsars} shows that in slow pulsars, the overall polarization fraction decreases as frequency increases, though some components can show an increase with frequency. This could be a consequence of a geometric process or involve orthogonal polarization modes. Overall, we find that the mean polarization fractions of linear and circular polarization do not show a clear trend with frequency. The exception is $\langle|V|\rangle/I$;  12 MSPs in this study have higher $\langle|V|\rangle/I$ values at 1500~MHz, while only six have a higher fraction at 820 MHz and four have identical fractions at both frequencies. This shows a hint of a correlation, and this correlation is opposite to that than observed for canonical pulsars. However, this is a very small sample and further study is needed to confirm if this is the case in all millisecond pulsars.  

As expected, many of the millisecond pulsars feature emission over a large portion of the profile (e.g. PSRs~J1614--2230 and J2302+4442). This supports the idea that millisecond pulsar beams are wider than those of canonical pulsars due to emission produced farther out in the magnetosphere.

\begin{figure*}[ht]
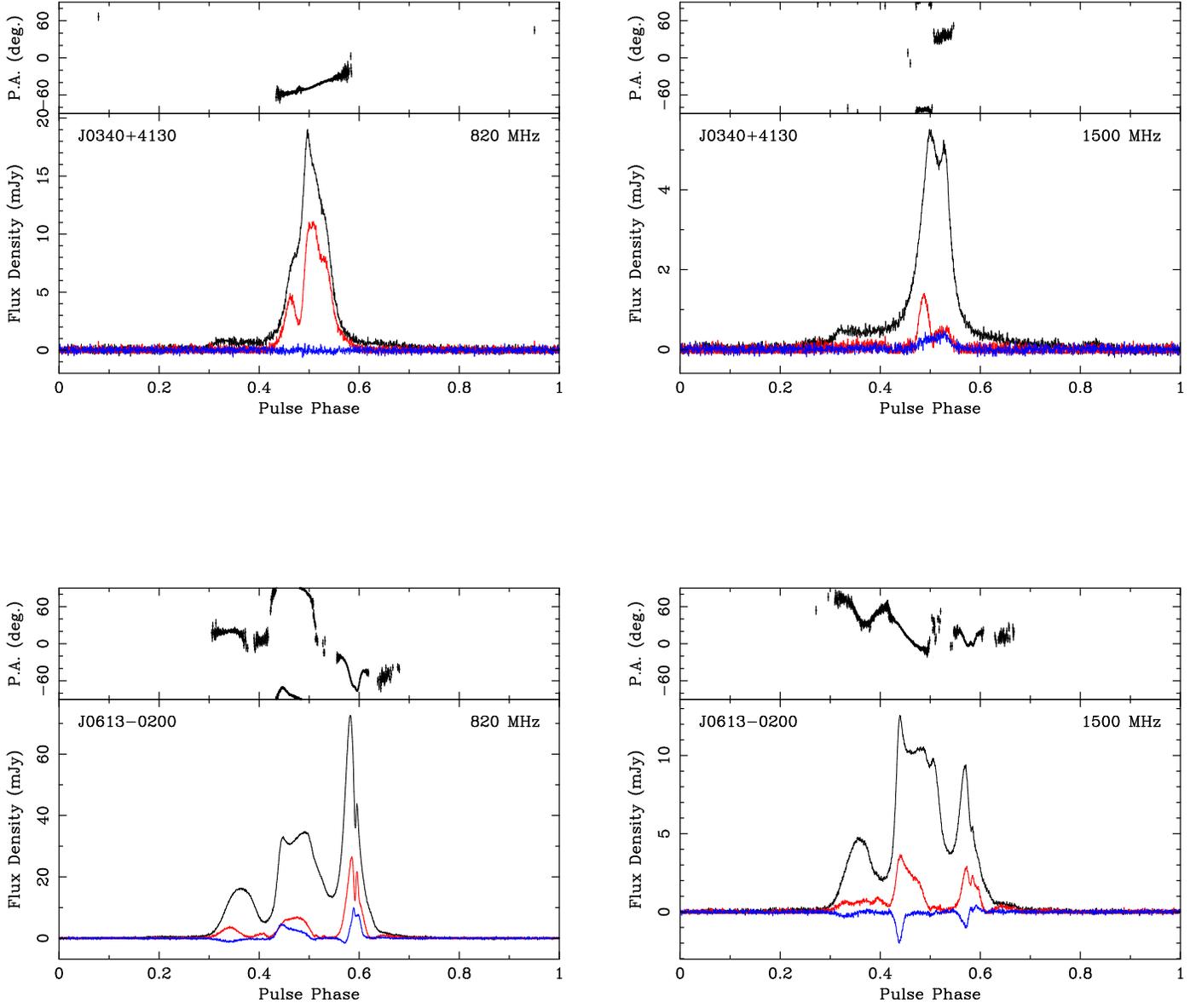

\begin{center}
\caption{Pulse profile for pulsars J0340+4130 and J0613--0200. The black line is the total intensity, red is the circular polarization, and blue is the circular polarization. The polarization position angle is shown in the top panel.}
 \vspace{1.35 cm}
\begin{tabular}{@{}ll@{}}
 \vspace{2.35 cm}
{\mbox{\includegraphics[height=87mm,angle=270]{J0340_820_composite.ps}}}& \ \ \ 
\hspace{0.2 cm}
{\mbox{\includegraphics[height=87mm,angle=270]{J0340_1500_composite.ps}}}\\
\vspace{2.35 cm}
{\mbox{\includegraphics[height=87mm,angle=270]{J0613_820_composite.ps}}}& \ \ \ 
\hspace{0.2 cm} 
{\mbox{\includegraphics[height=87mm,angle=270]{J0613_1500_composite.ps}}}\\
\end{tabular}
\label{fig-5}
\end{center}
\end{figure*}

\begin{figure*}[ht]
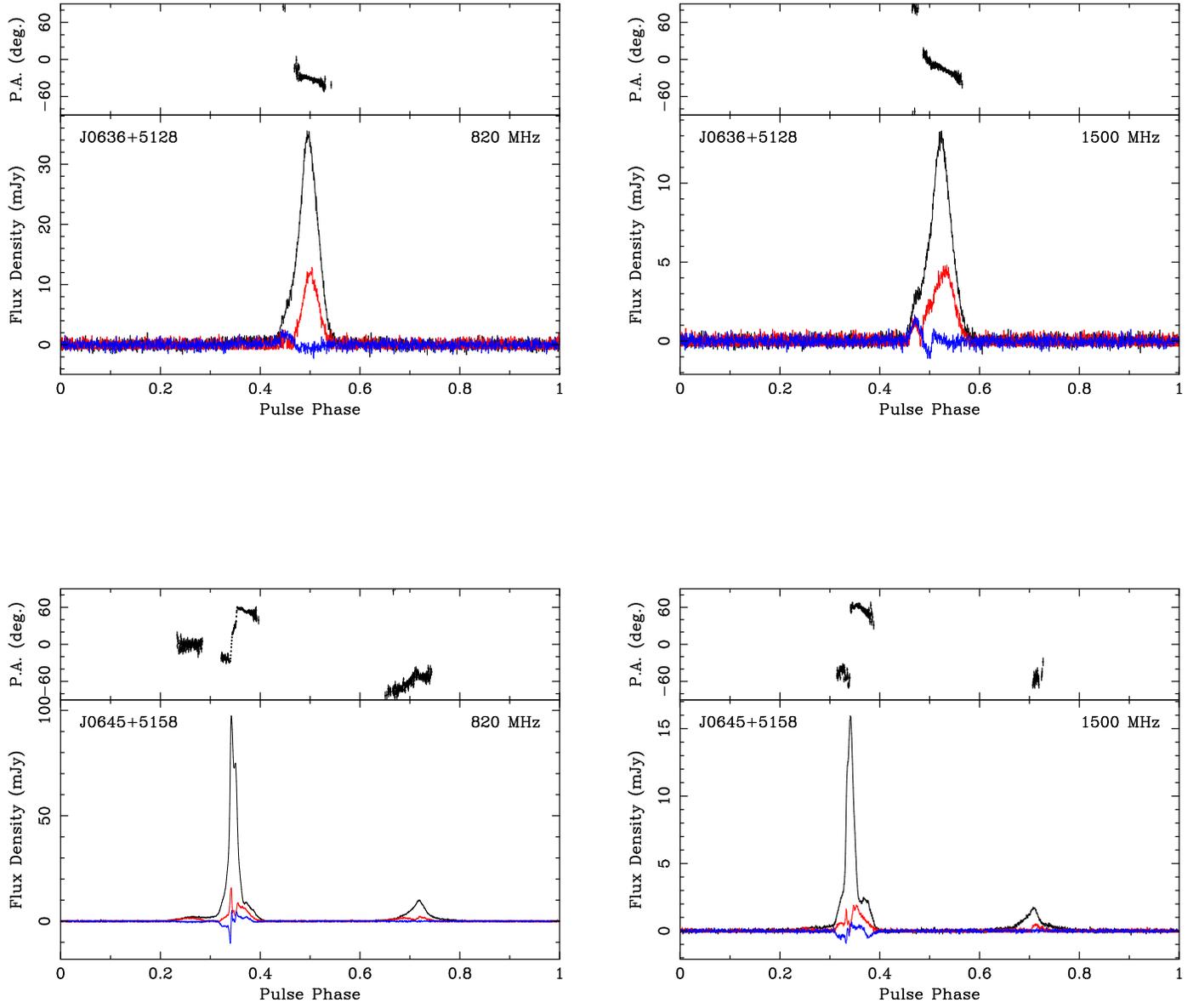

\begin{center}
\caption{The pulse profile for pulsars J0636+5128 and J0645+5158. The black line is the total intensity, red is the linear polarization, and blue is the circular polarization. The polarization position angle is shown in the top panel.}
 \vspace{1.35 cm}
\begin{tabular}{@{}ll@{}}
 \vspace{2.35 cm}
{\mbox{\includegraphics[height=87mm,angle=270]{J0636_820_composite.ps}}}& \ \ \ 
\hspace{0.2 cm}
{\mbox{\includegraphics[height=87mm,angle=270]{J0636_1500_composite.ps}}}\\
\vspace{2.35 cm}
{\mbox{\includegraphics[height=87mm,angle=270]{J0645_820_composite.ps}}}& \ \ \ 
\hspace{0.2 cm} 
{\mbox{\includegraphics[height=87mm,angle=270]{J0645_1500_composite.ps}}}\\

\end{tabular}
\label{fig-6}
\end{center}
\end{figure*}

\begin{figure*}[ht]
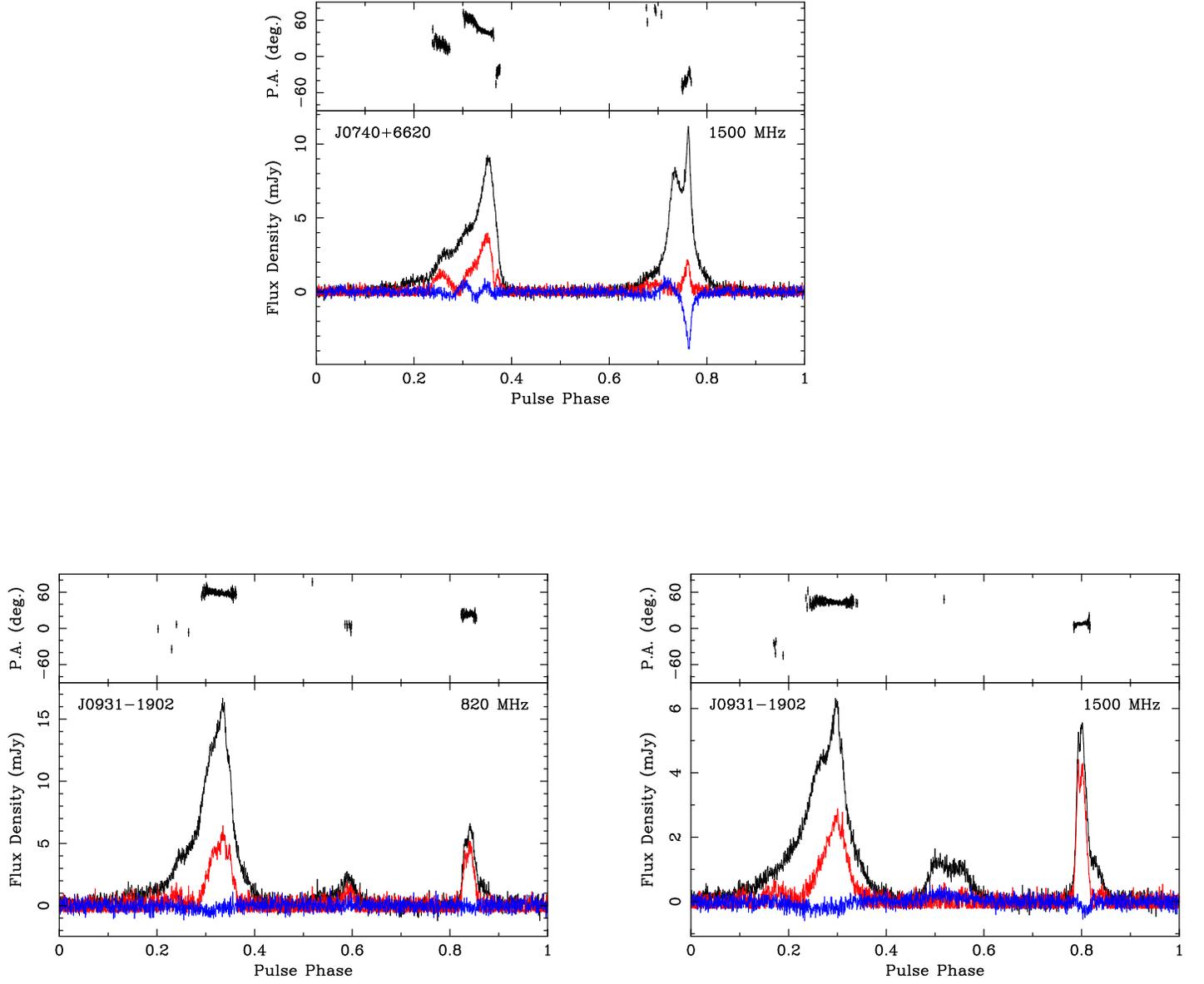

\begin{center}
\caption{Pulse profiles for pulsars J0740+6620 and J0931--1902. The black line is the total intensity, red is the linear polarization, and blue is the circular polarization. The polarization position angle is shown in the top panel.}
 \vspace{1.35 cm}
\begin{tabular}{@{}ll@{}}
 \vspace{2.35 cm}

\hspace{4 cm}
{\mbox{\includegraphics[height=87mm,angle=270]{J0740_1500_composite.ps}}}\\
\vspace{2.35 cm}
{\mbox{\includegraphics[height=87mm,angle=270]{J0931_820_composite.ps}}}& \ \ \ 
\hspace{-3.5 cm} 
{\mbox{\includegraphics[height=87mm,angle=270]{J0931_1500_composite.ps}}}\\
\end{tabular}
\label{fig-7}
\end{center}
\end{figure*}

\begin{figure*}[ht]
\begin{center}
\caption{Pulse profiles for pulsars J1012+5307 and J1024--0719 including microcomponents. The black line is the total intensity, red is the linear polarization, and blue is the circular polarization. The black arrow points to the location of the microcomponent in J1024--0719. The polarization position angle is shown in the top panel. The microcomponent plots for J1024--0719 have been plotted with fewer bins to increase the signal-to-noise. The polarization position angle is shown in the top panel.}
\begin{tabular}{@{}ll@{}}
{\mbox{\includegraphics[height=87mm,angle=270]{J1012_820_composite.ps}}}& \ \ \ 
\hspace{-1.6 cm}
{\mbox{\includegraphics[height=87mm,angle=270]{J1012_1500_composite.ps}}}\\
\hspace{-1.2 cm}
{\mbox{\includegraphics[height=115mm,angle=90]{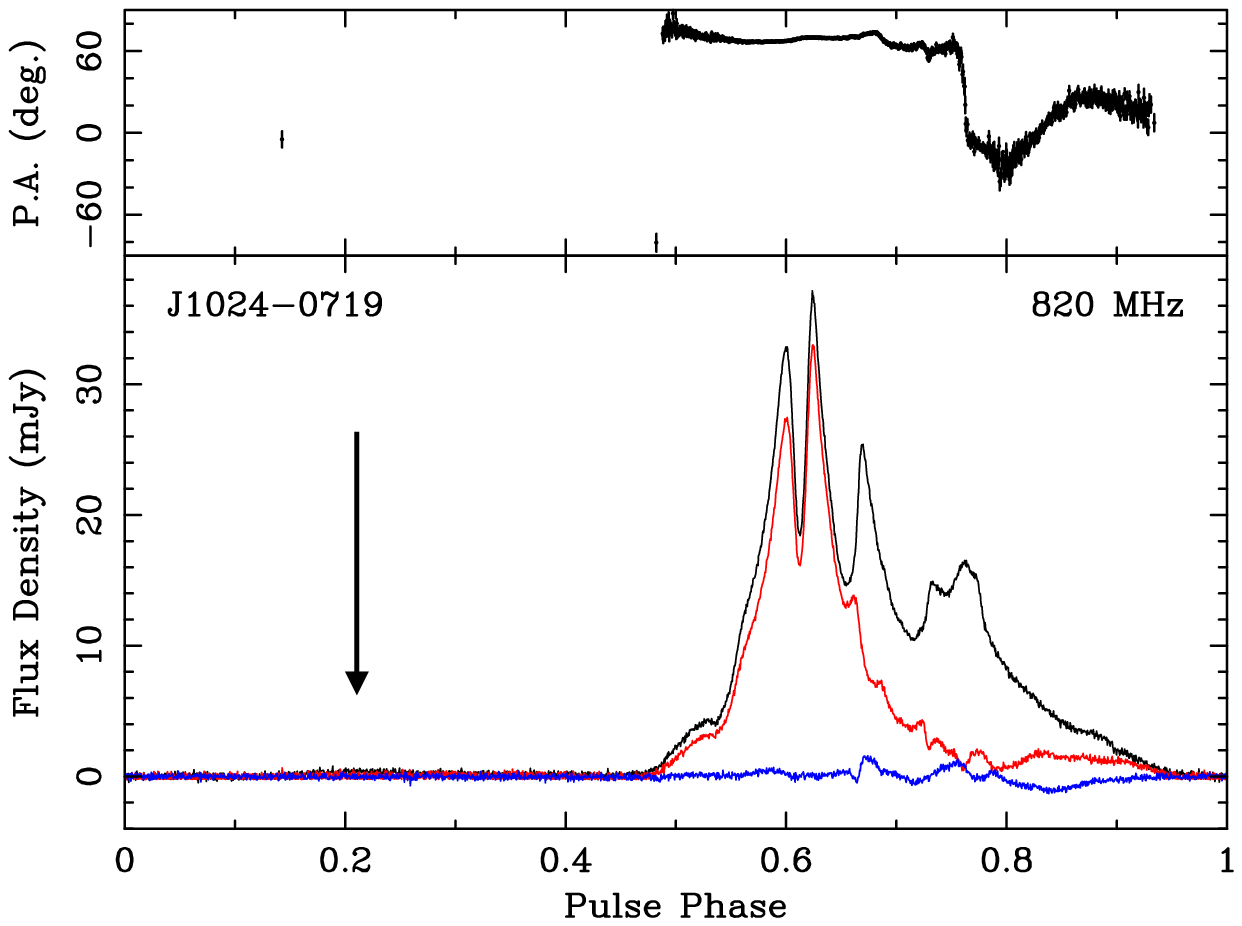}}}& \ \ \ 
\hspace{-2.7 cm}
{\mbox{\includegraphics[height=115mm,angle=90]{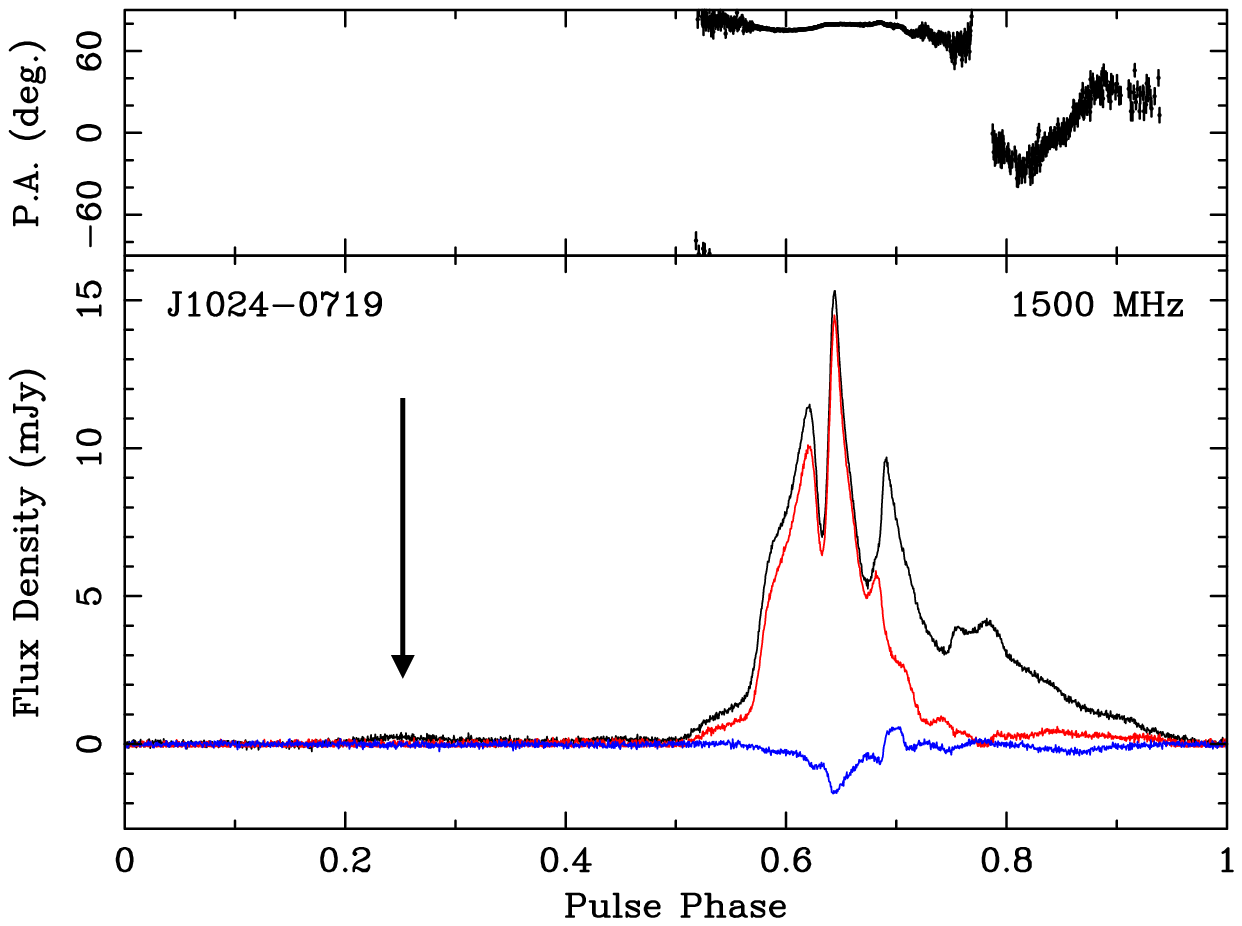}}}\\
{\mbox{\includegraphics[height=87mm,angle=270]{J1024_820_micro.ps}}}& \ \ \ 
\hspace{-1.6 cm}
{\mbox{\includegraphics[height=87mm,angle=270]{J1024_1500_micro.ps}}}\\
\end{tabular}
\label{fig-8}
\end{center}
\end{figure*}

\begin{figure*}[ht]
\begin{center}
\caption{Pulse profiles for pulsars J1125+7819 and J1455--3330 including microcomponents. The black line is the total intensity, red is the linear polarization, and blue is the circular polarization. The black arrow points to the location of the microcomponent of J1455--3330. The microcomponent plots for J1455--3330 have been plotted with fewer bins increase the signal-to-noise. The polarization position angle is shown in the top panel.}
\begin{tabular}{@{}ll@{}}
{\mbox{\includegraphics[height=87mm,angle=270]{J1125_820_composite.ps}}}& \ \ \ 
\hspace{-1.6 cm}
{\mbox{\includegraphics[height=87mm,angle=270]{J1125_1500_composite.ps}}}\\
\hspace{-1.2 cm}
{\mbox{\includegraphics[height=115mm,angle=90]{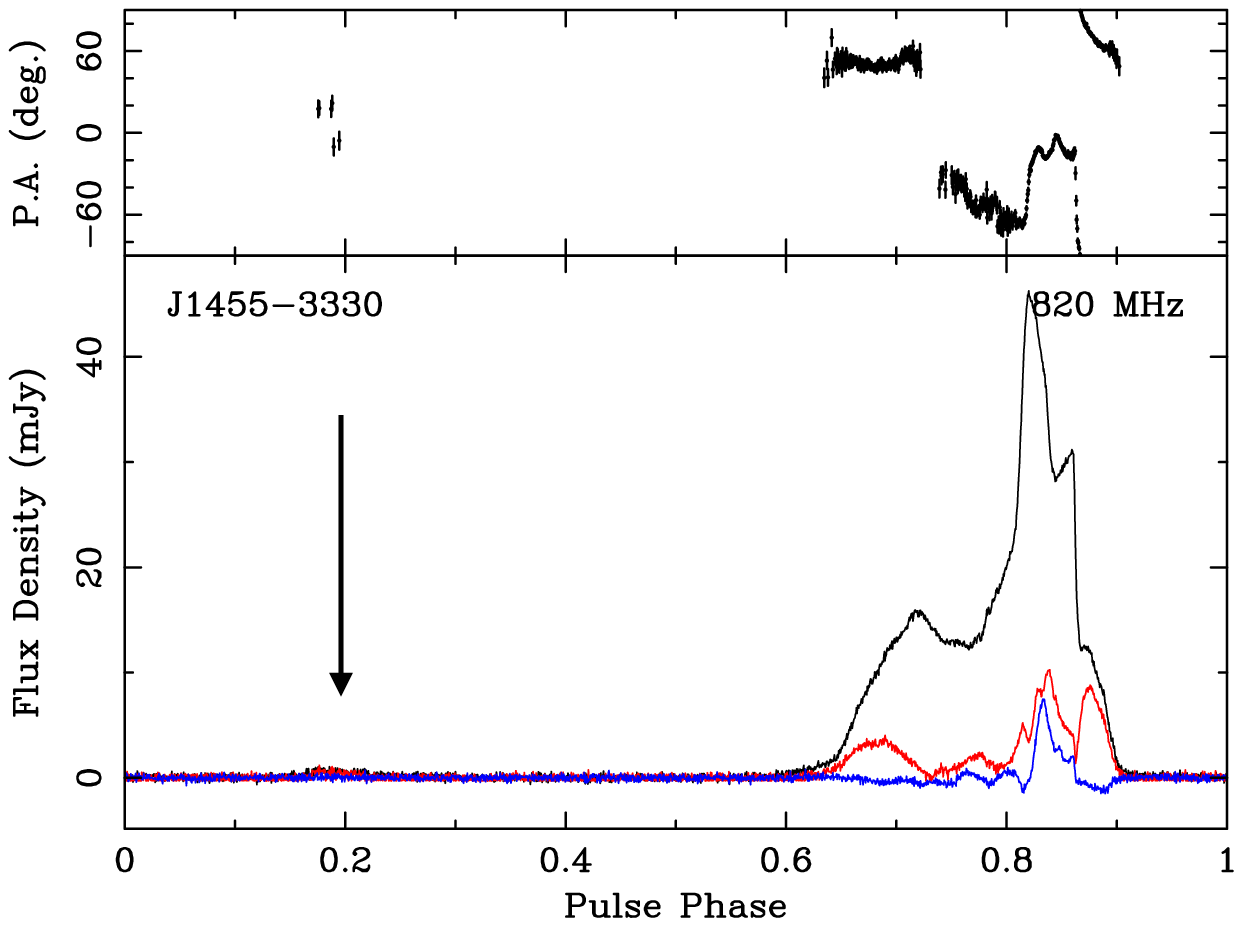}}}& \ \ \ 
\hspace{-2.7 cm}
{\mbox{\includegraphics[height=115mm,angle=90]{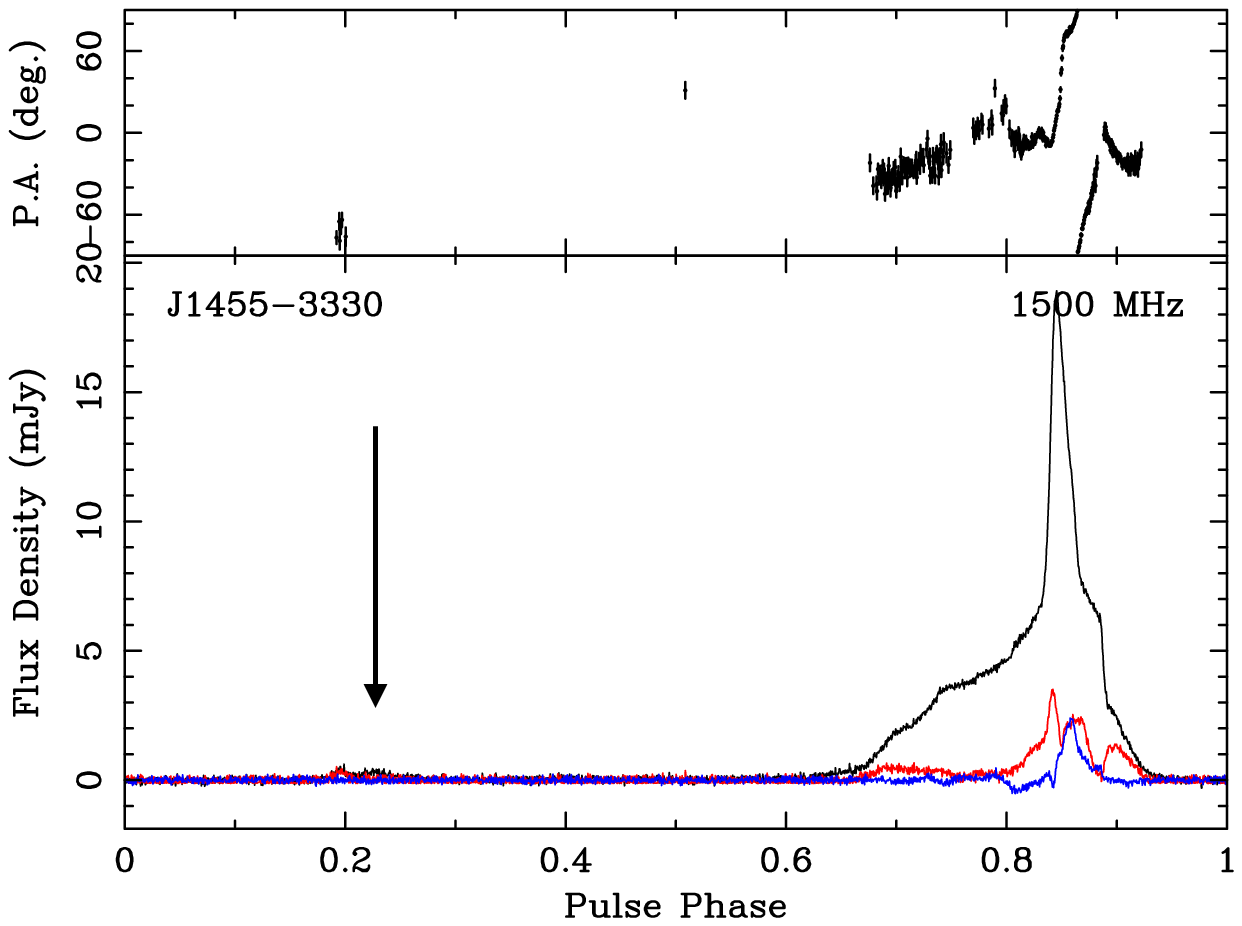}}}\\
{\mbox{\includegraphics[height=87mm,angle=270]{J1455_820_composite_micro.ps}}}& \ \ \ 
\hspace{-1.6 cm}
{\mbox{\includegraphics[height=87mm,angle=270]{J1455_1500_composite_micro.ps}}}\\
\end{tabular}
\label{fig-9}
\end{center}
\end{figure*}

\begin{figure*}[ht]
\begin{center}
\caption{Pulse profiles for pulsars J1600--3053 and J1614--3053 including microcomponents. The black line is the total intensity, red is the linear polarization, and blue is the circular polarization. The black arrow points to the location of the microcomponent of J1600--3053. The microcomponent plots for J1455--3330 have been plotted with fewer bins increase the signal-to-noise. The polarization position angle is shown in the top panel.}
\begin{tabular}{@{}ll@{}}
\hspace{-1.2 cm}
{\mbox{\includegraphics[height=115mm,angle=90]{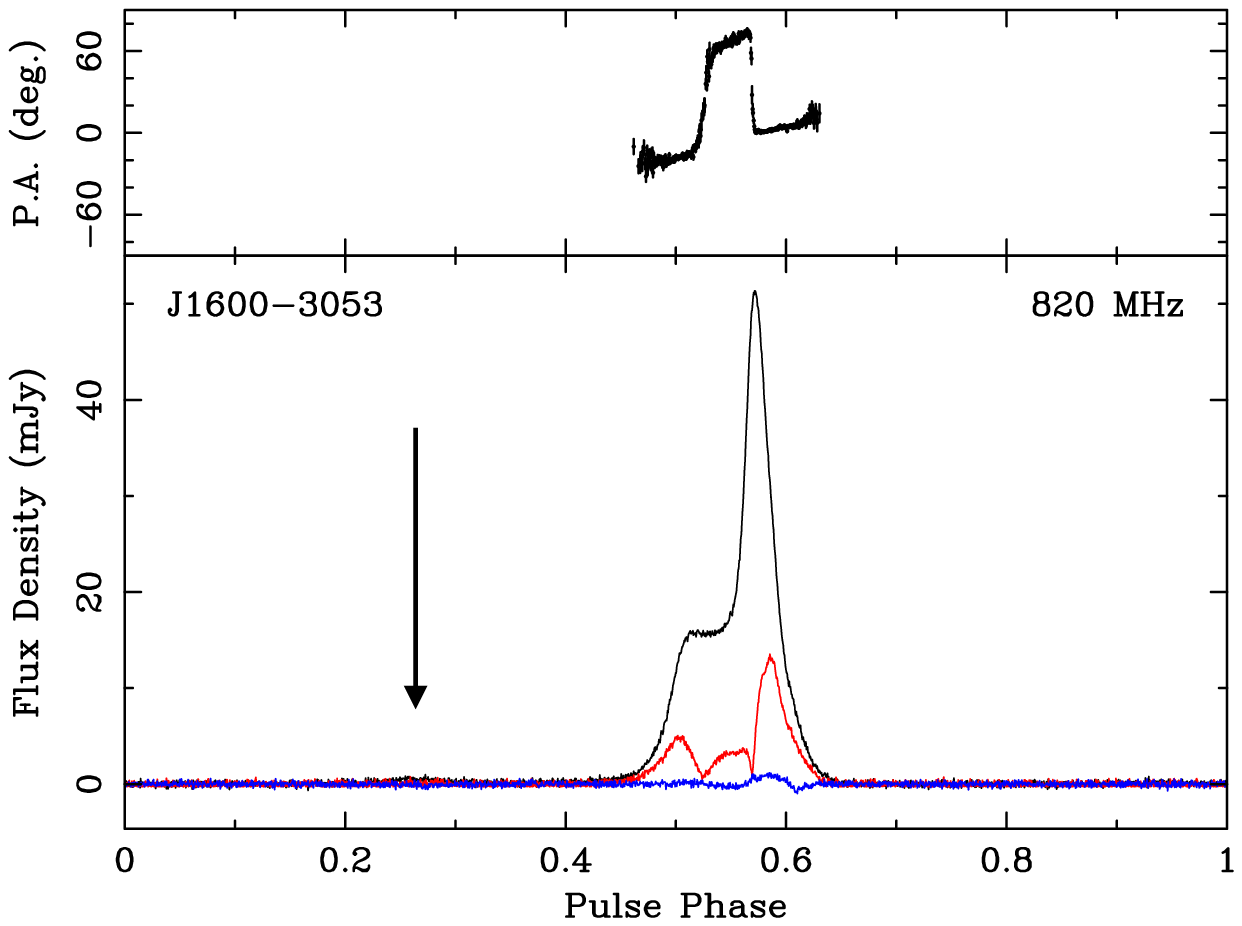}}}& \ \ \ 
\hspace{-2.7 cm}
{\mbox{\includegraphics[height=115mm,angle=90]{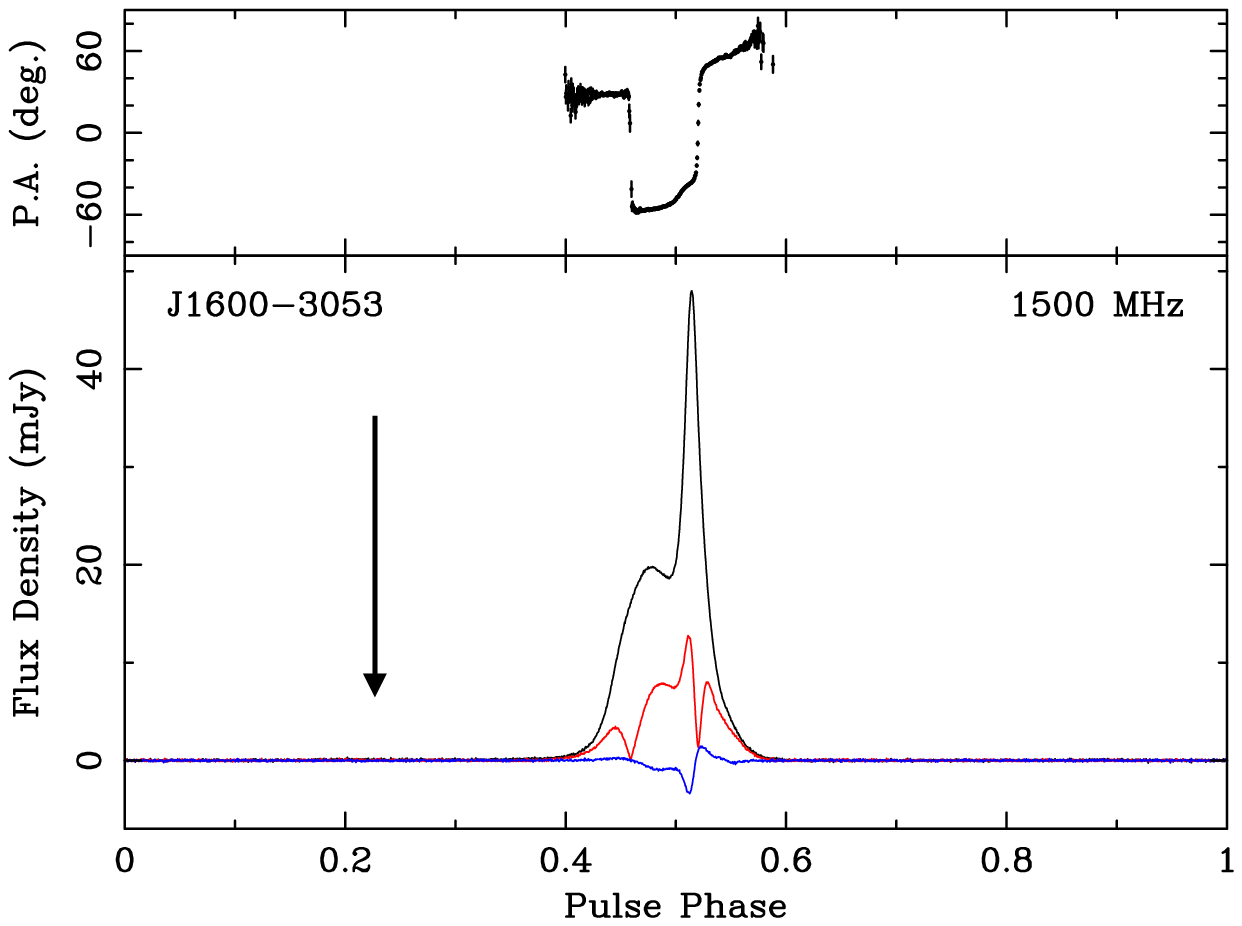}}}\\

{\mbox{\includegraphics[height=87mm,angle=270]{J1600_820_composite_micro.ps}}}& \ \ \ 
\hspace{-1.6 cm}
{\mbox{\includegraphics[height=87mm,angle=270]{J1600_1500_composite_micro.ps}}}\\
{\mbox{\includegraphics[height=87mm,angle=270]{J1614_820_composite.ps}}}& \ \ \ 

\hspace{-1.6 cm}
{\mbox{\includegraphics[height=87mm,angle=270]{J1614_1500_composite.ps}}}\\

\end{tabular}

\label{fig-10}
\end{center}
\end{figure*}

\begin{figure*}[ht]
\begin{center}
\caption{Pulse profiles for pulsars J1643--1224 and J1713+0747 including microcomponents. The black line is the total intensity, red is the linear polarization, and blue is the circular polarization. The black arrow points to the location of the microcomponent in each J1713+0747 profile. The microcomponent plot for J1713--0747  have been plotted with fewer bins to increase the signal-to-noise. The polarization position angle is shown in the top panel. Note: there is no detection of the microcomponent of J1713+0747 at 820\,MHz, the plot is just shown for comparison.}
\begin{tabular}{@{}ll@{}}

{\mbox{\includegraphics[height=87mm,angle=270]{J1643_820_newnewnew.ps}}}& \ \ \ 
\hspace{-1.6 cm}
{\mbox{\includegraphics[height=87mm,angle=270]{J1643_1500_newnewnew.ps}}}\\
\hspace{-1.2 cm}
{\mbox{\includegraphics[height=115mm,angle=90]{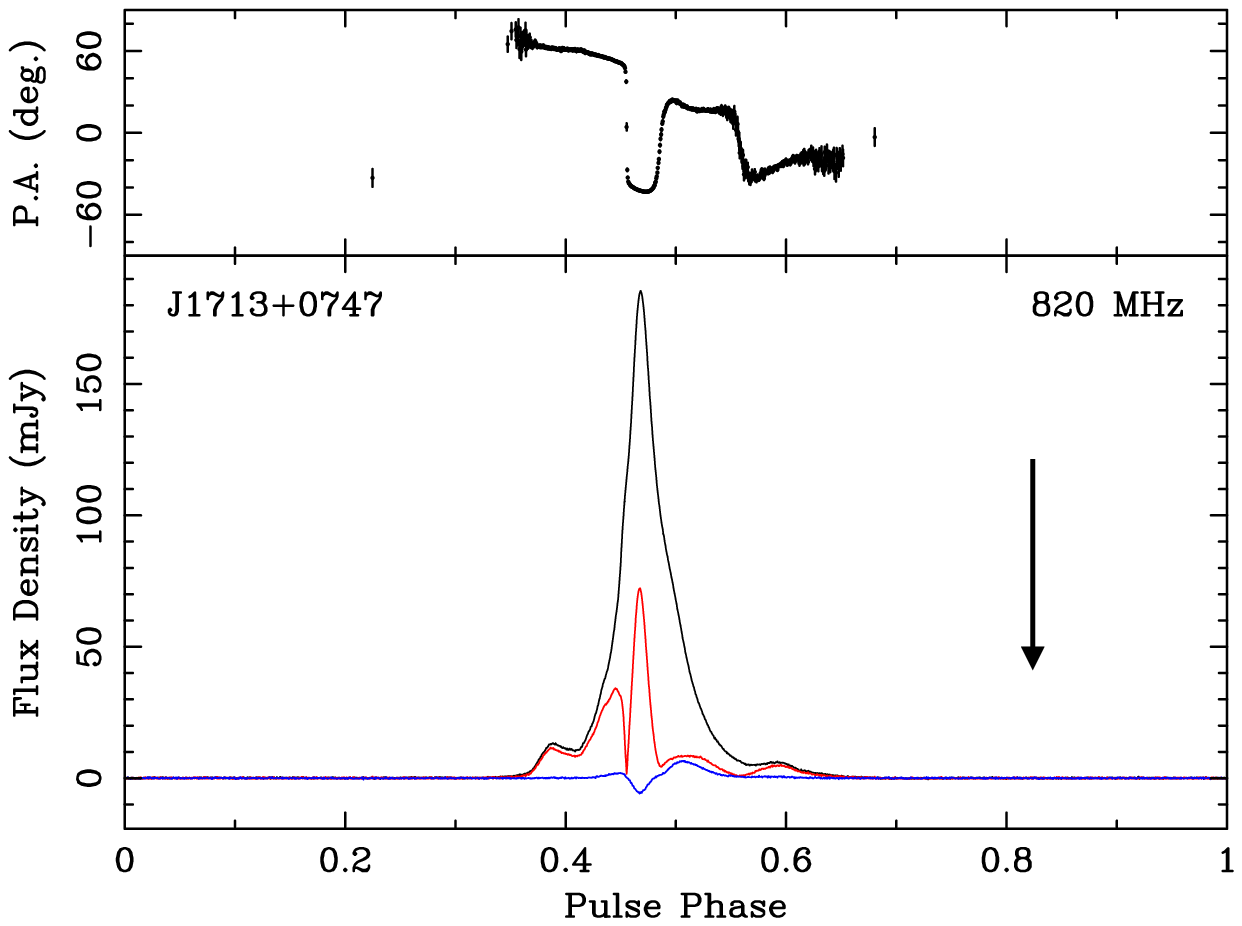}}}& \ \ \ 
\hspace{-2.7 cm}
{\mbox{\includegraphics[height=115mm,angle=90]{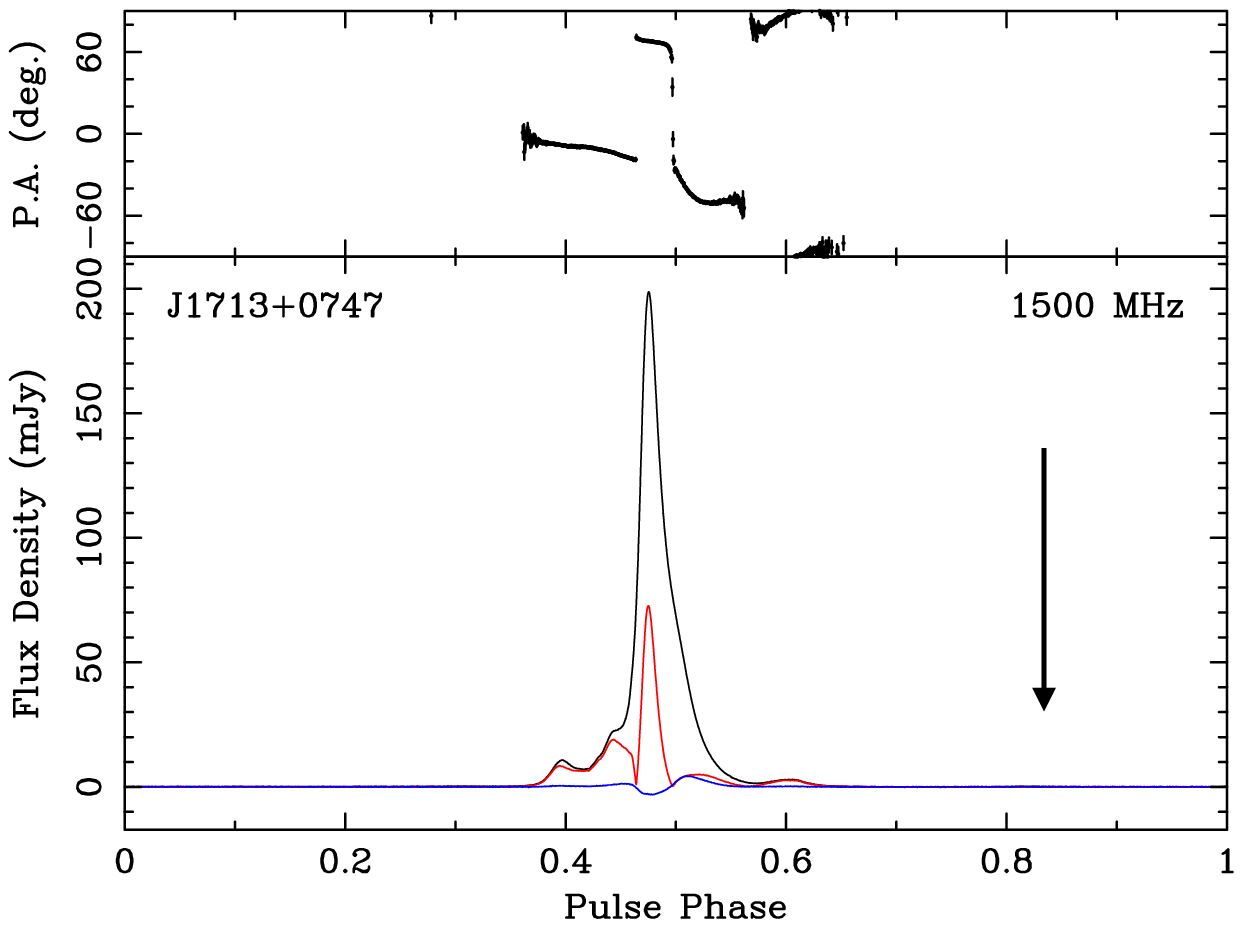}}}\\
{\mbox{\includegraphics[height=87mm,angle=270]{J1713_820_composite_micro.ps}}}& \ \ \ 
\hspace{-1.6 cm}
{\mbox{\includegraphics[height=87mm,angle=270]{J1713_1500_composite_micro.ps}}}\\
\end{tabular}
\label{fig-11}

\end{center}
\end{figure*}

\begin{figure*}[ht]
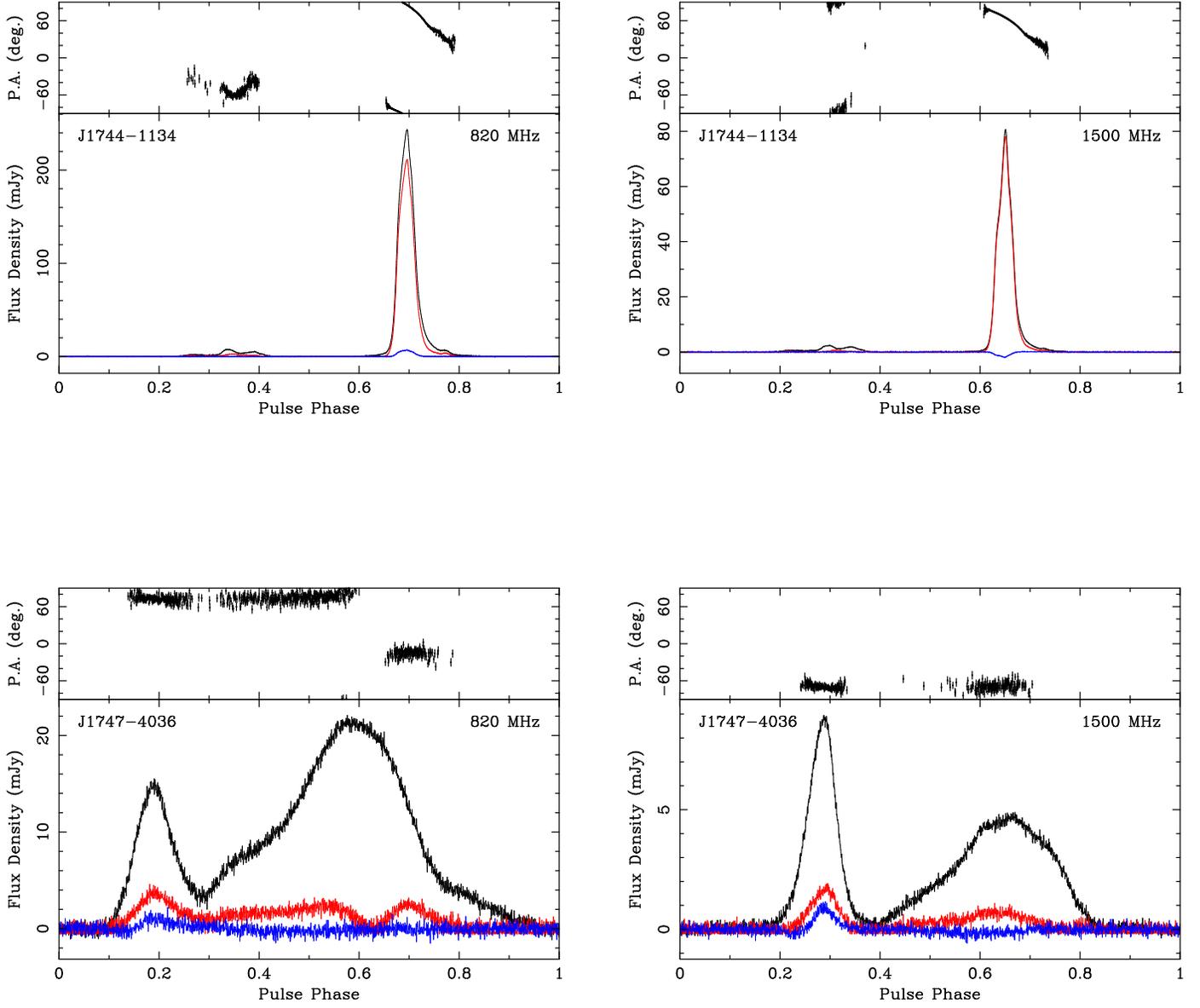

\begin{center}
\caption{Pulse profiles for pulsars J1744--1134 and J1747--4036. The black line is the total intensity, red is the linear polarization, and blue is the circular polarization.  The polarization position angle is shown in the top panel.}
 \vspace{1.35 cm}
\begin{tabular}{@{}ll@{}}

 \vspace{2.35 cm}
{\mbox{\includegraphics[height=87mm,angle=270]{J1744_820_composite.ps}}}& \ \ \ 
\hspace{0.2 cm} 
{\mbox{\includegraphics[height=87mm,angle=270]{J1744_1500_composite.ps}}}\\

\vspace{2.35 cm}

{\mbox{\includegraphics[height=87mm,angle=270]{J1747_820_composite.ps}}}& \ \ \ 
\hspace{0.2 cm} 
{\mbox{\includegraphics[height=87mm,angle=270]{J1747_1500_composite.ps}}}\\
\end{tabular}

\label{fig-12}
\end{center}
\end{figure*}

\begin{figure*}[ht]
\begin{center}
\caption{Pulse profiles for pulsars J1832--0836 and J1909--3744 including microcomponents. The black line is the total intensity, red is the linear polarization, and blue is the circular polarization. The black arrow points to the location of the microcomponent in each J1909--3744 profile. The polarization position angle is shown in the top panel.}
\begin{tabular}{@{}ll@{}}

{\mbox{\includegraphics[height=87mm,angle=270]{J1832_820_composite.ps}}}& \ \ \ 
\hspace{-1.6 cm}
{\mbox{\includegraphics[height=87mm,angle=270]{J1832_1500_composite.ps}}}\\

\hspace{-1.2 cm}
{\mbox{\includegraphics[height=115mm,angle=90]{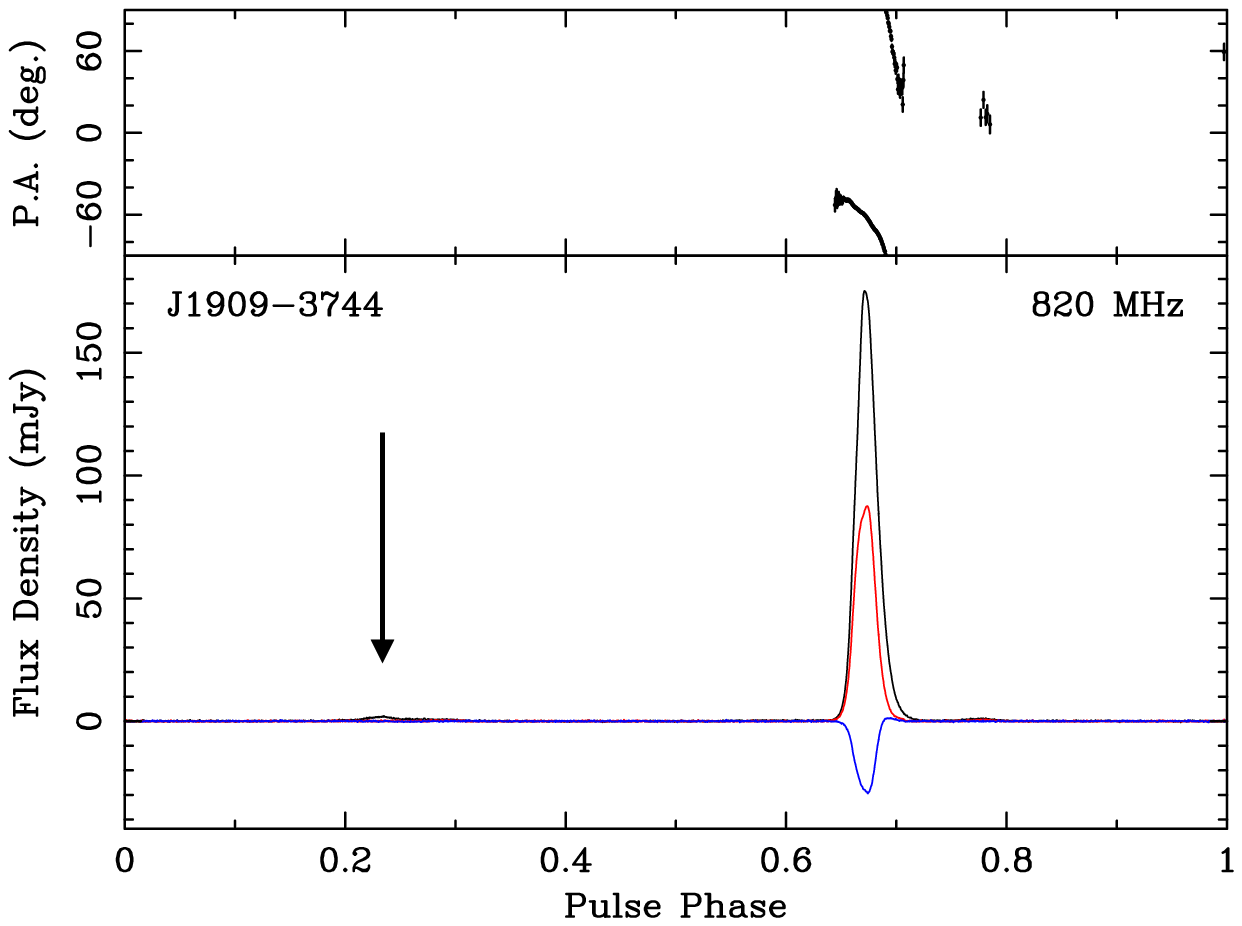}}}& \ \ \ 
\hspace{-2.7 cm}
{\mbox{\includegraphics[height=115mm,angle=90]{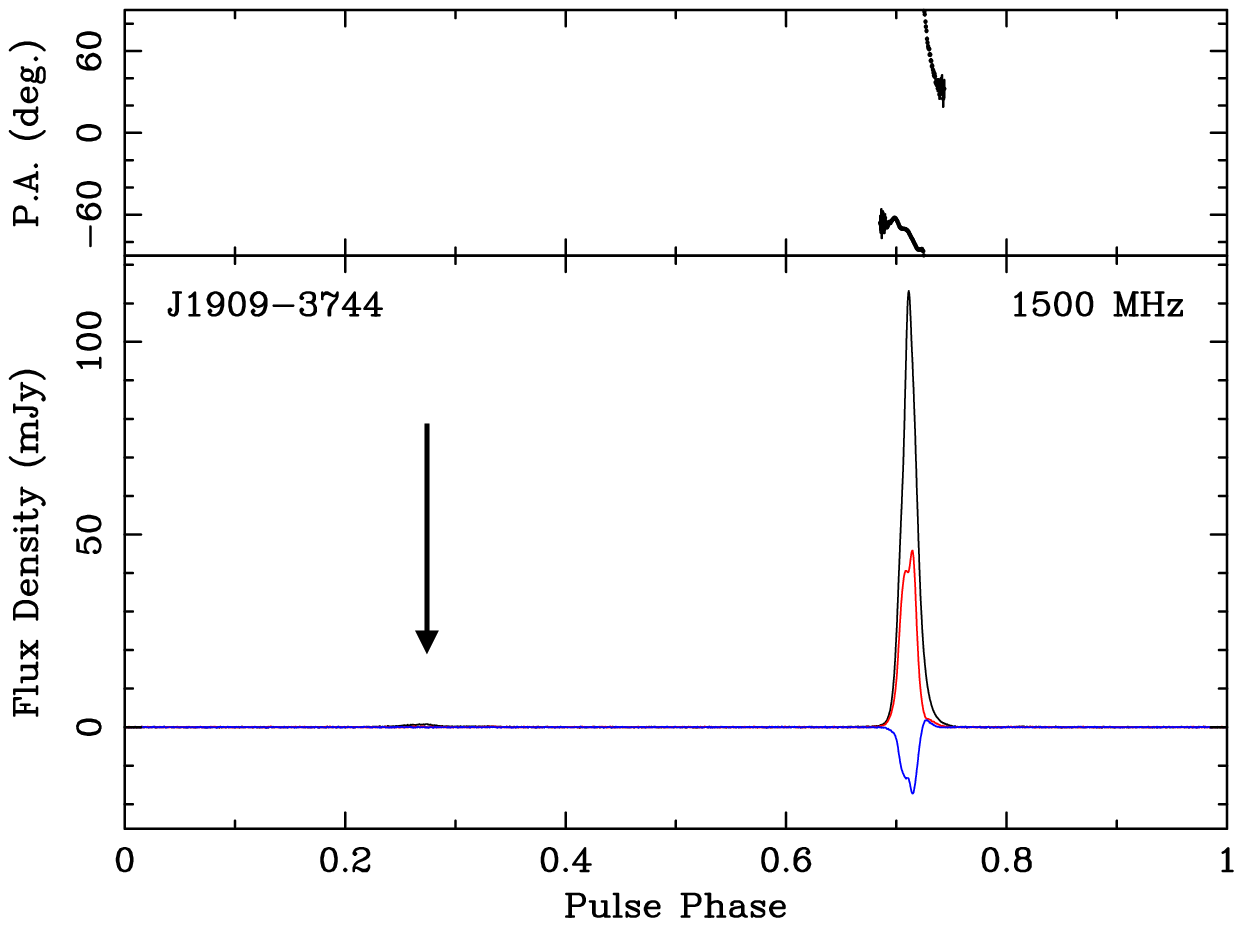}}}\\

{\mbox{\includegraphics[height=87mm,angle=270]{J1909_820_composite_micro.ps}}}& \ \ \ 
\hspace{-1.6 cm}
{\mbox{\includegraphics[height=87mm,angle=270]{J1909_1500_composite_micro.ps}}}\\

\end{tabular}

\label{fig-13}
\end{center}
\end{figure*}

\begin{figure*}[ht]
\begin{center}
\caption{Pulse profile for pulsars J1918--0642 and B1937+21 including microcomponents. The black arrow points to the location of the microcomponent in each B1937+21 profile. The black line is the total intensity, red is the linear polarization, and blue is the circular polarization. The polarization position angle is shown in the top panel.}
\begin{tabular}{@{}ll@{}}

{\mbox{\includegraphics[height=87mm,angle=270]{J1918_820_composite.ps}}}& \ \ \ 
\hspace{-1.6 cm}
{\mbox{\includegraphics[height=87mm,angle=270]{J1918_1500_composite.ps}}}\\

\hspace{-1.2 cm}
{\mbox{\includegraphics[height=115mm,angle=90]{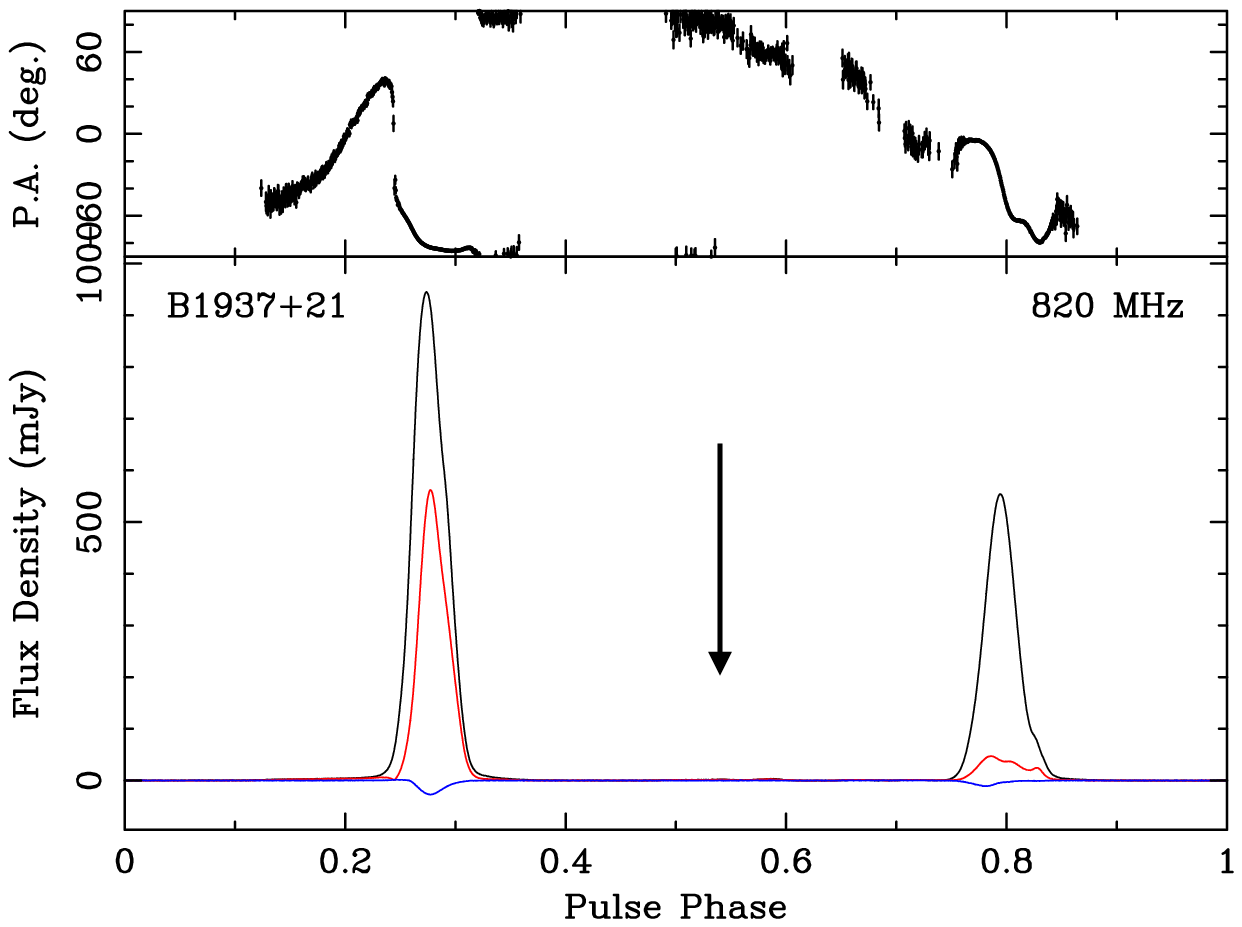}}}& \ \ \ 
\hspace{-2.7 cm}
{\mbox{\includegraphics[height=115mm,angle=90]{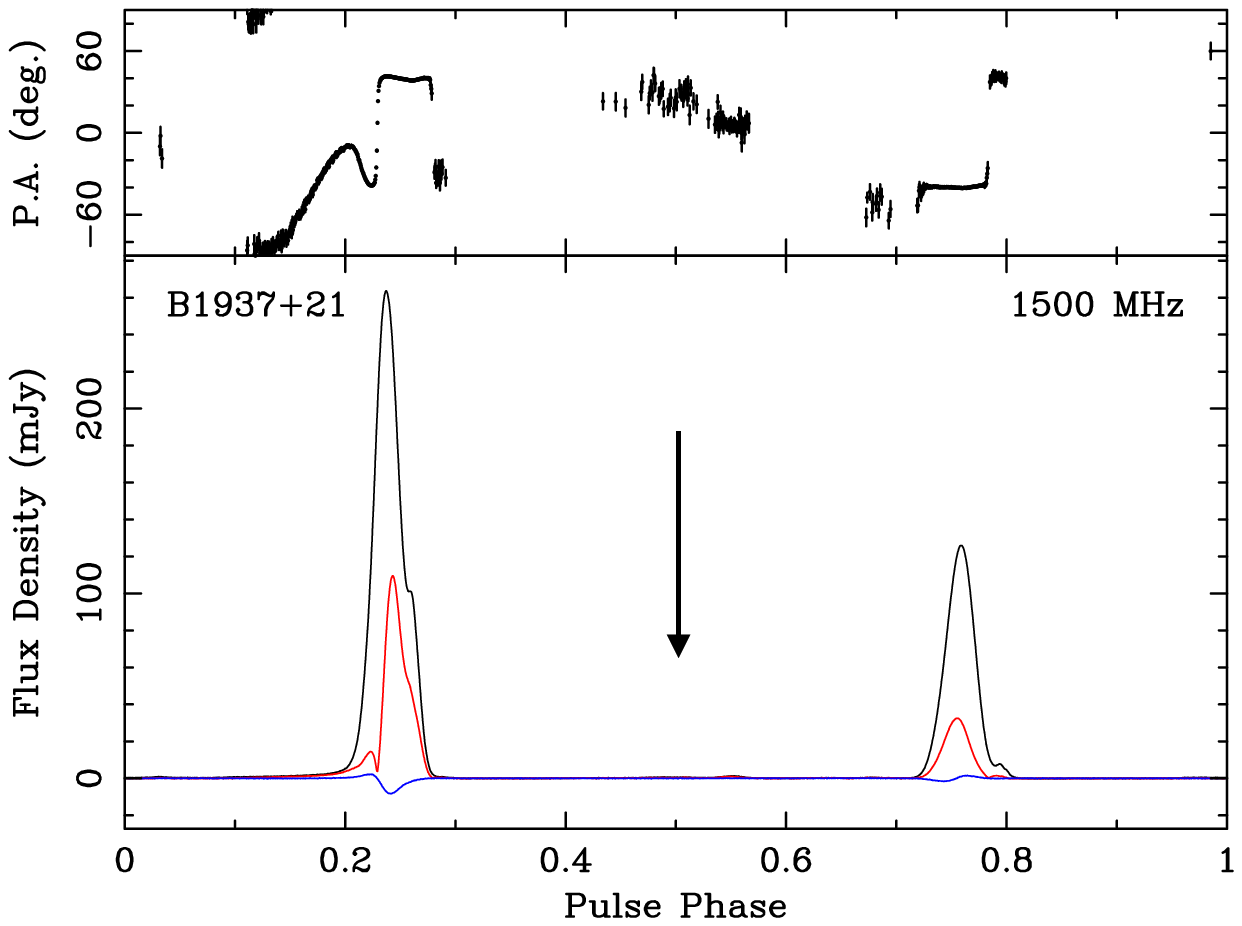}}}\\

{\mbox{\includegraphics[height=87mm,angle=270]{B1937_820_composite_micro.ps}}}& \ \ \ 
\hspace{-1.6 cm}
{\mbox{\includegraphics[height=87mm,angle=270]{B1937_1500_composite_micro.ps}}}\\

\end{tabular}
\label{fig-14}
\end{center}
\end{figure*}

\begin{figure*}[ht]
\begin{center}
\caption{Pulse profiles for pulsars J2010--1323 and J2145--0750 including microcomponents.  The black arrow points to the location of the microcomponent in each J2145--0750 profile.  The black line is the total intensity, red is the linear polarization, and blue is the circular polarization. The polarization position angle is shown in the top panel.}
\begin{tabular}{@{}ll@{}}

{\mbox{\includegraphics[height=87mm,angle=270]{J2010_820_composite.ps}}}& \ \ \ 
\hspace{-1.6 cm}
{\mbox{\includegraphics[height=87mm,angle=270]{J2010_1500_composite.ps}}}\\

\hspace{-1.2 cm}
{\mbox{\includegraphics[height=115mm,angle=90]{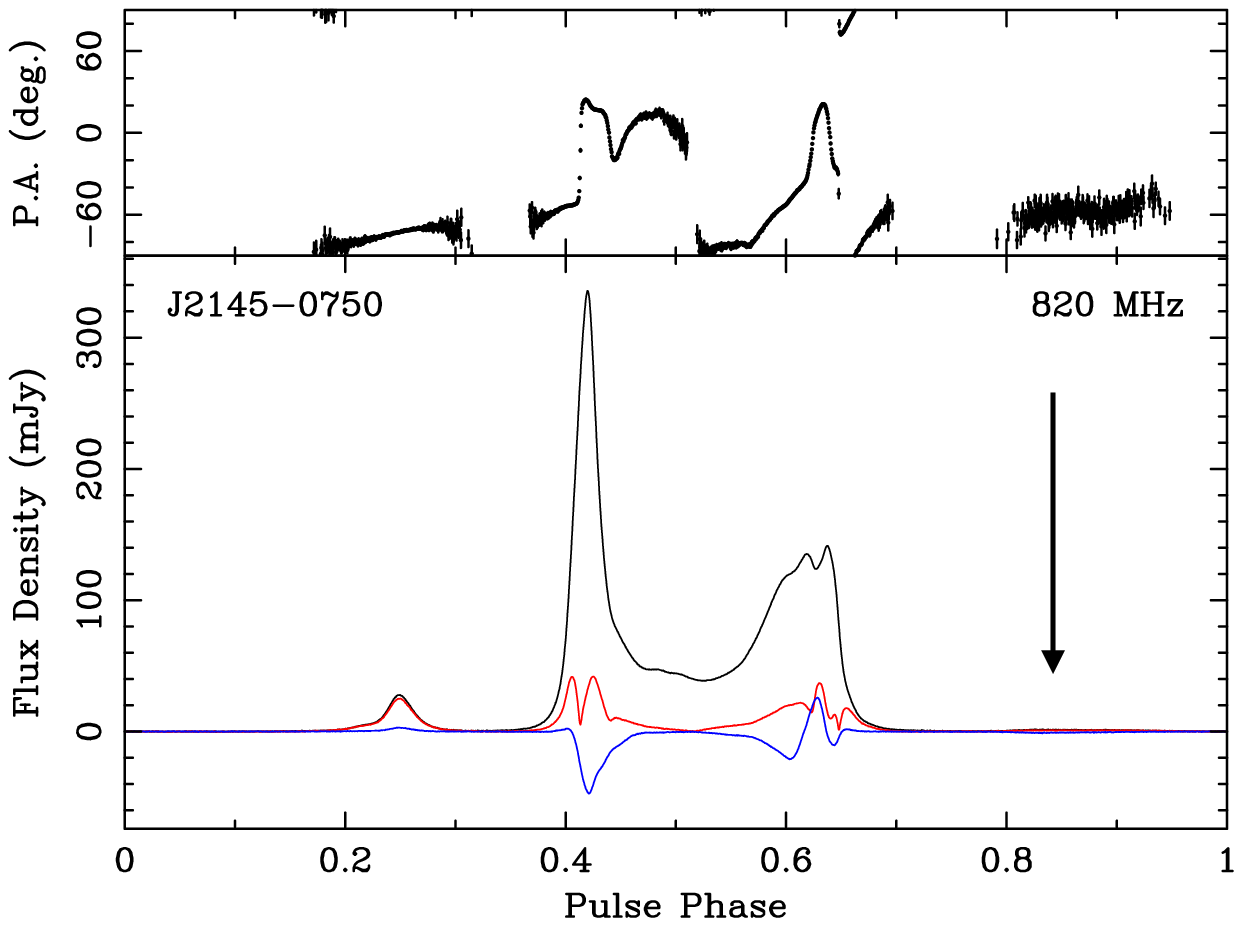}}}& \ \ \ 
\hspace{-2.7 cm}
{\mbox{\includegraphics[height=115mm,angle=90]{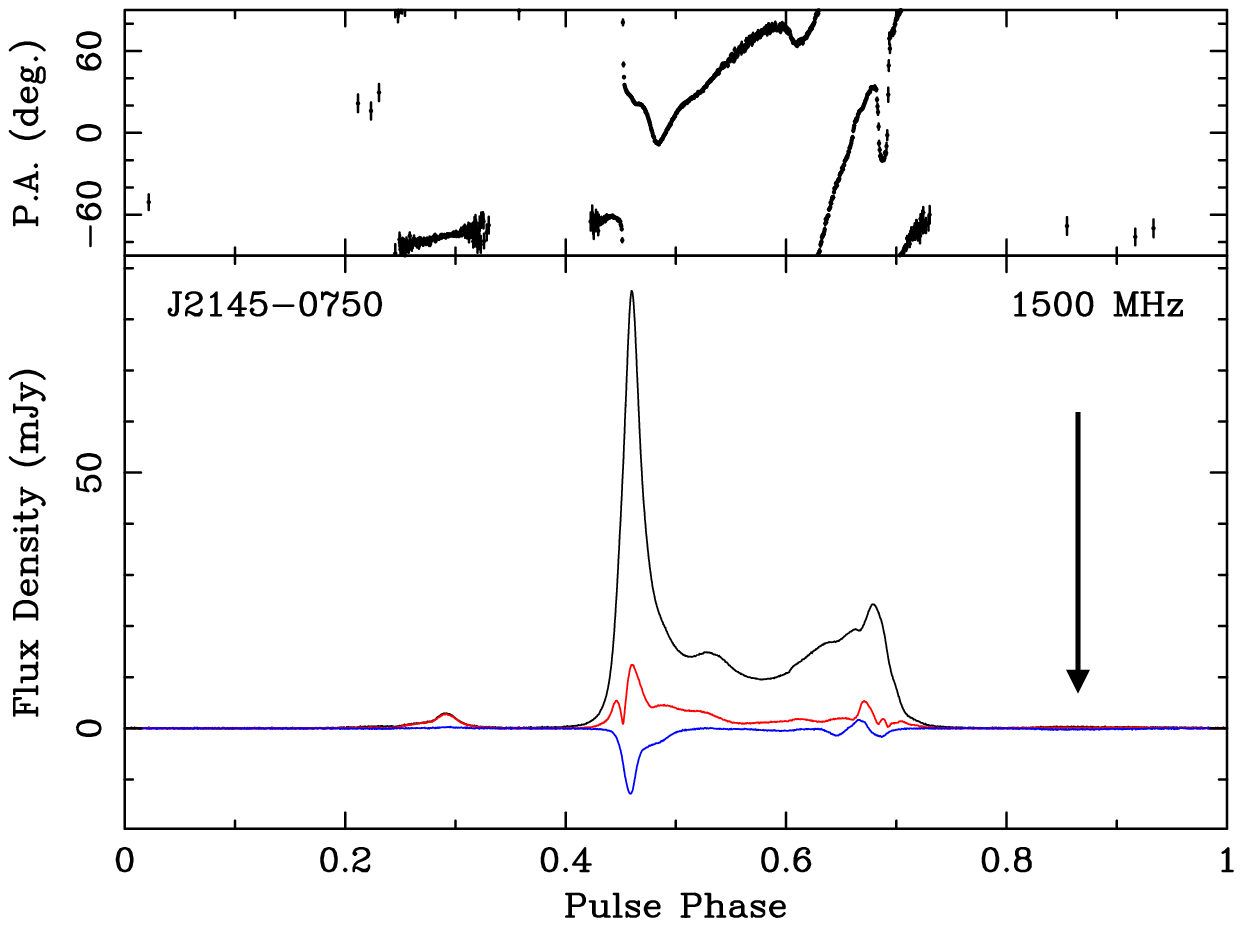}}}\\

{\mbox{\includegraphics[height=87mm,angle=270]{J2145_820_composite_micro.ps}}}& \ \ \ 
\hspace{-1.6 cm}
{\mbox{\includegraphics[height=87mm,angle=270]{J2145_1500_composite_micro.ps}}}\\

\end{tabular}
\label{fig-15}
\end{center}
\end{figure*}

\begin{figure*}[ht]
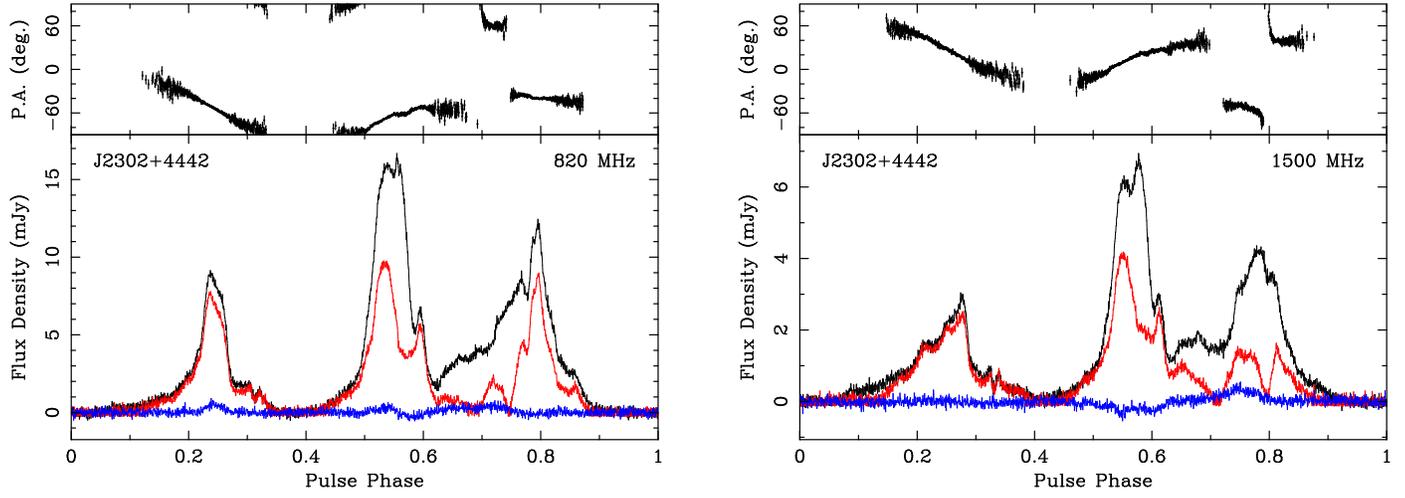

\begin{center}
\caption{Pulse profiles for pulsars J2302+4442. The black line is the total intensity, red is the linear polarization, and blue is the circular polarization. The polarization position angle is shown in the top panel.}
\begin{tabular}{@{}ll@{}}
 \vspace{2.35 cm}
{\mbox{\includegraphics[height=87mm,angle=270]{J2302_820_composite.ps}}}& \ \ \ 
\hspace{0.2 cm} 
{\mbox{\includegraphics[height=87mm,angle=270]{J2302_1500_composite.ps}}}\\

\end{tabular}

\label{fig-16}
\end{center}
\end{figure*}

The position angle (PA) sweep is shown in the top panel of Figures \ref{fig-5}--\ref{fig-16}; many of our pulsars (e.g. PSRs~J1455$-$3330, J1918$-$0642, and J2145$-$0750), show very complex PA sweeps, which would require a model more sophisticated than the RVM. Only two pulsars in our data set show a quasi-S-shaped curve in the PA. Using PSRCHIVE, we searched an 18 by 18 grid in which $\alpha$ (the angle between the spin axis and the magnetic axis) and $\zeta$ (the sum of $\alpha$ and the angle between the magnetic axis and sightline $\beta$) are varied from 5 to 175 degrees in steps of 10 degrees. We perform this fitting for both pulsars at each frequency. The only significant result is for the L-band observation of PSR~J1600$-$3053, where $\alpha$ = 162.8 $\pm$  5.9, $\beta$ = 2.35 $\pm$ 8.9 for a fit that has a  $\chi^2_{r}$  value of 11.05. This shows that PAs are very difficult to fit in MSPs and a more sophisticated model incorporating emission far from the polar caps and/or more complex magnetic field structures is required to fit the position angle sweeps.

\begin{deluxetable*}{l|c|c|c|c|c|c|c}
\tabletypesize{\footnotesize}
\tablewidth{0pt}
\tablecaption{Dispersion measure trends. The results of fitting a linear trend, a purely sinusoidal, and a sinusoidal + linear trend to the magnetic fields. A weighted least-squares fitting routine was performed and the periods of the sinusoidal fits first estimated with a Lomb-Scargle periodogram and then refined in the fitting routine.} Any period that had less than a 5\% false alarm probability was not considered significant. The trend reported is the one with the smallest $\chi^{2}_r$ value. The pre-fit $\chi^{2}_r$ value refers to the $\chi^{2}_r$ of fitting a horizontal line through the data.  \label{tbl-5}
\tablehead{
Pulsar &  Trend & dDM/dt &  Amplitude &Period & $\chi^{2}_r$ &Pre-Fit $\chi^{2}_r$ & Period FAP\\
& & (10$^{-4}$ pc cm$^{-3}$ yr$^{-1}$) & (pc cm$^{-3}$) & (days) &  &
}
\startdata
J1012+5307&Both&0.4(2)&1.6(2)&874(51)&0.8&1.6&0.7\% \\
J1713+0747&Both&--0.38(5)&0.61(7)&366(7)&3.1&6.7&0.3\%\\
J1744$-$1134&Both&0.2(1)&0.9(2)&425(16)&3.1&4.5&0.9\%\\
J1909$-$3744&Both&--5.3(2) $\times$ 10$^{-4}$&4(1) $\times$ 10$^{-5}$&568(46)&34&431&1.8\%\\
J2302+4442&Linear&--3.5(7)&---&---&---&2.5&---\\
\enddata
\end{deluxetable*}

\begin{deluxetable*}{l|c|c|c|c|c|c|c|c}
\tabletypesize{\footnotesize}
\tablewidth{0pt}
\tablecaption{Magnetic field trends. The results of fitting a linear trend, a purely sinusoidal, and a sinusoidal + linear trend to the magnetic fields. A weighted least-squares fitting routine was performed and the periods of the sinusoidal fits were first estimated with a Lomb-Scargle periodogram and then refined in the fitting routine. Any period that had less than a 5\% false alarm probability was not considered significant. The trend reported is the one with the smallest $\chi^{2}_r$ value.The pre-fit $\chi^{2}_r$ value refers to the $\chi^{2}_r$ of fitting a horizontal line through the data. \label{tbl-6}} 

\tablehead{
Pulsar & Frequency& Trend & dB/dt &  Amplitude &Period & $\chi^{2}_r$ &Pre-Fit $\chi^{2}_r$ & Period FAP\\
&(MHz)&&($\mu$G yr$^{-1}$) & ($\mu$G) & (days) &  &
}
\startdata
J1600$-$3053 & 1500 & Sine & --- & 0.041(9) & 366(14) & 11.21 & 16 & 3.60\% \\
J1643$-$1224 & 1500 & Both & 0.007(2) & 0.021(3) & 374(8) & 3.9 & 11.1 & 0.26\% \\
J1713+0747 & 820 & Both & 0.02(2) & 0.14(2) & 678(27) & 41.31 & 72 & 0.04\% \\
J1918$-$0642 & 820 & Linear & 0.14(2) & --- & --- & 105.62 & 190 & ---\\
B1937+31 & 820 & Sine & --- & 0.025(5) & 366(11) & 4.5 & 6.03 & 1.50\% \\
\enddata
\end{deluxetable*}

\subsection{Variations in Measured Values}

For each of the three parameters (ionosphere-corrected RM, DM, and $\langle B_{\parallel} \rangle$), we performed a least-squares fit weighted by the uncertainties for a purely linear trend, a purely sinusoidal trend, and a combination of the two for all pulsars for which we have greater than one year of data. We only performed a sinusoidal and combination fit if a significant period with a false alarm probability (FAP) less than 5\% was first identified through a Lomb-Scargle periodogram analysis. This FAP was calculated using the formula from \cite{scargle}, which uses the length of the dataset and power spectral density to determine the probability that the period of the Lomb-Scargle periodogram is detected by random chance. That period was then used as the initial guess for the fitting. The reduced chi-squared ($\chi^2_{r}$) values were calculated for each fit; the trend reported for each pulsar is the model with the smallest $\chi^2_{r}$ value. 

The parameters for the trends are reported in Tables \ref{tbl-5} and \ref{tbl-6} and the data containing the best fit trend lines are shown in Figures \ref{fig-17}$-$\ref{fig-21}. We plot the two frequencies separately in order to gauge which trends are truly astrophysical. In addition, one frequency may be more sensitive than another due to the pulsar's spectral index, DM, or RFI, so we may only see the trend significantly in one.

\begin{figure*}[ht] 
\centering
\includegraphics[height=80mm,angle=0,  clip=true]{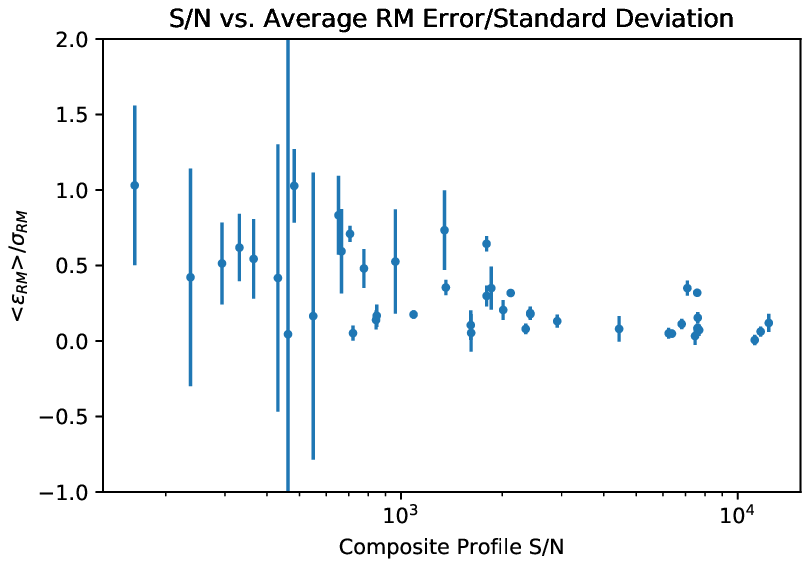}
\caption{The S/N of the composite profiles plotted against the average error divided by the standard deviation for each pulsar. The slope downward toward a higher S/N shows that the errors are either underestimated at high S/N or there is significant astrophysical variation. Note: the full error bars for PSR J0636+5128 are not shown.} 
\label{fig-4}
\end{figure*}

In the absence of astrophysical variations, we would expect the root-mean-square deviation of RMs to equal roughly the average 1-sigma error on those measurements. In Figure \ref{fig-4}, following \citet{caleb}, we plot the ratio of the average RM error to the standard deviation vs. S/N. We find that these values are typically smaller than one, indicative of either real astrophysical variations or underestimated errors. We see more variation in RM values at higher S/N values. This seems to indicate that in the moderately high S/N regime RM errors are accurate, but that in the very high S/N regime, RM errors may be underestimated. Note that \citet{caleb} found that RM errors measured for very low S/N profiles ($\lessapprox$ 17) were also underestimated. There are likely systematic effects that are not taken into account at high S/N, as shown by the lack of significant trends in the plots for bright pulsars such as J1713+0747.

\subsubsection{DM Variations}

Dispersion measure trends are shown in Table \ref{tbl-5} and the second panel of Figures \ref{fig-17}--\ref{fig-21}. We detect significant trends in five pulsars (PSRs J1012+5307, J1713+0747, J1744$-$1134, J1909$-$3744, and J2302+4442). They all exhibit some kind of linear trend, though four exhibit a sinusoidal trend combined with a linear trend. Our results are similar to those of \citet{Jones2017} for the NANOGrav 9-year data set which used most of the same observational data underlying the present work. There are twelve pulsars that overlap between our data sets, and \citet{Jones2017} finds significant trends in eleven, whereas we only find trends in three. For those pulsars in which we do both find significant trends, the slopes are roughly the same magnitude and the trends are the same for two of them (we find an extra sinusoidal trend in PSR J1012+5307). The differences in our results can be attributed to a lack of overlap in the data sets. \citet{Jones2017} is sensitive to longer term trends because they fit nine years of data, whereas we only include four years in our analysis.

\citet{Yan2011RM} also fit DM trends with one year of data observed at 1.4~GHz from \citet{you2007}, though they fit only for linear trends. Their data are not sensitive enough for high-precision DM measurements, and they therefore only report upper limits on the slope of DM variations. For the pulsars that overlap between their paper and this one, the upper limits for only three (PSRs J0613$-$0200, J1024$-$0719, and B1937+21) are significant. We find no significant trends in the DM of any of those pulsars. Discrepancies could be caused by our longer baselines. Though \citet{Yan2011RM} predict that the slopes they measure are believed to be representative of the longer-term gradients, the linear trends they see are most likely fitted out over longer data sets (which is seen in our analysis).

In addition, \citet{donner2020} analyzed DM variations in 36 MSPs at a frequency of 150 MHz in a data set  that spans 2012$-$2020. Nine pulsars overlap between our data sets. While they report linear trends for all nine of the pulsars, we find linear trends for only four of them. Their much lower observational frequency make them more sensitive to DM variations. This, combined with their longer datasets, likely explain this discrepancy; for the four pulsars for which we both measure trends, ours are generally of the same order of magnitude and are all of the same sign to those of \citet{donner2020}.

\subsubsection{Variations in Measured $\langle B_{\parallel} \rangle$}\label{5.1.3}

\begin{figure*}[ht]
\begin{center}

\caption{Dispersion measure and magnetic field changes over time for pulsars J0340+4130, J0613--0200, J0645+5158, and J1012+5307. The uncertainties on the DM come from those on the DMX value and uncertainties on the magnetic field are a combination of the uncertainties on the ionosphere-corrected RM (which are a combination of those of fitting for Faraday rotation and from the ionospheric correction) and the DM. No trendlines are shown because the lowest $\chi^2_{r}$ value for the fits was that of a horizontal line with a slope of zero. }
\begin{tabular}{@{}ll@{}}
\hspace{-1 cm}
{\mbox{\includegraphics[height=70mm,angle=0]{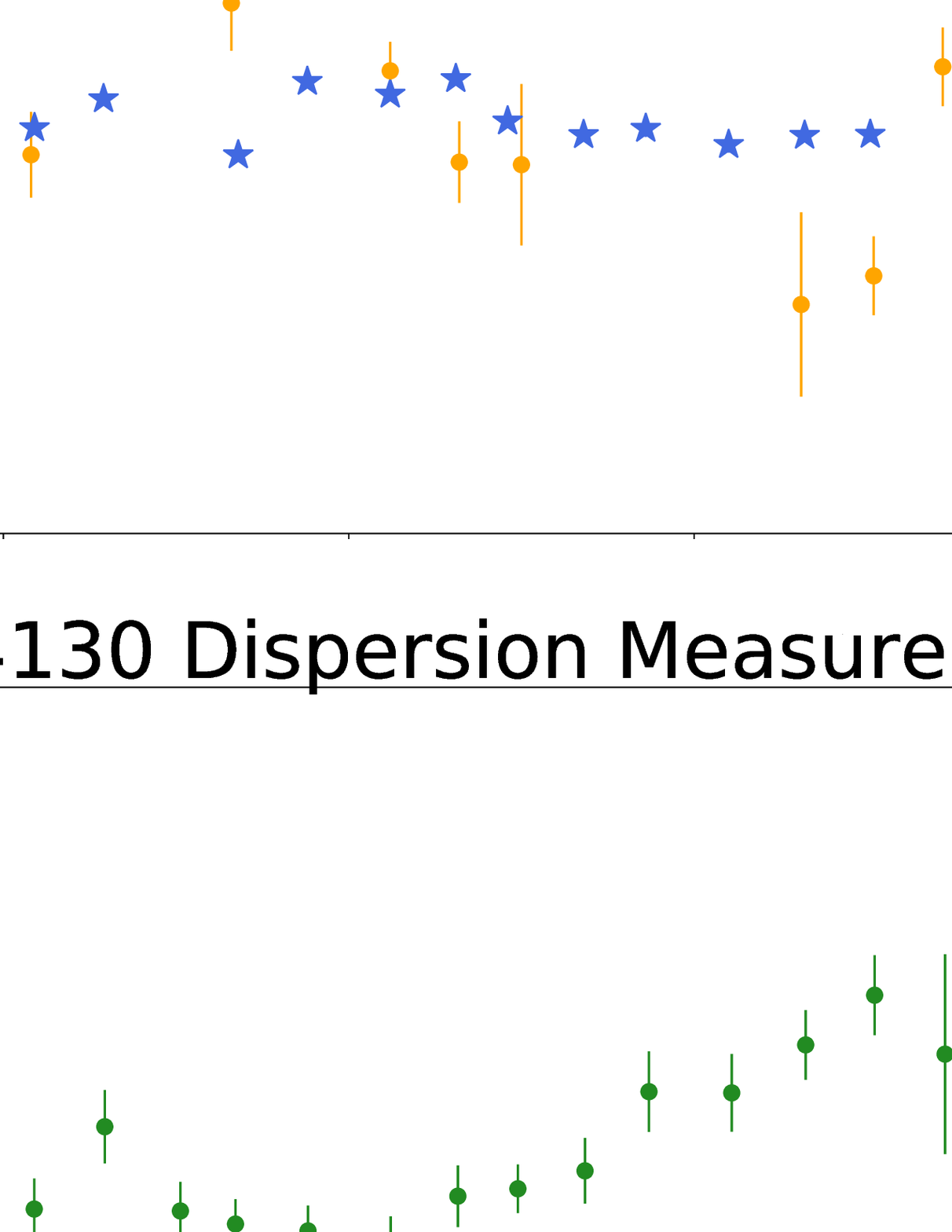}}}   & \ \ \ 
\hspace{-1.3 cm}
{\mbox{\includegraphics[height=70mm,angle=0]{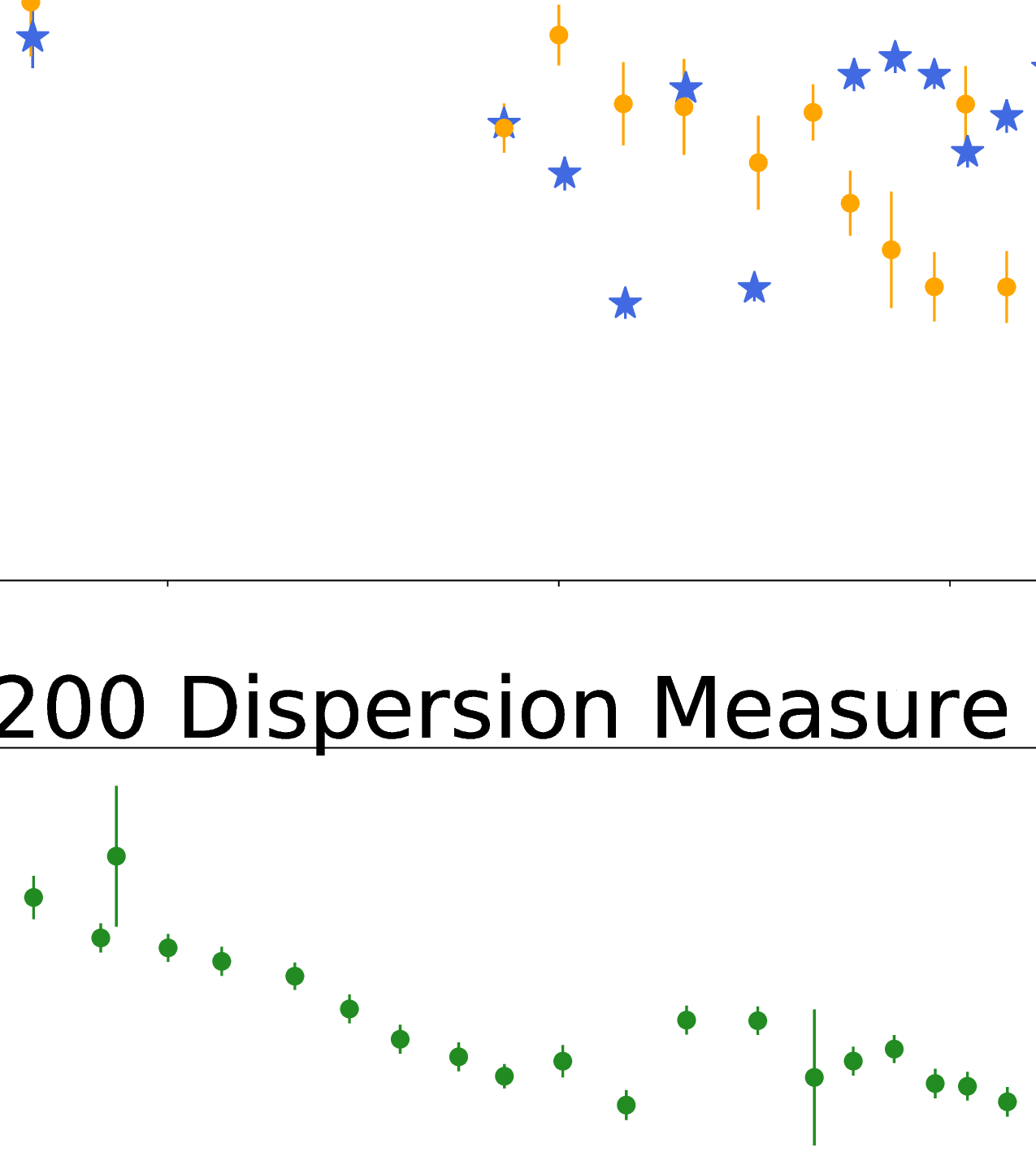}}}\\
\hspace{-1 cm}
{\mbox{\includegraphics[height=70mm,angle=0]{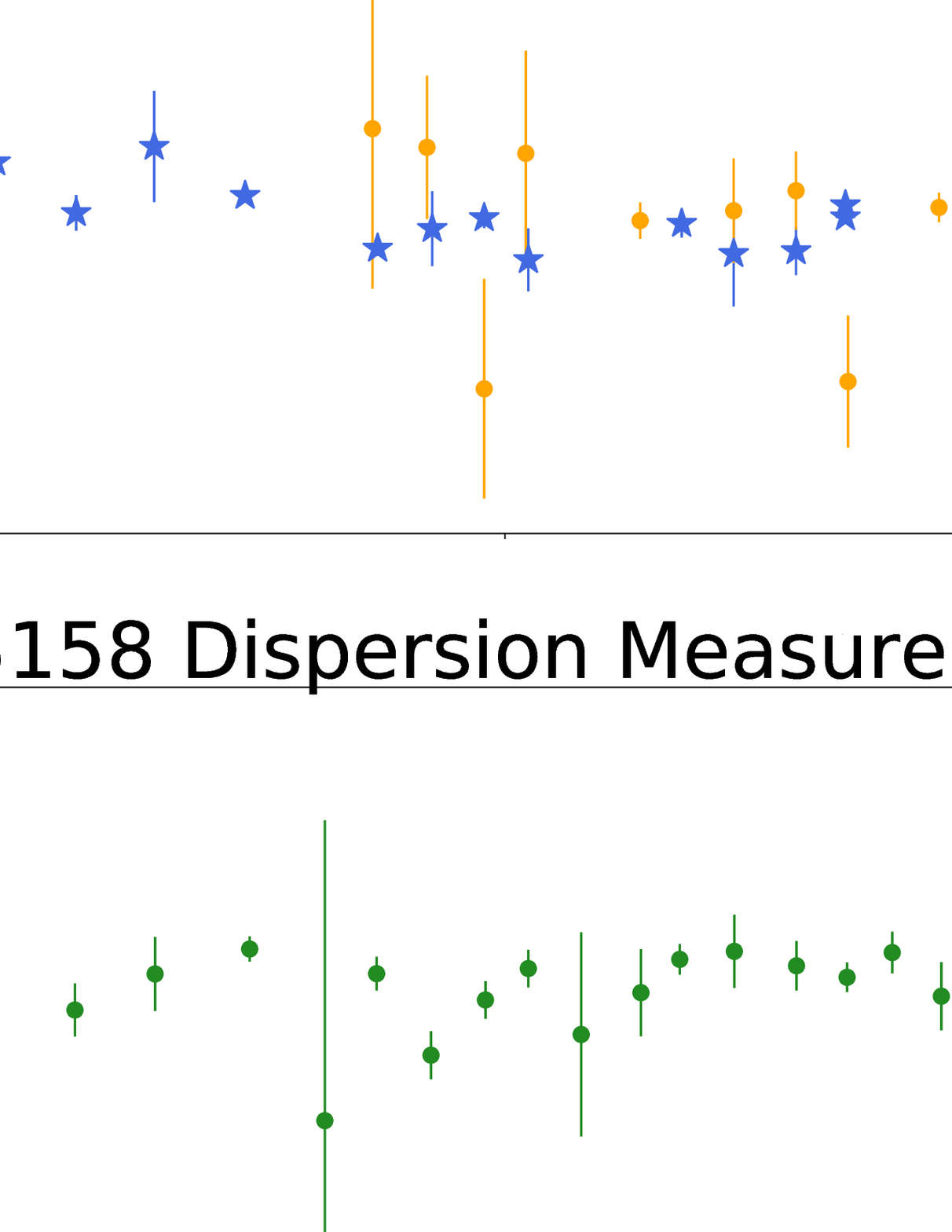}}}   & \ \ \ 
\hspace{-1.3 cm}
{\mbox{\includegraphics[height=70mm,angle=0]{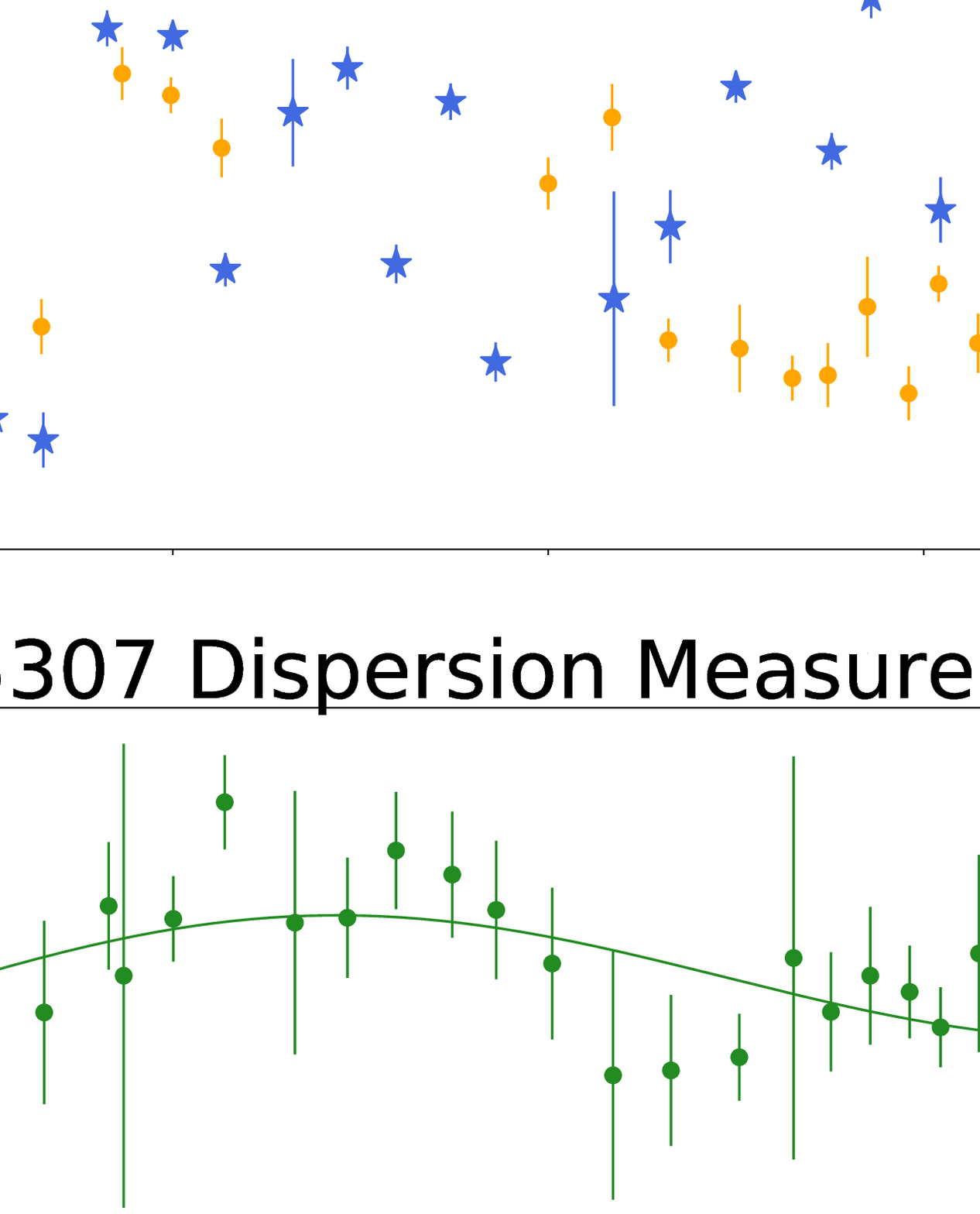}}}\\

\end{tabular}
\label{fig-17}
\end{center}
\end{figure*}

\begin{figure*}[ht]
\begin{center}
\caption{Dispersion measure and magnetic field variations over time for pulsars J1024$-$0719, J1455$-$3330, J1600$-$3053, and J1614$-$2230. The uncertainties on the DM come from those on the DMX value and uncertainties on the magnetic field are a combination of the uncertainties on the ionosphere-corrected RM (which are a combination of those of fitting for Faraday rotation and from the ionospheric correction) and the DM.  Any trendlines shown represent the trend with the lowest $\chi^2_{r}$ value. If no trendlines are shown then the lowest $\chi^2_{r}$ value for the fits was that of a horizontal line with a slope of zero. Note: the plots for J1614--2203 contain two outliers at epochs of small ecliptic angle (less than 3 degrees) (MJDs 55892 and 55893, as discussed in Section \ref{5.1.3}). These points are excluded from the fitting and the mean RM and B calculation but included in the plot to show the spike in RM, DM, and B when the pulsar is close to the Sun.}

\begin{tabular}{@{}ll@{}}
\hspace{-1 cm}
{\mbox{\includegraphics[height=70mm,angle=0]{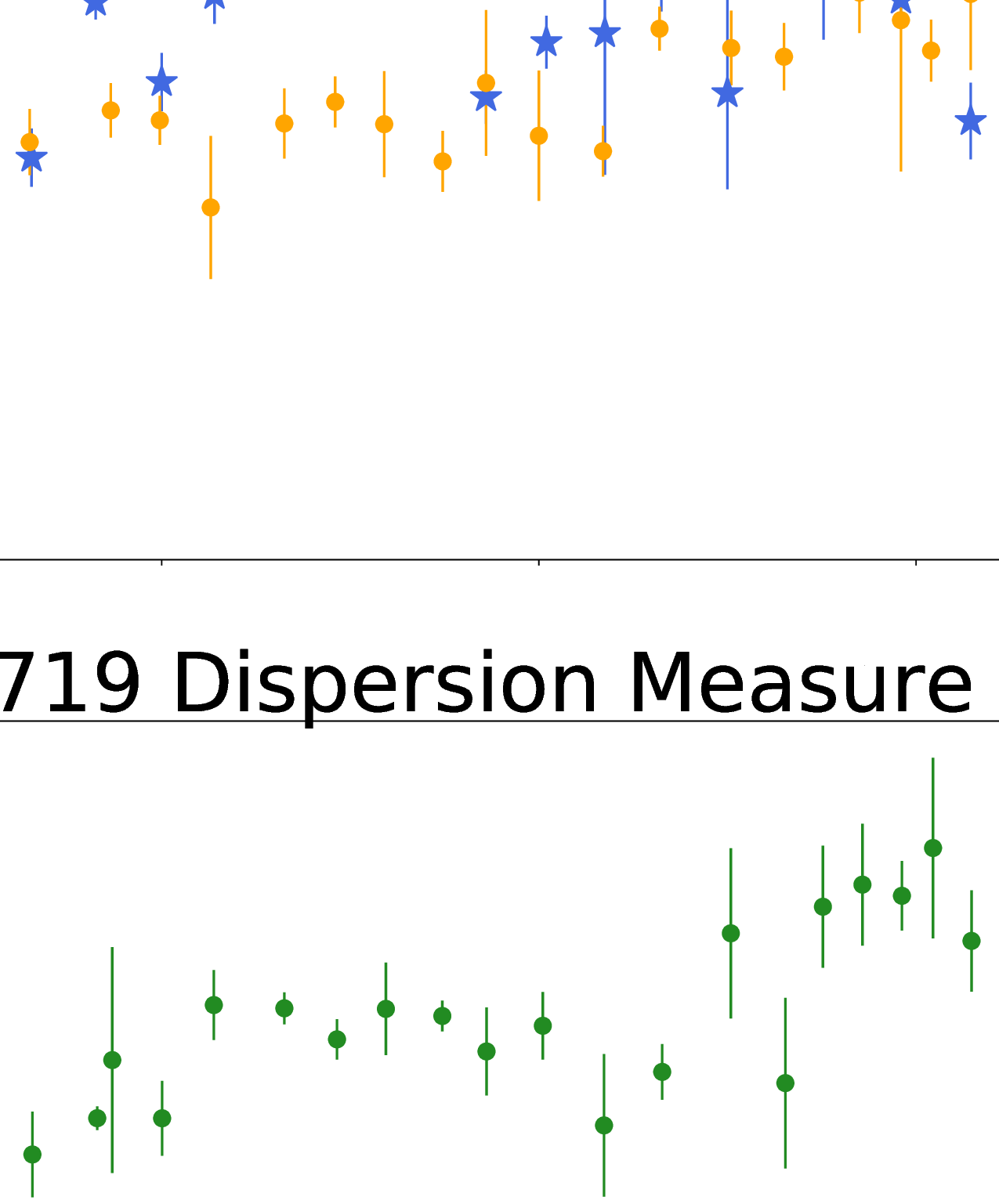}}}   & \ \ \ 
\hspace{-1.3 cm}
{\mbox{\includegraphics[height=70mm,angle=0]{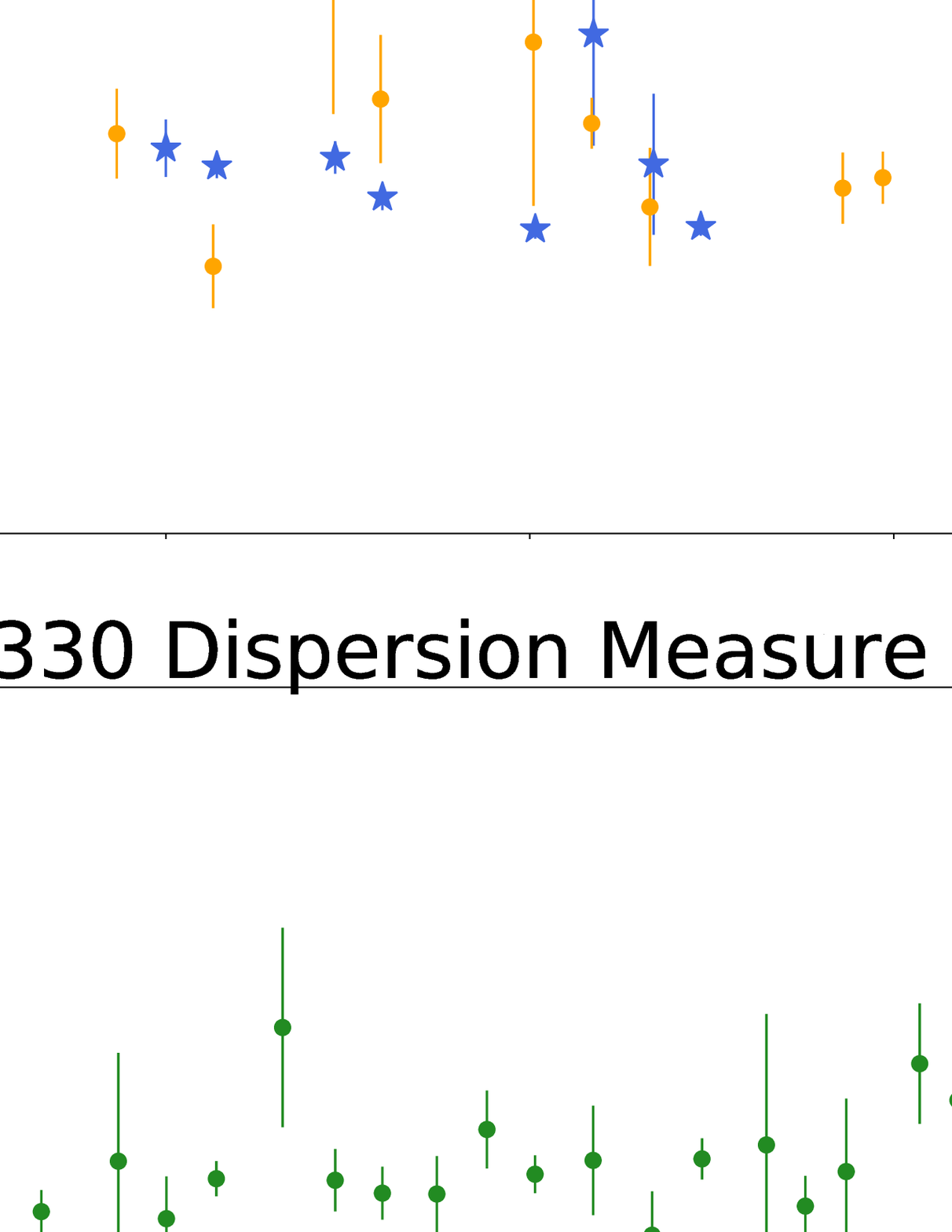}}}\\
\hspace{-1 cm}
{\mbox{\includegraphics[height=70mm,angle=0]{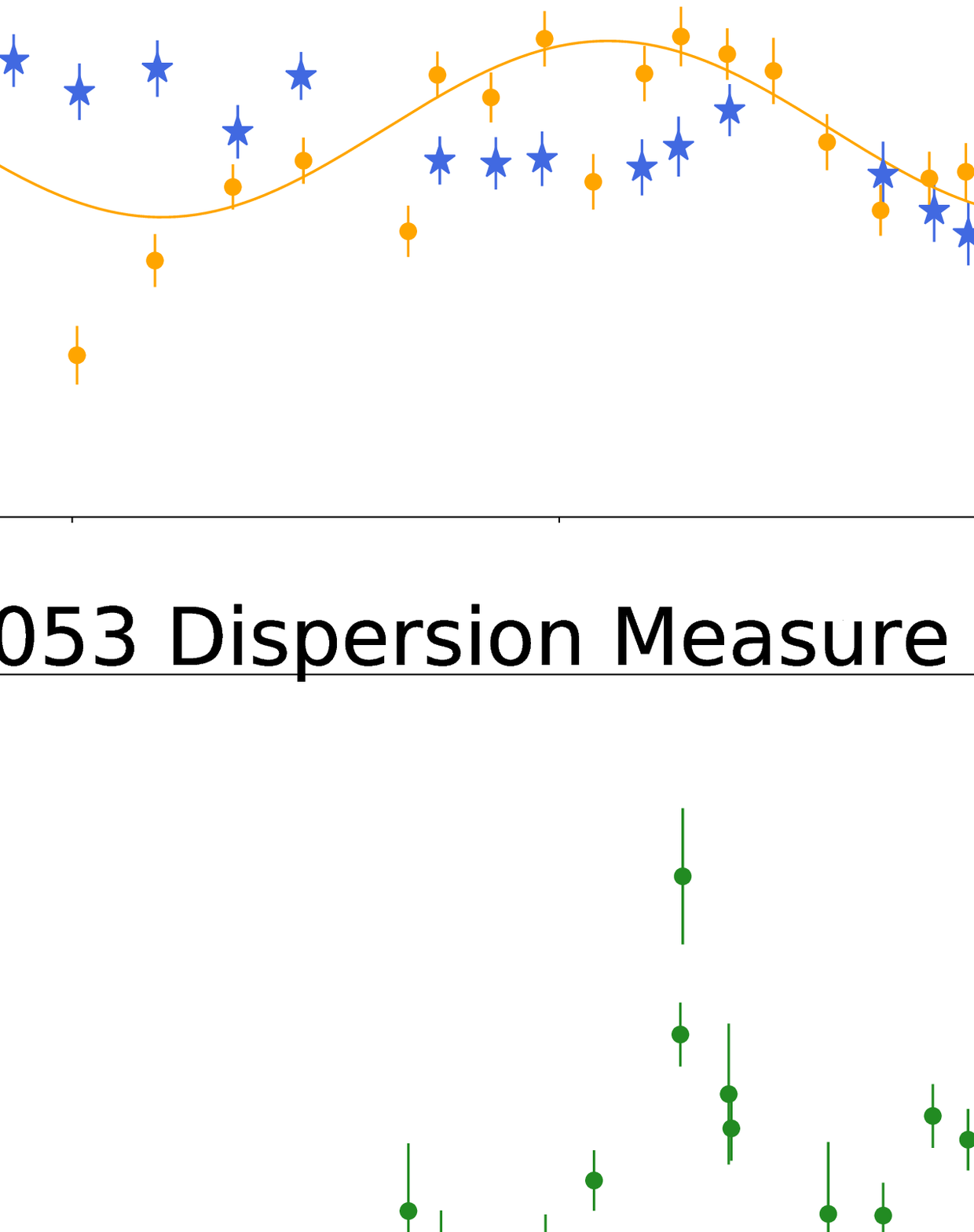}}}   & \ \ \ 
\hspace{-1.3 cm}
{\mbox{\includegraphics[height=70mm,angle=0]{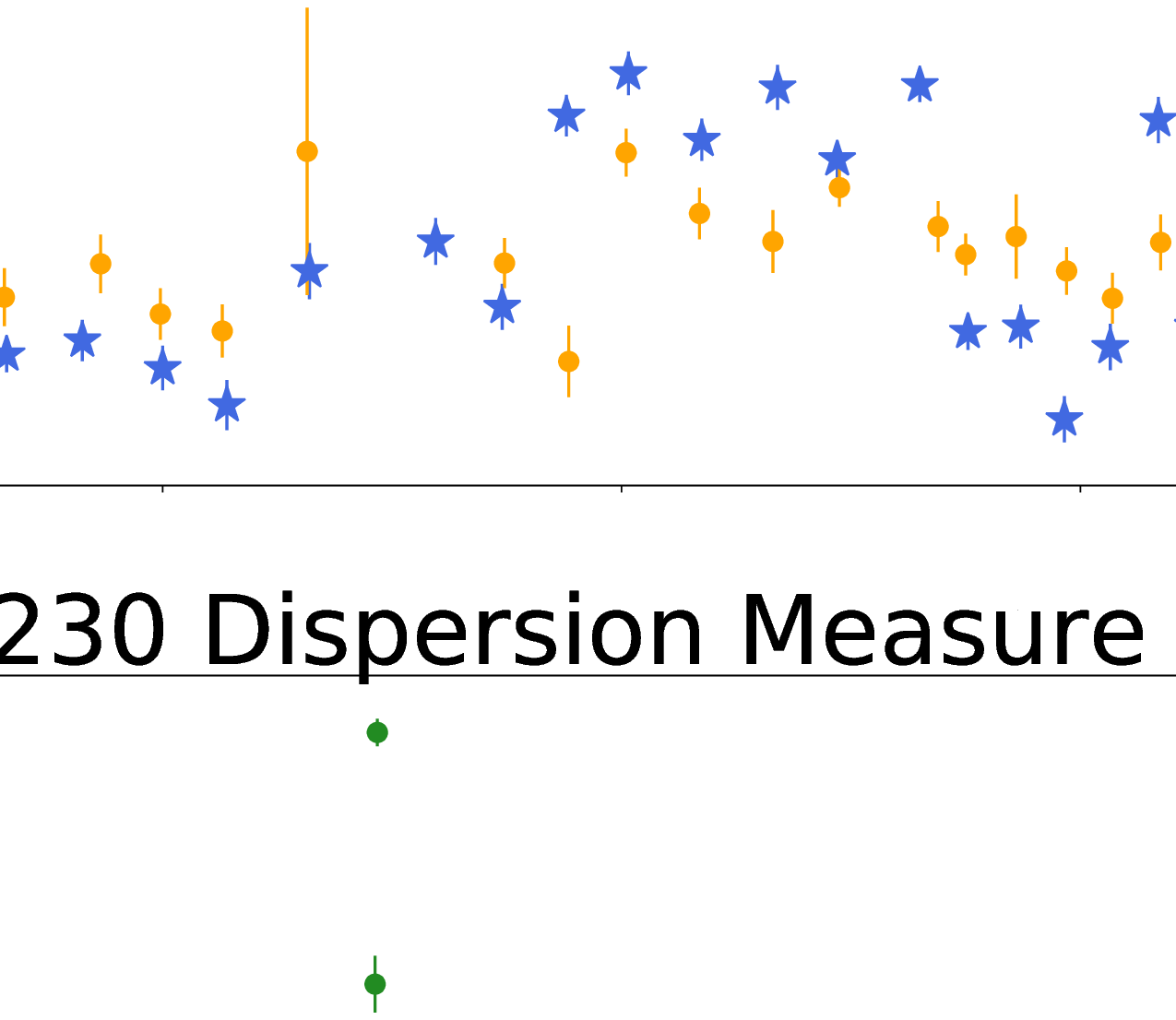}}}\\

\end{tabular}
\label{fig-18}
\end{center}
\end{figure*}

\begin{figure*}[ht]
\begin{center}
\caption{Dispersion measure and magnetic field changes over time for pulsars J1643$-$1224, J1713+0747, J1744$-$1134, and J1747$-$4036.The uncertainties on the DM come from those on the DMX value and uncertainties on the magnetic field are a combination of the uncertainties on the ionosphere-corrected RM (which are a combination of those of fitting for Faraday rotation and from the ionospheric correction) and the DM.  Any trendlines shown represent the trend with the lowest $\chi^2_{r}$ value. If no trendlines are shown then the lowest $\chi^2_{r}$ value for the fits was that of a horizontal line with a slope of zero.}
\begin{tabular}{@{}ll@{}}
\hspace{-1 cm}
{\mbox{\includegraphics[height=70mm,angle=0]{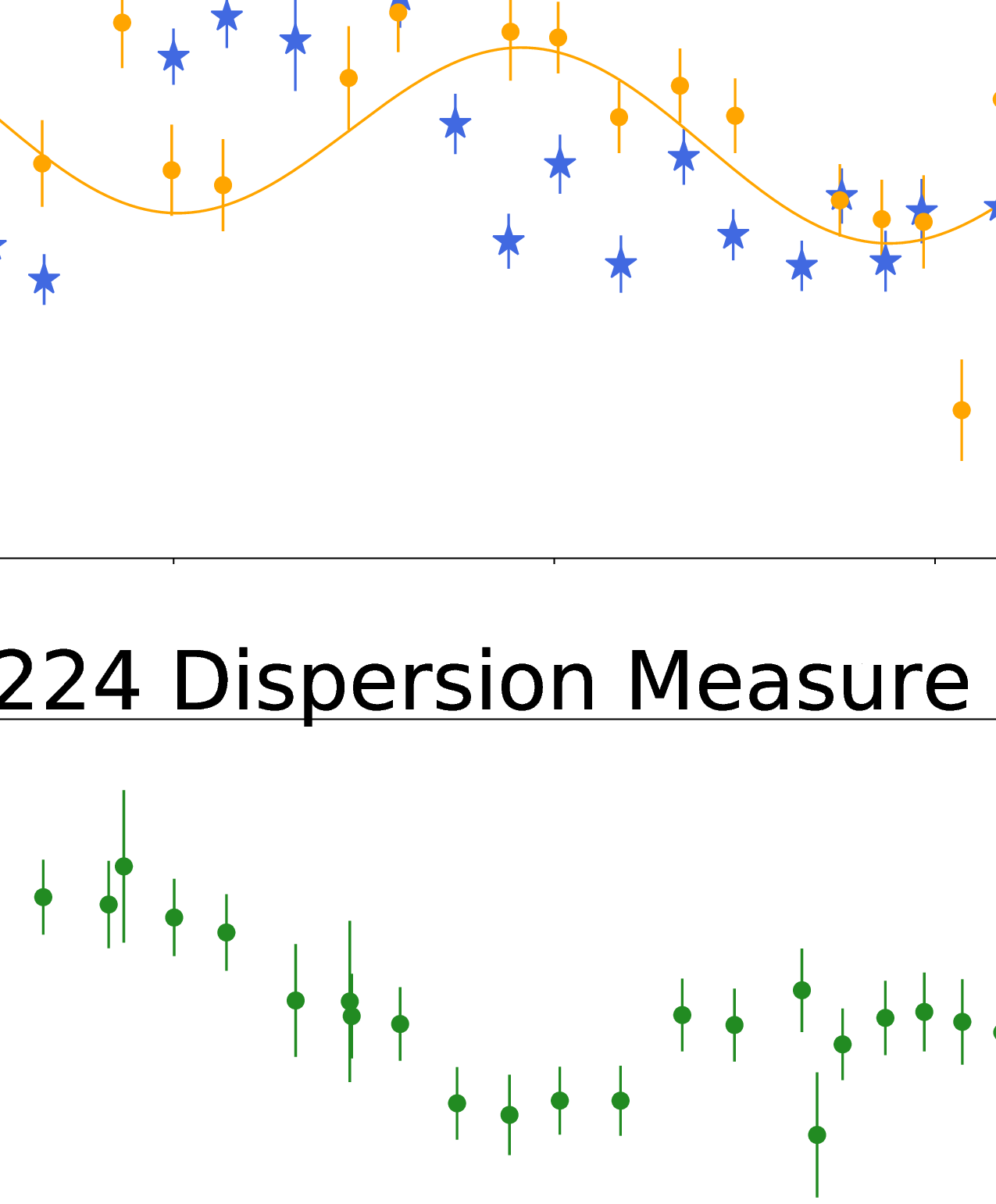}}}
\hspace{-1.1 cm}
{\mbox{\includegraphics[height=70mm,angle=0]{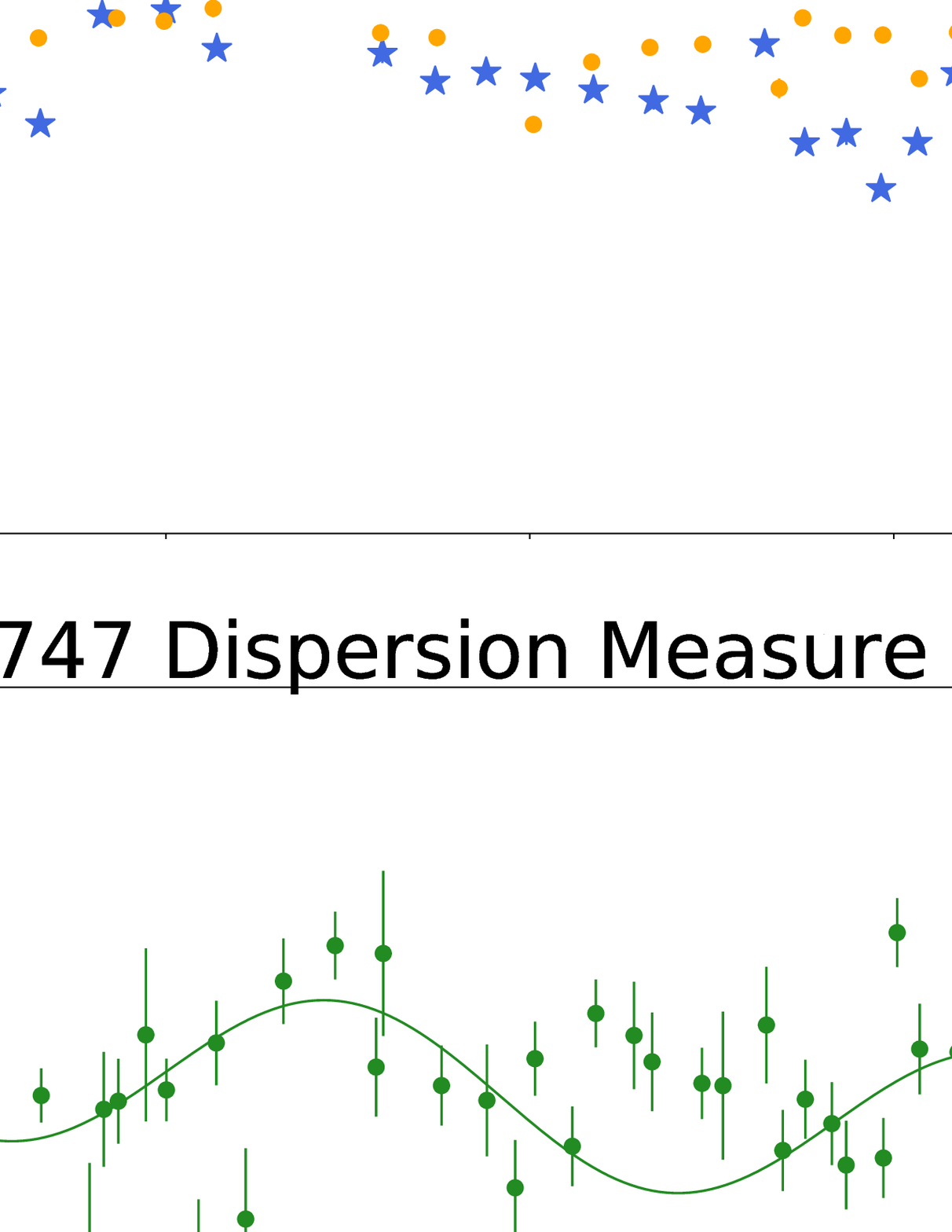}}}\\
\hspace{-1 cm}
{\mbox{\includegraphics[height=70mm,angle=0]{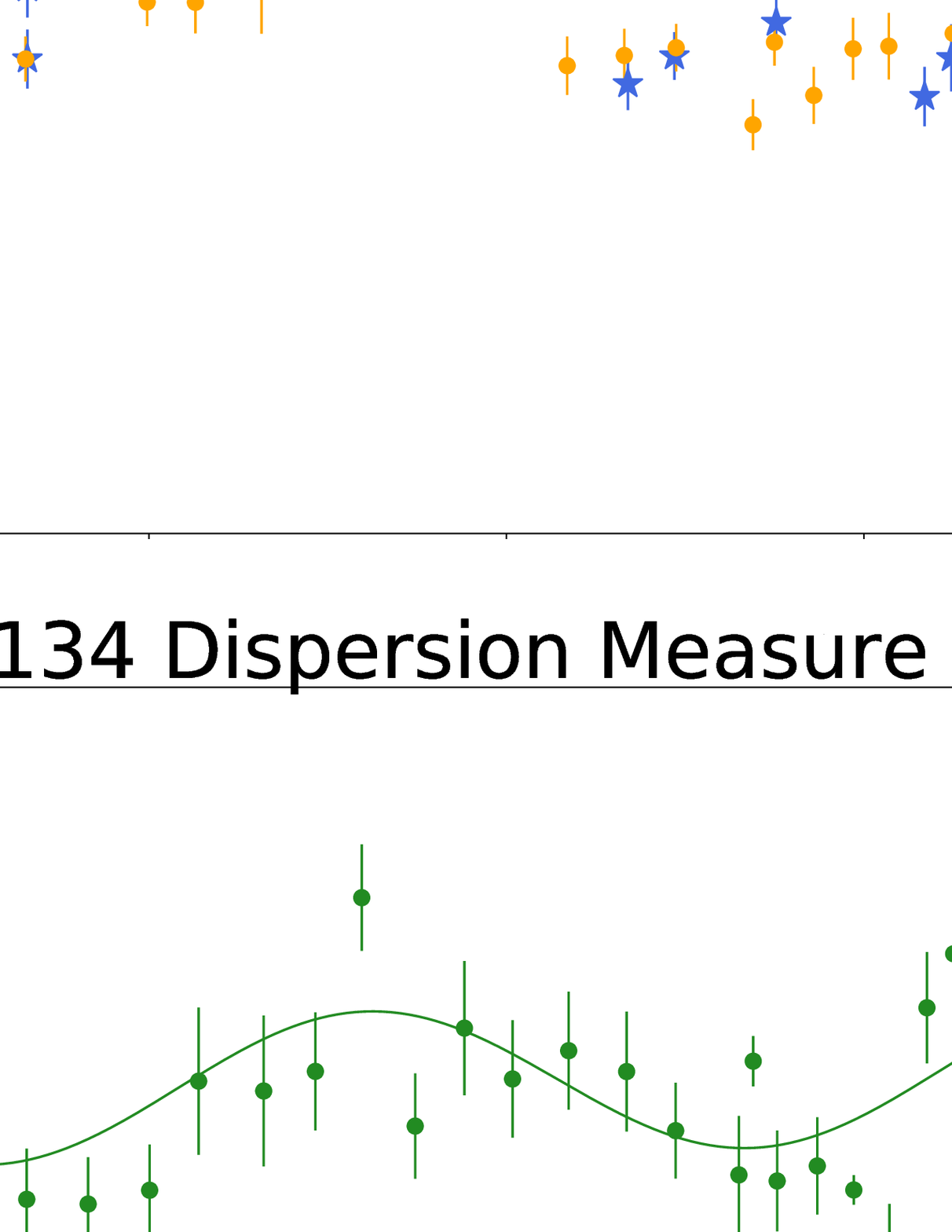}}}
\hspace{-1.1 cm}
{\mbox{\includegraphics[height=70mm,angle=0]{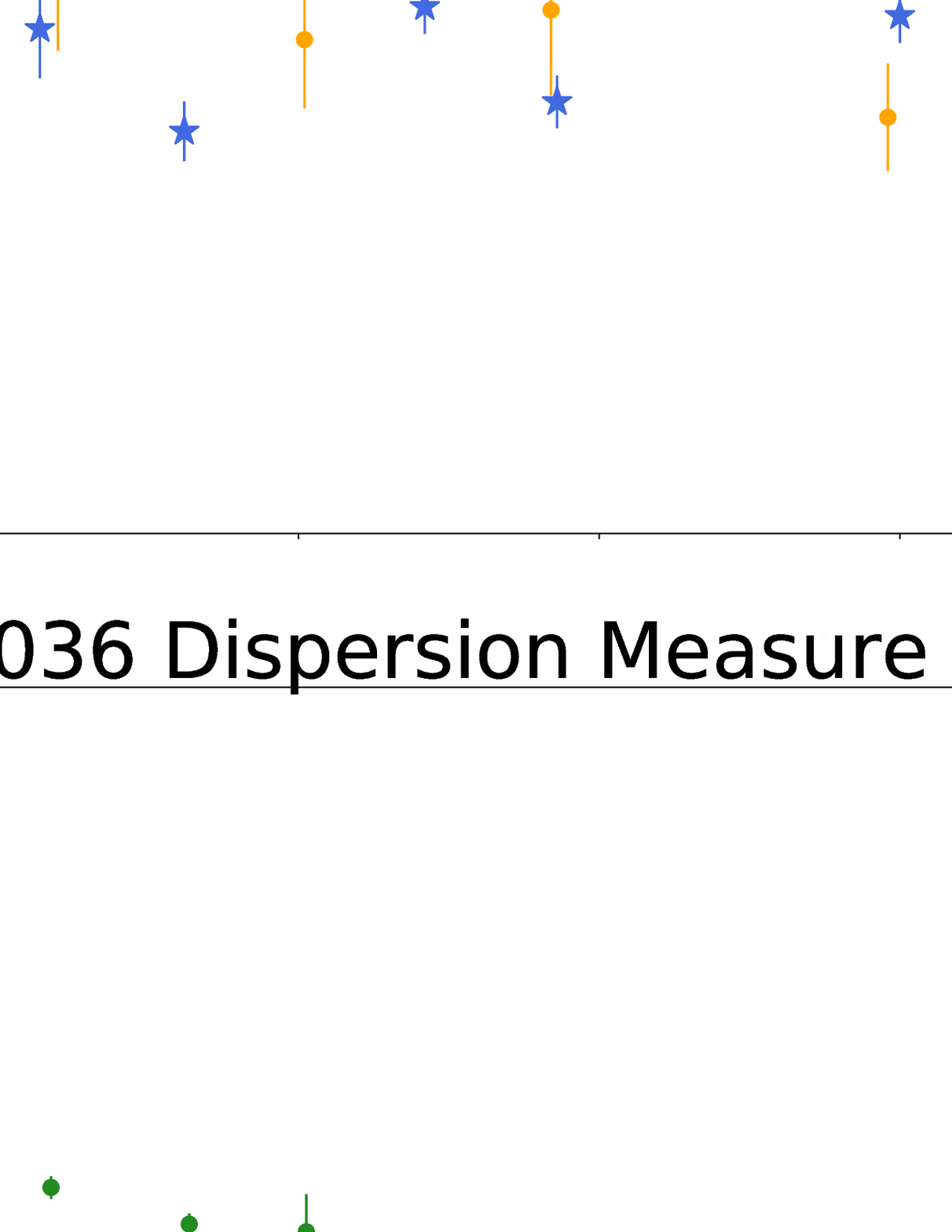}}}\\

\end{tabular}
\label{fig-19}
\end{center}
\end{figure*}

\begin{figure*}[ht]
\begin{center}
\caption{Ionosphere-corrected rotation measure, dispersion measure, and magnetic field changes over time for pulsars J1909$-$3744, J1918$-$0642, B1937+21, and J2010$-$1323. The uncertainties on the DM come from those on the DMX value and uncertainties on the magnetic field are a combination of the uncertainties on the ionosphere-corrected RM (which are a combination of those of fitting for Faraday rotation and from the ionospheric correction) and the DM.  Any trendlines shown represent the trend with the lowest $\chi^2_{r}$ value. If no trendlines are shown then the lowest $\chi^2_{r}$ value for the fits was that of a horizontal line with a slope of zero.}
\begin{tabular}{@{}ll@{}}
\hspace{-1 cm}
{\mbox{\includegraphics[height=70mm,angle=0]{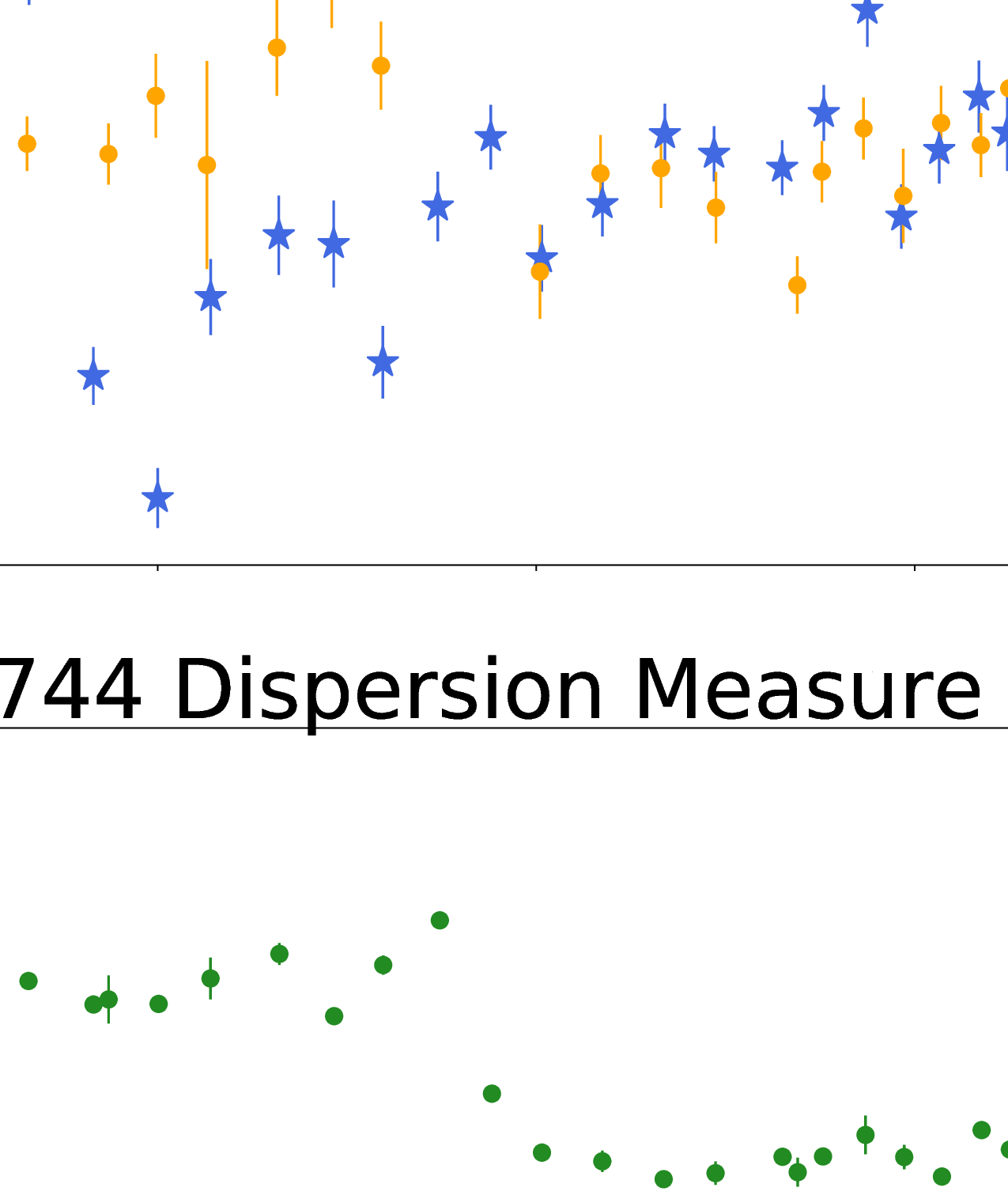}}} \ \ \
\hspace{-1.3 cm}
{\mbox{\includegraphics[height=70mm,angle=0]{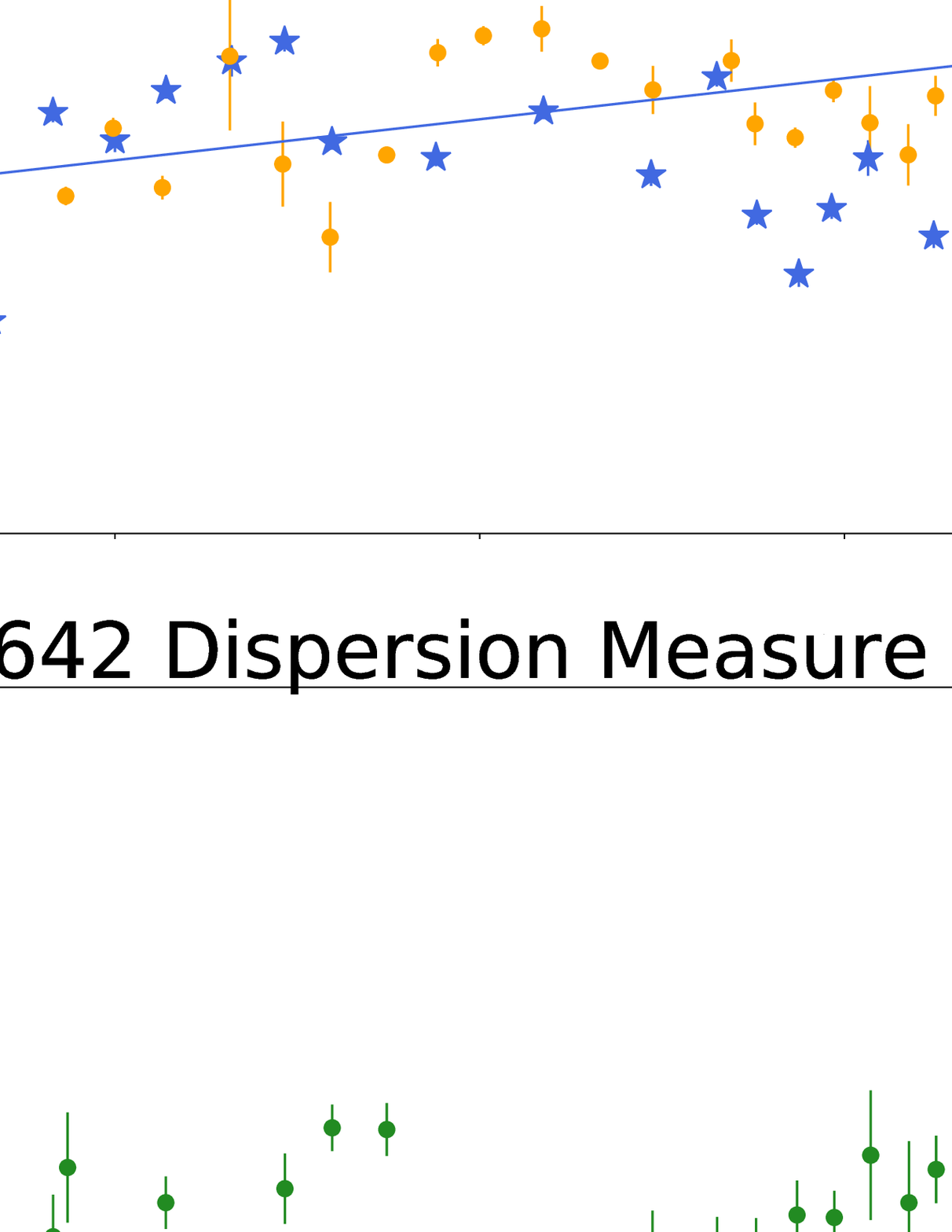}}}\\
\hspace{-1 cm}
{\mbox{\includegraphics[height=70mm,angle=0]{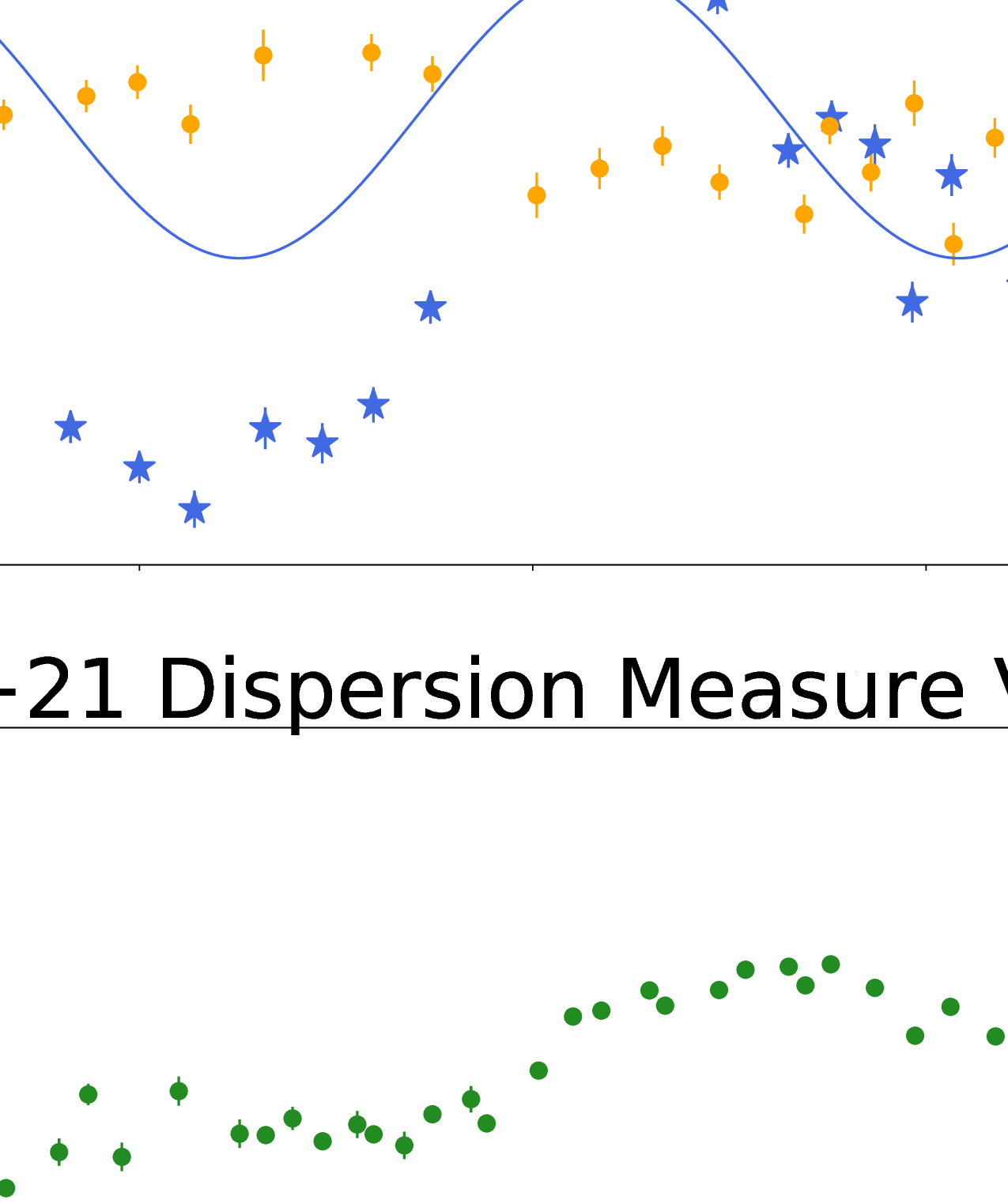}}} \ \ \
\hspace{-1.3 cm}
{\mbox{\includegraphics[height=70mm,angle=0]{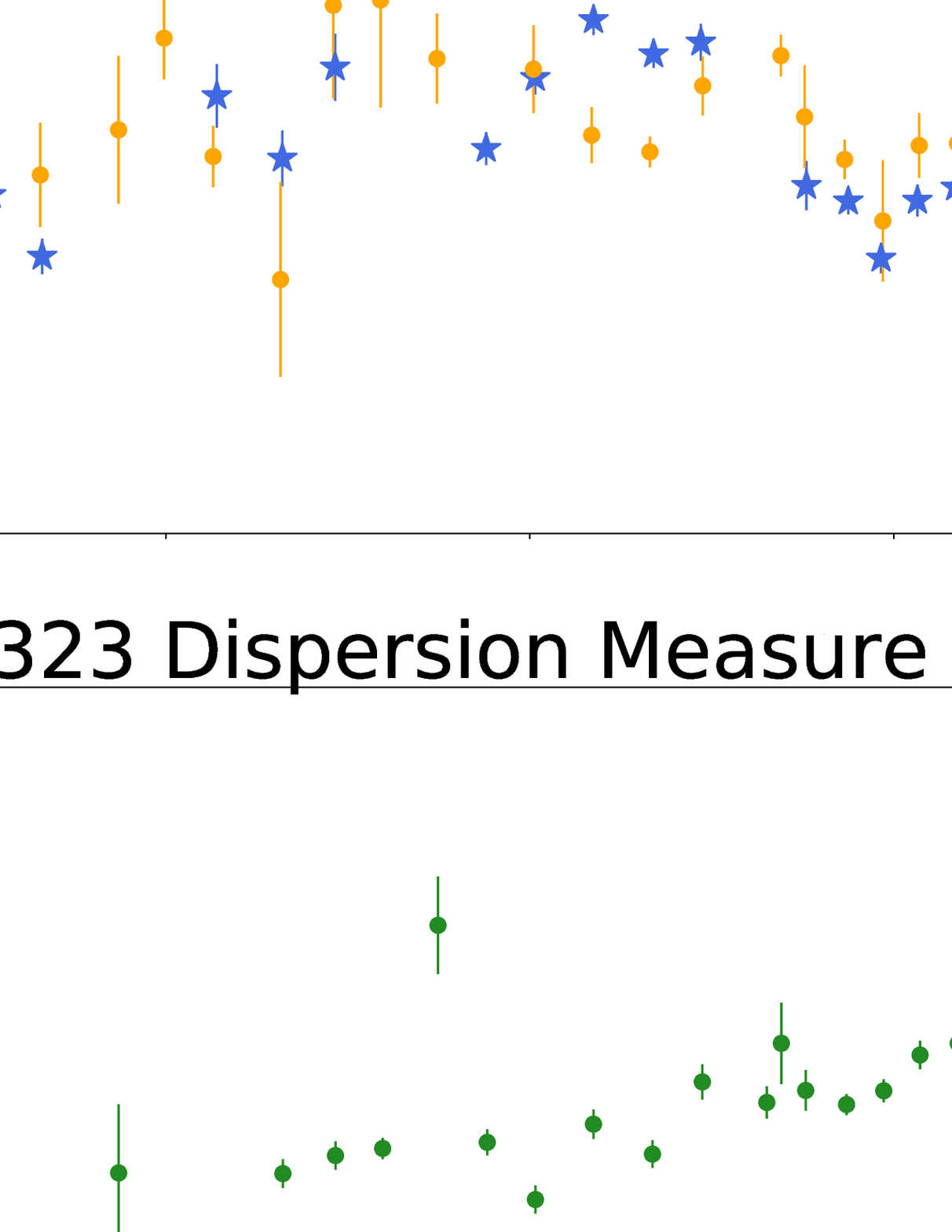}}}\\

\end{tabular}
\label{fig-20}
\end{center}
\end{figure*}

\begin{figure*}[ht]
\begin{center}
\caption{Dispersion measure, and magnetic field changes over time for pulsars J2145$-$0750 and J2302+4442. The uncertainties on the DM come from those on the DMX value and uncertainties on the magnetic field are a combination of the uncertainties on the ionosphere-corrected RM (which are a combination of those of fitting for Faraday rotation and from the ionospheric correction) and the DM. No trendlines are shown because the lowest $\chi^2_{r}$ value for the fits was that of a horizontal line with a slope of zero.}
\begin{tabular}{@{}ll@{}}
\hspace{-1 cm}
{\mbox{\includegraphics[height=70mm,angle=0]{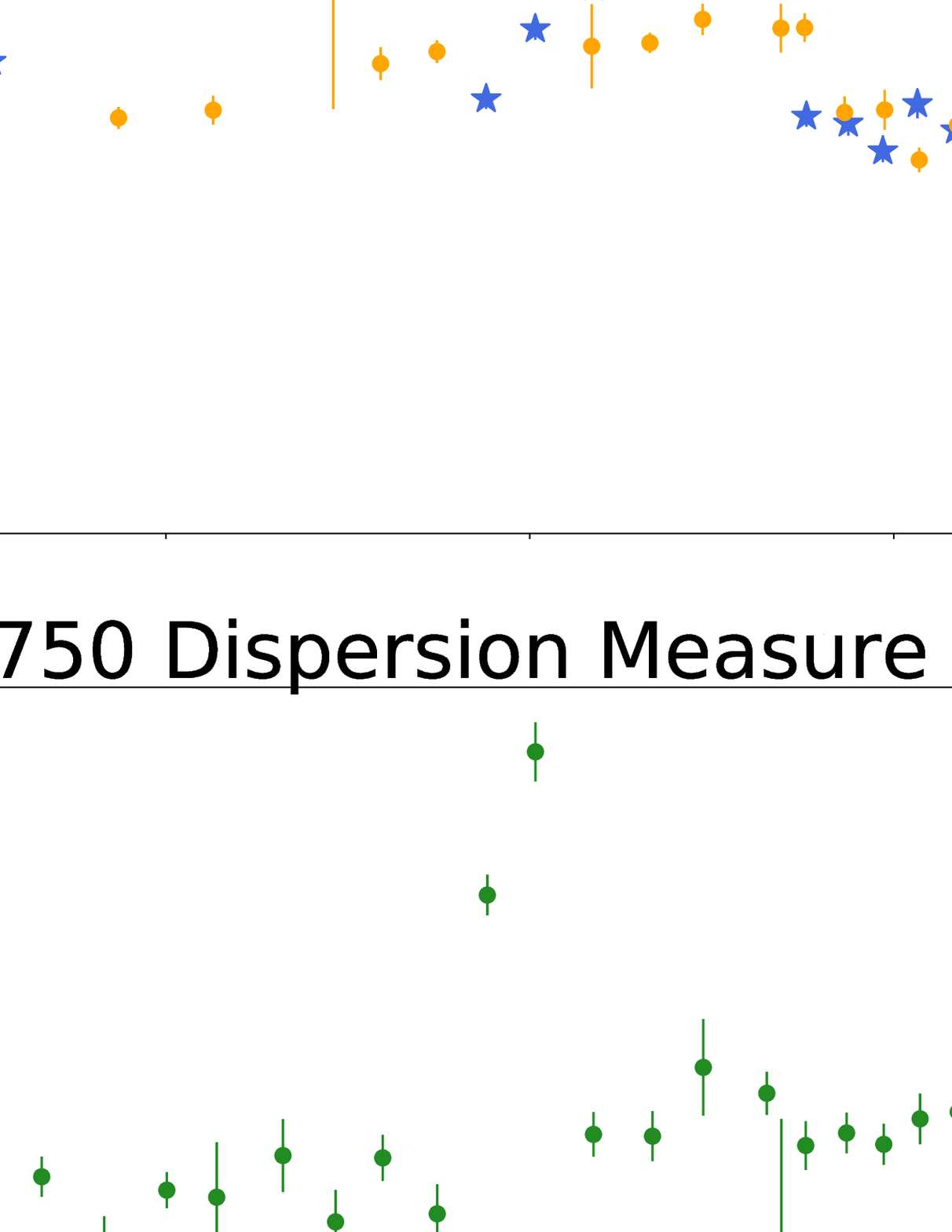}}} \ \ \
\hspace{-1.3 cm}
{\mbox{\includegraphics[height=70mm,angle=0]{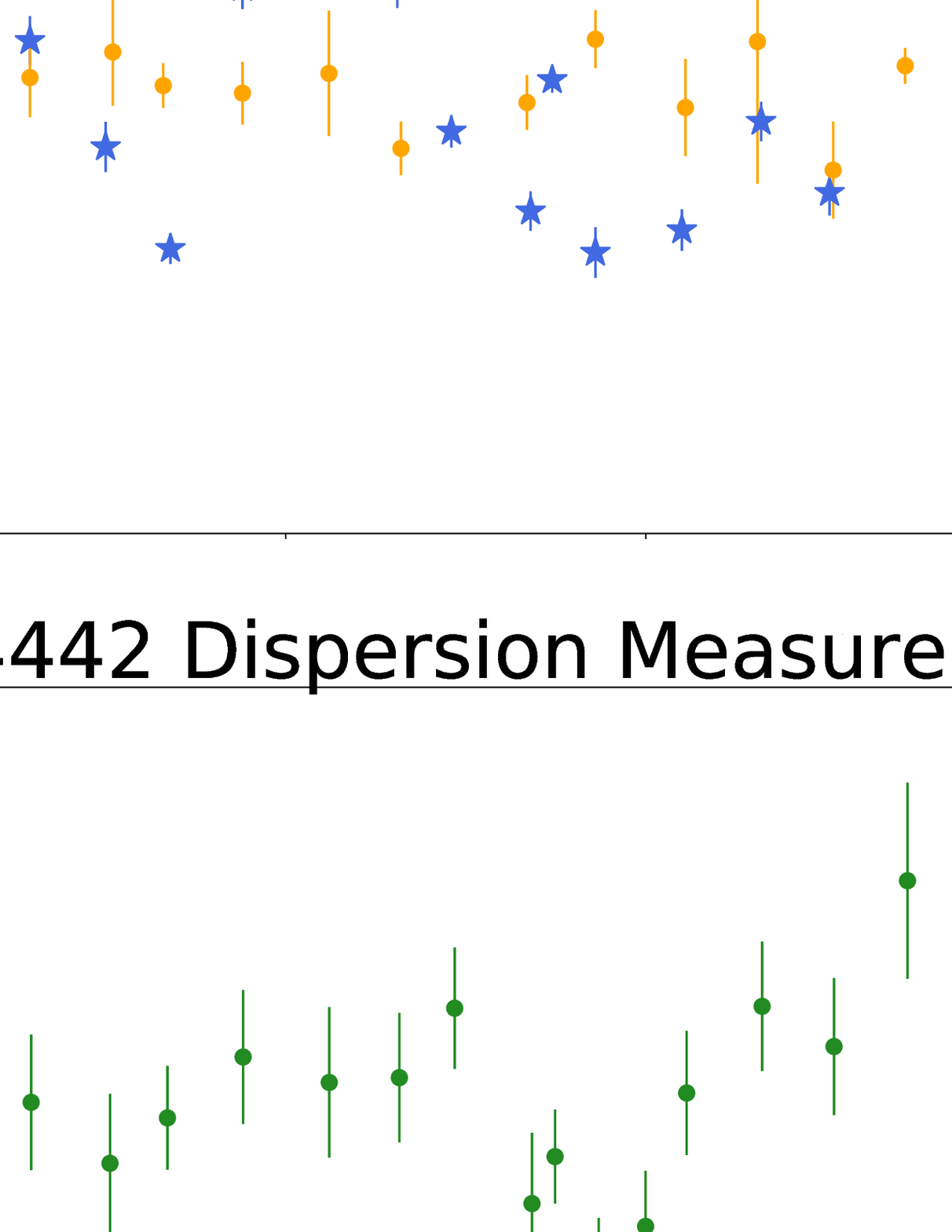}}}\\

\end{tabular}
\label{fig-21}
\end{center}
\end{figure*}

The magnetic field variations are shown in Table \ref{tbl-6} and the top panels of Figures \ref{fig-17}--\ref{fig-21}. We find five pulsars with significant trends (J1600$-$3053, J1643$-$1224, J1713+0747, J1918$-$0642, and B1937+21). Four pulsars show a trend with a sinusoidal component, and three of the periods are consistent with one year, the other with a period of almost 700 days. Periods consistent with one year point to either contributions from the solar wind or magnetized clumps of material along our line of sight to the pulsar. We only see two to three full periods in the data set, so these are likely due to stochastic processes and are not true periodicities. As previously noted, because of corrupted data after the sampler board switch, we used a maximum of four years of data for each pulsar.

Two pulsars, PSRs J1643$-$1224 and J1918$-$0642, show significant linear  trends in magnetic field. Assuming that this is due to movement along the line of sight through a region of increasing or decreasing Galactic magnetic field, we can use the timescale and slope of the trends to calculate the ambient magnetic field over the distance the pulsars have traversed over the timespan of these observations. Local magnetic fields of roughly 400~mG for PSR~J1643$-$1224 and 3200~mG for PSR~J1918$-$0642 would be required over the distances of roughly 100 $\mu$pc traveled by the pulsars over the timespan of our observations in order to produce the changes in average magnetic field observed. This is much larger than ambient and/or local magnetic fields expected in the Milky Way.

\citet{vano} measured the time variability of the RMs of PSRs B1556$−$44 and B1727$−$47 and found that local magnetic fields of 2 $\mu$G and 16 $\mu$G, respectively, were required to explain the observed variations. The latter was attributed to motion through irregularities within a nearby HII region. \citet{rankin1988} observed RM and DM variations towards the Crab pulsar for two years and calculated a local magnetic field of $\sim$170 $\mu$G, consistent with its dense magnetic environment. \citet{hamilton} observed another pulsar in a supernova remnant, the Vela pulsar, and found that the RM is increasing and found the magnetic field along the line of sight to be 22 $\mu$G, which was attributed to a magnetized cloud moving out of the line-of-sight to the source. Most recently, \citet{johnston2021} used the ultra-wideband on the Parkes radio telescope to observe pulsars over two years. They measured the RM and DM and found that PSR J1825$-$1446 showed significant RM and DM changes, with the magnetic field along the line of sight changing by 0.2 $\mu$G in 2 years, which is due to the pulsar passing behind a magnetised filament in a supernova remnant.

Our ambient magnetic fields are much larger than any measured values, including the 1 mG fields sampled by PSR B1959$-$63 as it travels through the disk of its companion star \citep{johnston2005}. This shows that the linear trends in magnetic fields we observe are much too large to be explained due to pulsar movement solely through an over-dense region along our line of sight, and are more likely due to our line of sight traversing variations in Galactic magnetic field structure in the transverse direction.

\citet{Yan2011RM} point out similarly large ($\sim$0.1 mG) derived local magnetic fields for pulsars for which they measure linear changes in Galactic magnetic field with time (specifically PSRs J0613$-$0200, J1909$-$3744, and J2129$-$5721). Their slopes, however, are one to two orders of magnitude larger than ours. They use a different technique, relying on the slope of the RM divided by the slope of the DM to calculate the ambient magnetic field.

Their method, along with that of \citet{hamilton}, \citet{rankin1988}, and \citet{vano}, assumes that the entire change in magnetic field is due to a small clump of material  with a discrete RM and DM contribution into our line of sight, and does not account for the pulsar's movement along the line of sight. Our equation calculates the ambient magnetic field assuming that the magnetic field changes are due to the pulsar moving closer or further away from us in a region of dense magnetic field.  If we make this assumption, the magnetic field for J1713+0747 (the only pulsar that shows a linear trend in both RM and DM) is 9 mG, which is more comparable to previous estimates but still large. 

\citet{Yan2011RM} point to statistical fluctuations due to random spatial and temporal variations in the interstellar electron density and $\langle B_{\parallel} \rangle$ to explain RM variations. Our numbers show that the magnetic field changes cannot entirely come from the motion of the pulsar through the interstellar medium.

\citet{you2012} explored the effects of the Sun on pulsar RM values by observing PSR~J1022+1001 when its line-of-sight passed close to the Sun. They found significant effects when the line-of-sight to the pulsar passed below 10\,R$_\odot$, which corresponds to $\sim$3$^{\circ}$ of elongation. 

We also checked the outliers for large changes in RM, DM, and B when the pulsars were close to minimum elongation. We found that PSR J1614$-$2230 experiences an increase in all three parameters when it came within 1.3$^{\circ}$ of the Sun (which corresponds to $\sim$4.5\,R$_\odot$). The increase in RM and DM at minimum elongation corresponds to a solar Galactic magnetic field contribution of 12(1) mG. This is consistent with \citet{you2012} and \citet{ord_bfield}, who report Galactic magnetic fields of the same order of magnitude at similar distances from the Sun.

\subsection{Correlations with Pulsar Spin-Down Parameters}

Studies such as \citet{params1} have examined the correlation between polarization fraction and spin-down parameters, but none have been conclusive. Using the wealth of polarization information in this study, we examine the relationship between fractional linear and circular polarization and five parameters: spin period, age, surface dipole magnetic field, spin down energy loss, and proper motion. We find no conclusive evidence of any correlations between these parameters and linear or circular polarization fraction 820\,MHz or 1500\,MHz. If a relation did arise, it would give information about the magnetosphere, pointing to the fact that MSPs, for instance, with different spin periods have different sized magnetospheres. However, our sample size is fairly small, covering only a small range of distances, inclination angles, and other parameters. A larger sample size is needed for this analysis for any definitive conclusions to be drawn.

\subsection{Timing Implications}

Effects of polarization calibration on timing have been explored in many studies in the past decade, including \citet{Desvignes2016}, \citet{manchester2013}, and \citet{cab2016}.  \citet{vanstraten2013} used matrix template matching to polarization calibrate PSR J1022+1001. They found that the RMS residuals decreased by a factor of two when polarization calibration was applied.

Pulsar time-of-arrival measurements calculated from data which have not been corrected for telescope polarization distortions, such as the NANOGrav 12.5-year data set \citep{12halfyr}, are susceptible to systematic timing uncertainties. These uncertainties will be higher for pulsars with larger polarization fractions (see Table \ref{tbl-3}). Correction of these data using the Mueller matrix formulation, as in the present paper, has the potential to improve the timing accuracy of such data sets. Also note that while incorrect polarization calibration could lead to higher levels of noise in the dataset, it would not show the spatial correlations expected for a gravitational wave signature.

\citet{rogers} analyzed the effect of different combinations of polarization calibration meth- ods on the Parkes Pulsar Timing Array (PPTA) data using three techniques: Scalar Template Matching (STM), Measurement Equation Template Matching (METM), and Matrix Template Matching (MTM). STM, which is NANOGrav’s method for calibrating profiles, models the transformation uses only the total intensity Stokes parameter. MTM, the method used in this paper, uses all four Stokes parameters to model the transformation between calibrated timing templates and the uncalibrated observations. METM, the method used by \citet{Gentile2018}, relies on a bright pulsar as a standard source and produces a template/Mueller matrix solution for each day by forcing the observation of that pulsar to look like the template, obtaining a solution for each day. The work also relies on the Ideal Feed Assumption (IFA), which assumes that the receivers are perfectly orthogonal, the reference source is 100\% polarized, and that the noise diode illuminates both receivers equally).

\citet{rogers} calculated the TOAs for five millisecond pulsars using data calibrated with combinations of these techniques: IFA/STM, IFA/MTM, METM/MTM, and METM/STM. Both the METM combined with MTM and IFA combined with MTM method resulted in significantly more precise and accurate TOAs and timing residuals with smaller amounts of red and white noise, with the METM/MTM showing slightly better improvement overall.  When compared to NANOGrav's method of IFA/STM, the combination of IFA/MTM used in \citet{rogers} improved the RMS of the post-fit residuals and the white noise residuals an average of 21\% and 48\% respectively, with a white noise residual improvement of above 60\% in two pulsars. 

Though METM produces TOAs with less red and white noise, it relies on the assumption that the pulsars do not have any intrinsic polarization variability. It also removes any sensitivity to variability in the pulsars used as templates. However this work indicated the IFA/MTM method is just as effective as MTM/METM.  Future work will apply the methods outlined in this paper to NANOGrav data to determine the effect of polarization-calibrated profiles on timing.

\section{Conclusions} \label{sec:conclusions}
In this work, we presented polarization-calibrated profiles for 23 millisecond pulsars timed by the NANOGrav collaboration, which represent the first published polarization profiles for nine pulsars. NANOGrav's high S/N observations allowed for the discovery of very low intensity average profile components (microcomponents) in four pulsars. These are the highest S/N polarization profiles ever published for these millisecond pulsars and are made publicly available to the community to facilitate sensitive modeling of MSP emission mechanisms and geometries. We found that our MSPs are consistent with previous studies in that they evolve and behave differently than canonical pulsars. 

We fit for Faraday rotation on each epoch and used the rotation measure and dispersion measure to calculate the magnetic field parallel to the line of sight of the pulsar. After fitting for a linear, sinusoidal, and sinudoisal + linear trend, we found a significant linear trend in three pulsars. Calculation of the ambient magnetic field produced large values on the order of microGauss, which showed that the magnetic field changes cannot be entirely due to the motion of the pulsar along the line of sight and must be due to transverse motion through the large-scale Galactic magnetic field structure. Recent literature shows that this method of polarization calibration is likely to greatly improve the timing precision of our pulsars, which will be examined in future work. 

These data only represent a portion of those obtained by the NANOGrav timing campaign. New ultra-wideband receivers on the GBT will provide more sensitivity. Also, the Canadian HI Mapping Experiment (CHIME) telescope will provide complementary frequency coverage to track how the polarization and microcomponents behave at lower frequencies.

\section{Acknowledgements}

We thank the double pulsar team (M. Kramer, N. Pol, R. Ferdman, A. Possenti, P. Freire) for sharing their B1929+10 data with us. We also thank Dr. James McKee for useful conversations regarding this work. This research was made possible by NASA West Virginia Space Grant Consortium, NASA Agreement \#80NSSC20M0055. This work was also supported by NSF Award OIA-1458952. The Green Bank Observatory is a facility of the NSF operated under cooperative agreement by Associated Universities, Inc. The NANOGrav project receives support from National Science Foundation (NSF) Physics Frontiers Center award number 1430284. Work on NANOGrav at NRL is supported by ONR 6.1 basic research funding. TD and ML acknowledge NSF AAG award number 2009468. KC is supported by a UBC Four Year Fellowship (6456).

\textit{Author Contributions}: HMW carried out the analysis and prepared the text, figures, and tables. MAM helped with the development of the framework and the text. PAG helped with the development of the framework. MLJ helped with the analyses and code. RS assisted with the analysis and the discussion. All authors contributed to the collection and analysis of the NANOGrav 12.5-year data set; see \citet{12halfyr} for further details.

\textit{Software}: numpy \citep{numpy1,numpy2}, matplotlib \citep{matplotlib}, PSRCHIVE \citep{PSRCHIVE_software}

\bibliography{MAIN_TEXT}{}
\bibliographystyle{aasjournal}

\newpage

\newpage

\newpage

\newpage

\newpage

\newpage

\newpage

\newpage

\newpage

\newpage

\newpage

\newpage

\newpage

\newpage

\newpage

\end{document}